\documentclass[12pt,a4paper]{article}
\usepackage{ragged2e}
\usepackage{subcaption} 
\usepackage{amsfonts}
\usepackage{amsmath}
\usepackage{array}
\usepackage{arydshln}
\usepackage{rotating}
\usepackage{amssymb}
\usepackage{bm}
\usepackage{geometry}
\usepackage{setspace}
\usepackage{graphicx}
\usepackage{lscape}
\usepackage{verbatim}
\usepackage{natbib}
\usepackage{xcolor}
\usepackage{calrsfs}
\usepackage{mathrsfs}
\usepackage{algorithm2e}
\usepackage{algpseudocode}
\usepackage{tablefootnote}
\usepackage{tabularx}
\usepackage{threeparttablex}
\usepackage{ragged2e}
\definecolor{winered}{rgb}{0.5,0,0}
\usepackage[bookmarks=true, bookmarksnumbered=true, allbordercolors={1 1 1}]{hyperref}
\hypersetup{
  colorlinks   = true, 
  urlcolor     = blue, 
  linkcolor    = blue, 
  citecolor    = winered,
}
\usepackage[open=true, numbered=true]{bookmark}
\usepackage{float}
\usepackage{fancyhdr}
\usepackage[toc,page]{appendix}
\usepackage{scalefnt}
\usepackage{afterpage}
\usepackage[left]{lineno}
\usepackage{caption}
\usepackage{subcaption}
\usepackage{varioref}
\usepackage{authblk}
\usepackage{cleveref}
\usepackage{amsthm}

\DeclareFontFamily{OT1}{pzc}{}
\DeclareFontShape{OT1}{pzc}{m}{it}{<-> s * [0.900] pzcmi7t}{}
\DeclareMathAlphabet{\mathpzc}{OT1}{pzc}{m}{it}

\DeclareMathOperator*{\argmax}{arg\,max}

\makeatletter
\def\blfootnote{\xdef\@thefnmark{}\@footnotetext}
\makeatother

\emergencystretch=\maxdimen
\theoremstyle{definition}

\usepackage{thmbox} 

\newtheorem[S, bodystyle=\normalfont\noindent]{defiS}{Definition}[section]
\newtheorem[M]{defiM}{Definition}[section]
\newtheorem[M]{defiL}{Lemma}[section]

\usepackage{calrsfs}
\DeclareMathAlphabet{\pazocal}{OMS}{zplm}{m}{n}

\newcommand{\mycirc}[1][black]{\Large\textcolor{#1}{\ensuremath\bullet}}

\definecolor{PastInflGrey}{rgb}{0.76863,0.76863,0.76863}
\definecolor{DomBCGreen}{rgb}{0.55686,0.86275,0.36863}
\definecolor{GlobBCGreen}{rgb}{0.33725,0.65882,0.56078}
\definecolor{EnergyOrange}{rgb}{0.94902,0.57647,0.2}
\definecolor{FoodOrange}{rgb}{0.93725,0.71373,0.16078}
\definecolor{MetalYellow}{rgb}{0.87843,0.84314,0.23137}
\definecolor{EBPBlue}{rgb}{0.28235,0.48627,0.86667}
\definecolor{PolicyBlue}{rgb}{0.18824,0.71373,0.94118}

\begin{document}

\title{\LARGE{Mixing it up: Inflation at risk}\thanks{
I would like to thank Sylvia Kaufmann, Sylvia Frühwirth-Schnatter, Hilde Bjørnland, Silvia Miranda-Agrippino, Dimitris Korobilis, Leif A. Thorsrud, Jim Griffin, Fabio Canova, Michele Lenza, Franceso Ravazzolo, Francesco Furlanetto, Domenico Giannone, Giorgio Primiceri, Herman van Dijk, Ørjan Robstad, Lennart Brandt, Frank Schorfheide, and participants at the following conferences for useful comments and discussions: 12th European Seminar on Bayesian Econometrics in Glasgow, 3rd Sailing the Macro Workshop in Siracusa, 3rd Dolomiti Macro Meeting in San Candido, Junior Workshop in Econometrics and Applied Economics in Rome, 31st SNDE Symposium 2024 in Padova, Workshop in Empirical Macroeconomics in Innsbruck. I would also like to thank seminar participants at the following institutions: BI Norwegian Business School; Norges Bank. 
The views expressed are those of the authors and do not necessarily reflect those of Norges Bank or any of the affiliated institutions. This paper is part of the research activities at the Centre for Applied
Macroeconomics and Commodity Prices (CAMP) at the BI Norwegian Business School.} }
\author[1]{Maximilian Schr\"{o}der} 
\affil[1]{{\footnotesize BI Norwegian Business School and Norges Bank}}

\date{\today}

\maketitle
\begin{abstract} \noindent 

Assessing the contribution of various risk factors to future inflation risks was crucial for guiding monetary policy during the recent high inflation period. However, existing methodologies often provide limited insights by focusing solely on specific percentiles of the forecast distribution. In contrast, this paper introduces a comprehensive framework that examines how economic indicators impact the entire forecast distribution of macroeconomic variables, facilitating the decomposition of the overall risk outlook into its underlying drivers. Additionally, the framework allows for the construction of risk measures that align with central bank preferences, serving as valuable summary statistics. Applied to the recent inflation surge, the framework reveals that U.S. inflation risk was primarily influenced by the recovery of the U.S. business cycle and surging commodity prices, partially mitigated by adjustments in monetary policy and credit spreads. 


\noindent 

\bigskip

\bigskip \medskip

\end{abstract}
\thispagestyle{empty} 

\newpage
\onehalfspace
\setcounter{page}{1}

\section{Introduction \& Motivation}

\noindent Contemporary monetary policy is forward-looking and relies on economic forecasts, which are inherently uncertain. This uncertainty has led policymakers to adopt risk management strategies \citep{greenspanRiskUncertaintyMonetary2004,kilianCentralBankerRisk2008}. However, managing risks becomes particularly challenging when large policy tradeoffs emerge, as exemplified by the recent inflation surge. Persistent supply chain disruptions, geopolitical tensions and threats to shipping routes in the black and red sea, and the post-lockdown recovery of demand intensified pressures on prices. Crucially, it remained unclear how much supply and demand side effects would contribute to future inflation and whether the individual drivers would prove transitory or persistent. Without this knowledge, however, policy makers faced the dilemma of whether to raise policy rates swiftly to address inflation risks derived from persistent price pressures or to proceed cautiously to avoid stalling the post-pandemic economic recovery by overreacting to mostly transitory forces. The uncertainty about the appropriate monetary policy stance fuelled vigorous policy debates, questioning the policy responses of central banks as inflation reached historic highs in the first half of 2021. This underscores the critical need for precise measures of macroeconomic risk and its drivers, serving as critical guides for effective policy-making especially in turbulent economic environments.



This paper contributes by introducing a unified framework for assessing macroeconomic risk, tailored to provide granular insights into the composition of risk as it arises. Without imposing restrictions on the shape of the forecast distribution, the new methodology allows to decompose the complete risk outlook into its underlying drivers and to attribute changes in probability mass within regions of the forecast distribution to economic variables. The entire risk outlook hence becomes interpretable, allowing new insights in research and policy making. 

To illustrate the framework, this paper investigates the role of key economic predictors as drivers of inflation risks during the recent high-inflation period in the U.S., with a particular emphasis on commodity prices and the U.S. and global business cycle. Empirically, this question connects to the literature on inflation risk, including \cite{korobilisTimeVaryingEvolutionInflation2021}, \cite{lopez-salidoInflationRisk2022}, \cite{tagliabracciAsymmetryConditionalDistribution2020} and to a series of recent studies that investigate the drivers of inflation during the recent inflation surge, by \cite{ballUnderstandingInflationCOVID2022}, \cite{blanchard2023caused}, \cite{haInflationPandemicWhat2021}, \cite{hallDriversSpilloverEffects2023}, and \cite{labelle2022global}. 

In the empirical exercises, this paper finds that post-pandemic inflation risk is primarily driven by increasing commodity prices and the recovery of the domestic U.S. business cycle. Additionally, monetary policy contributed mostly negatively during this period, mitigating some of the increase in inflation risk.

In practice, obtaining these results involves addressing three modeling challenges. These challenges are relevant to all research questions regarding the drivers of risk and serve as a solid foundation to introduce the proposed framework and highlight its contributions to various strands of the literature.


As a first challenge, the economic uncertainty literature provides ample evidence that economic aggregates have non-standard forecast distributions that are characterized by skewness, fat tails, and potential multimodality. For inflation, distributional asymmetries are documented by \cite{banerjeeInflationRiskAdvanced2020}, \cite{korobilisTimeVaryingEvolutionInflation2021}, \cite{korobilisQuantileRegressionForecasts2017}, \cite{lopez-salidoInflationRisk2022}, and \cite{tagliabracciAsymmetryConditionalDistribution2020}. Similarly, \cite{adrianVulnerableGrowth2019a} find negatively skewed forecast distributions of GDP, particularly during recessions. In this strand of the literature, quantile regressions (QRs) \citep{koenkerRegressionQuantiles1978}, which do not require parametric assumptions on the forecast distribution, have emerged as the workhorse tool. 

The framework proposed in this paper contributes by redirecting attention from specific percentiles to the forecast distribution as a whole, while maintaining the flexibility of QRs. In particular, the forecast distribution is estimated using a Gaussian mixture model, which is effective in capturing real-world data features such as multimodality, skewness, kurtosis, and unobserved heterogeneity \citep{fruhwirth-schnatterFiniteMixtureMarkov2006}. In the economics literature, mixture models have been applied by \cite{gewekeSmoothlyMixingRegressions2007} and \cite{villaniRegressionDensityEstimation2009} to model earnings and returns data and to study non-linearities in U.S. inflation. In this paper, the mixture model is specified as a density regression and leverages the modeling strategy in \cite{rigonTractableBayesianDensity2020a}. This specification results in a conditionally Gaussian structure and facilitates the implementation of fast estimation algorithms.

The second challenge lies in designing summary statistics that effectively capture the risk outlook and reflect the information contained in the complete forecast distribution. The macroeconomic uncertainty literature typically derives such risk measures directly from the predicted quantiles of economic variables. However, as this approach is quantile-specific, it potentially discards valuable information on macroeconomic risk contained in other parts of the distribution. This might be particularly crucial in a policy-making context. Additionally, in practice, central banks likely perceive larger deviations of economic aggregates from their targets as more costly. Therefore, a comprehensive summary statistic should not only accurately reflect the entire forecast distribution but also integrate the central bank's risk preferences \citep{kilianQuantifyingRiskDeflation2007,kilianCentralBankerRisk2008,machinaRisk1987}. 

Drawing from the finance literature, \cite{kilianQuantifyingRiskDeflation2007, kilianCentralBankerRisk2008} construct preference consistent risk measures based on probability-weighted central bank target deviations. They establish an equivalence between risk managing and utility maximizing central banks, grounding this approach theoretically and linking it to the macroeconomics and the central banking literature. However, \cite{kilianQuantifyingRiskDeflation2007, kilianCentralBankerRisk2008} obtain their probability weighting from models with symmetric error distributions that cannot capture the asymmetries highlighted by the macroeconomic uncertainty literature, potentially misrepresenting tail risks vital for policy considerations. Conversely, approximating the forecast distribution with a grid of independent quantile regressions, is likely prone to approximation error \citep{koenkerRegressionQuantiles1978, chernozhukovQuantileProbabilityCurves2010}. 

In contrast, because the density regression model in this paper yields the complete forecast distribution, this allows the construction of preference consistent risk measures without sacrificing the flexibility of quantile regressions. This paper hence effectively unifies the literature on macroeconomic uncertainty with the central banking and risk management literature.

The third challenge revolves around measuring the drivers of risk, integral for answering the research question. To study the drivers of risk comprehensively, changes in the shape of the entire forecast distribution have to be considered. However, a framework that attributes local changes in probability mass to economic predictors is so far missing in the literature. Standard time-series regression models and QRs, for example, only capture how changes in the predictor variables affect specific features of the forecast distribution, such as the conditional mean or quantiles. 

To close this important gap in the literature, this paper developments two algorithms that facilitate the interpretation of both the forecast distribution and the associated risk measures in terms of economic predictors. By exploiting the non-linearity of the mixture model, changes in the predictor variables can be directly linked to shifts in probability mass across the forecast density and to changes in the risk measures. In addition, rooted in game theory \citep{shapley1953value} and drawing inspiration from the explainable machine learning literature \citep{strumbelj2010efficient}, these algorithms are universal. They can be applied to interpret forecast densities and risk measures generated with any other non-linear, semi-parametric, or mixture model and are not constraint to the specific model in this paper.

In addition, this paper illustrates that once risk preferences are taken into account, the interpretation of the drivers of risk can no longer be considered independently. The theoretical results in the risk management literature hence not only apply to the risk measures themselves, but also extend to their underlying drivers.

The empirical section can be split into three main exercises. The first exercise decomposes the forecast distribution of inflation into its underlying drivers for the recent high inflation period, demonstrating the first algorithm. It reveals that rising commodity prices and the U.S. business cycle redistribute probability mass towards high inflation realizations, while monetary policy increases the likelihood of inflation events that more closely align with the policy target.  

The second exercise illustrates risk measures constructed from these forecast densities for different central bank preferences. Historically, the highest levels of inflation risk are attained during the 1970s and 1980s as well as during the recent inflation surge. Additionally, these risk measures differ notably from those derived using homoskedastic and heteroskedastic time series models, particularly for longer forecast horizons and during periods characterized by extreme deflation and inflation risk, such as the Volcker era, the recent sample, or recessions. 

The third exercise computes decompositions of these risk measures, illustrating the second algorithm. Overall, the U.S. and global business cycle are prominent drivers of downside risks, especially during recessions and the Covid-19 lockdown. Conversely, commodity prices drive upside risks during the first Gulf war, the energy price shocks of 2007 and 2008, the oil price shocks of 2011 and 2017, and the recent high inflation period. Monetary policy and credit spreads typically act as counter-cyclical drivers of risks, driving upside pressure during recessions and downward pressure during periods of high inflation risk. 

The remainder of the paper is organized as follows. The next section outlines the density regression framework, the estimation algorithm, and introduces methods to decompose the forecast density. Section 3 describes the data, presents a model validation exercise, and the decompositions of the forecast density. Section 4 introduces risk measures, the in-sample exercise, and methods for analyzing the drivers of risk, as well as the remaining empirical results. Section 6 concludes.

\section{Model and Forecast Density Decompositions}

\begin{figure}
\centering
\caption{Annualized Quarterly Inflation} \label{fig:density_fit_samples}
\vspace{-2mm}
\includegraphics[width=0.6\textwidth, trim={7cm 6cm 5cm 0cm}]{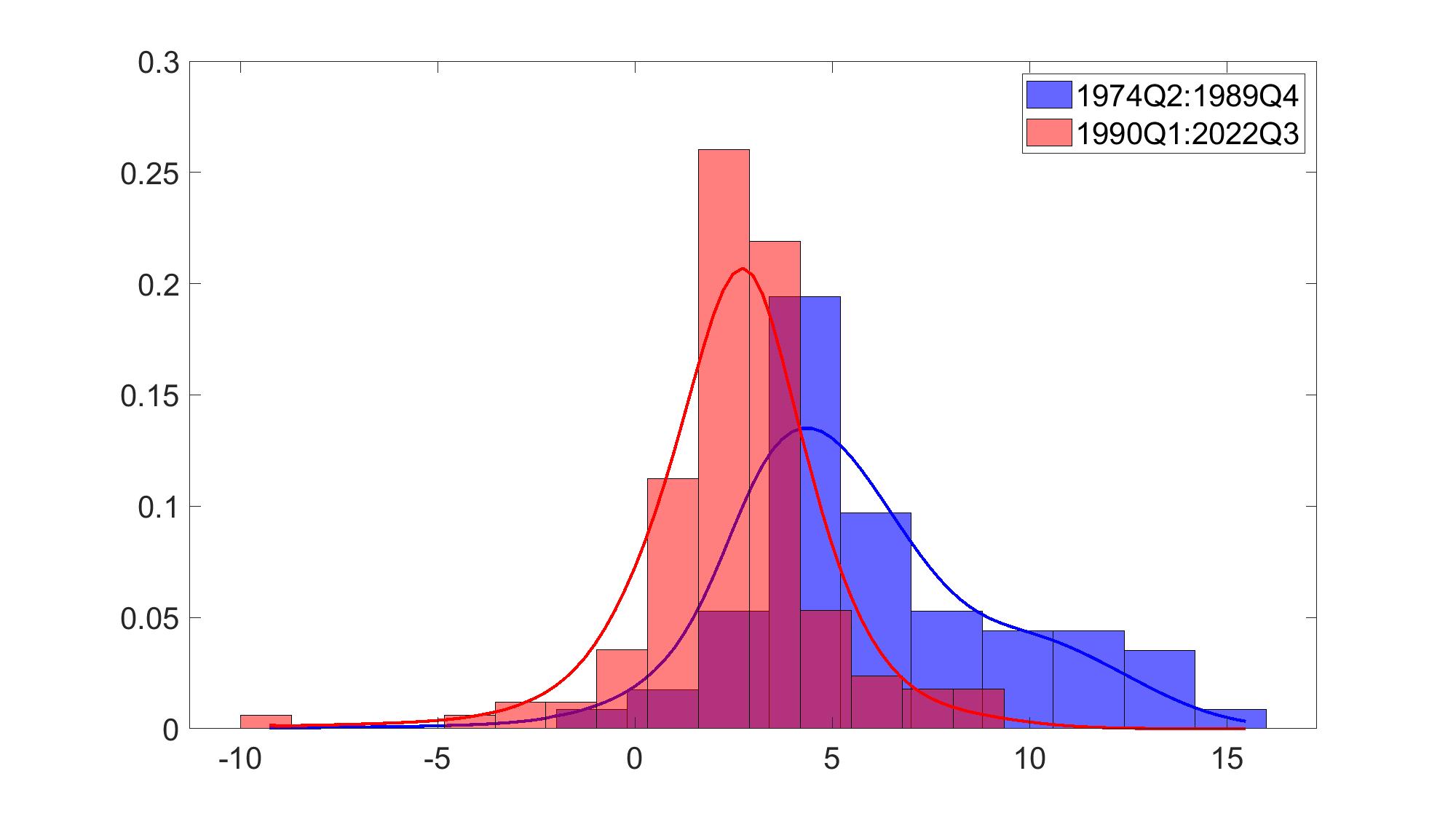}
\begin{flushleft}
\footnotesize \singlespacing \textit{Notes: The histograms display the observed values for annualized quarterly inflation from 1974Q1 to 1889Q4 (blue) and from 1990Q1 to 2022Q3 (red). The corresponding lines show the density estimates over these sub-samples, obtained with the density regression model estimated over the full sample with MCMC, as introduced in \autoref{sec:estimation}. }
\end{flushleft}
\end{figure}

As a starting point, \autoref{fig:density_fit_samples} shows histograms of the annualized quarterly inflation rate as well as fitted distributions over two sub-samples: from 1974Q1 to 1989Q4 and from 1990Q1 to 2022Q3. In both sub-samples the distributions are strongly skewed and fat-tailed, with varying characteristics that do not align with standard parametric families. Whereas this behaviour is readily observed for these two samples, the macroeconomic uncertainty literature provides ample evidence that the same characteristics generally apply to the conditional forecast distribution of economic aggregates. For inflation, asymmetries and skewness are documented by \cite{banerjeeInflationRiskAdvanced2020}, \cite{korobilisTimeVaryingEvolutionInflation2021}, \cite{korobilisQuantileRegressionForecasts2017}, \cite{lopez-salidoInflationRisk2022}, and \cite{tagliabracciAsymmetryConditionalDistribution2020} and for GDP by \cite{adrianVulnerableGrowth2019a} and others. In this literature, QRs have emerged as workhorse tools to forecast the conditional percentiles of economic variables. To capture the entire forecast distribution, the literature estimates a grid of quantile regressions and overlays a known parametric density, such as a skewed-t distribution \citep{lopez-salidoInflationRisk2022}. However, since the quantile regressions are estimated independently and the chosen target distribution may not match the true forecast distribution, e.g. if multimodality is present, this approach is susceptible to approximation error  \citep{koenkerRegressionQuantiles1978, chernozhukovQuantileProbabilityCurves2010}. In addition, this literature often provides risk measures in the form of extreme percentiles or interquartile ranges, discarding valuable information about the risk outlook that is contained in the remainder of the forecast density.

\subsection{Density Regression}

This paper takes a different approach by directly estimating the complete conditional forecast distribution, considering the non-standard characteristics highlighted in the macroeconomic uncertainty literature. In particular, it applies the density regression framework introduced in \cite{rigonTractableBayesianDensity2020a}, adapted to time series and economic settings. In addition, this paper expands upon both, the model and estimation algorithms, as detailed in \autoref{sec:priors} and \autoref{sec:estimation}, respectively. 

For an economic variable denoted by $y$, let the conditional forecast distribution for forecast horizon $h$, estimated with data up to point $t$, be given by 

$$
f(y_{t+h}) = \int K(y_{t+h}; \bm\theta)p(d\bm \theta),
$$

\noindent where $K(y_{t+h}; \bm \theta)$ denotes a parametric kernel, $\bm \theta$ collects the kernel specific parameters, and $p$ denotes a probability measure. Without loss of generality \citep{goodfellowDeepLearning2016}, this distribution can alternatively be represented as an infinite mixture model

$$
f(y_{t+h}) = \int K(y_{t+h}; \bm\theta)p(d\bm \theta) = \sum^{\infty}_{c=1} w_c  K(y_{t+h}; \bm \theta_c).
$$

To simplify inference and facilitate the integration of flexible prior distributions, this paper adopts a stick-breaking representation\footnote{The stick-breaking representation is an alternative representation of the Dirichlet process.} on $p$, as in \cite{ishwaranGibbsSamplingMethods2001}, where $w_c = \nu_c  \prod^{c-1}_{l=1}\left(1-\nu_l \right)$ and $w_1=\nu_1$, denote the stick breaking weights, for every $c\geq 1$.

In contrast to traditional time series models, data on economic predictors such as lags of $y$, the interest rate, or the state of the business cycle can be included in the individual kernels as well as the mixture component weights \citep[also see][]{gewekeSmoothlyMixingRegressions2007,villaniRegressionDensityEstimation2009}. Methodological advancements in this direction are introduced by \cite{deiorioANOVAModelDependent2004}, \cite{gelfandBayesianNonparametricSpatial2005}, and \cite{delacruz-mesiaSemiparametricBayesianClassification2007} and \cite{dunsonKernelStickbreakingProcesses2008}, \cite{griffinOrderBasedDependentDirichlet2006}, and \cite{gutierrezTimeDependentBayesian2016}, respectively. This yields the mixture representation  

\begin{equation}\label{mixturemodel}
f_x(y_{t+h}) = \int K_{\bm x}(y_{t+h}; \bm \theta)p_{\bm x}(d \bm\theta) = \sum^{\infty}_{c=1} w_c(\bm x_t)  K_{\bm x}(y_{t+h}; \bm \theta_c),
\end{equation}

\noindent where $w_c(\bm x_t) = \nu_c(\bm x_t) \prod^{c-1}_{l=1}\left(1-\nu_l(\bm x_t) \right)$.

Intuitively, the model bears similarity to a regime switching model with infinitely many regimes. Here, both the regimes and switching probabilities directly depend on economic predictors. This non-linear relationship links the shape of the forecast distribution directly to changes in the economic environment, without imposing any specific structural constraints. However, this flexibility is accompanied by substantial computational cost. Recently, \cite{rigonTractableBayesianDensity2020a} propose a model formulation building on the work of \cite{polsonBayesianInferenceLogistic2013}, \cite{renLogisticStickBreakingProcess2011},  \cite{rodriguesNonparametricBayesianModels2011}, and \cite{tutzSequentialModelsCategorical1991} that allows for the derivation of computationally simple and efficient estimation algorithms. The key innovation in \cite{rigonTractableBayesianDensity2020a} is to relate the stick-breaking weights, $\nu_c(\bm x_t)$, to a function of the covariates, $\bm x_t$, via the logistic link. This results in a Logit stick-breaking prior (LSBP).

This approach rests on three main results, which are restated here for convenience. The individual steps follow \cite{rigonTractableBayesianDensity2020a}. First, the stick-breaking prior admits a continuation-ratio parameterization as in \cite{tutzSequentialModelsCategorical1991}. Rewriting the model in Equation (\ref{mixturemodel}) in its hirarchical representation, yields 

\begin{equation}\label{sequentialproblem1}
(y_{t+h}|\bm G_t = c, \bm x_t) \sim K_{\bm x_t}(y_{t+h}; \bm \theta_c) \text{,   with    } \text{pr}(G_t=c|\bm x_t) = w_c(\bm x_t) = \nu_c(\bm x_t) \prod_{l=1}^{c-1} (1-\nu_l(\bm x_t)),
\end{equation}

\noindent for each $t = 1,\dots,T$, with $\bm \theta_c$ distributed independently for each mixture component, $c$. $G_t$ denotes a categorical variable that stores the mixture component allocation at time $t$ and has probability mass function $f(G_t| \bm x_t) = \prod_{c=1}^{\infty} w_c(\bm x_t)^{\mathbb{I}{(G_t = c) }} $, where $\mathbb{I}(\cdot)$ denotes the indicator function. This allows to rewrite the stick-breaking weights, $\nu_c(\bm x_t)$, as a function of the mixture probabilities $w_c(\bm x_t)$

\begin{equation}
\nu_c(\bm x_t) = \frac{w_c(\bm x_t)}{1- \sum_{l=1}^{c-1} w_h(\bm x_t)} = \frac{\text{pr}(G_t = c | \bm x_t)}{\text{pr}(G_t > c -1| \bm x_t)}, \text{ with } f\in \mathbb{N}.
\end{equation}

\noindent Now the stick-breaking weights, $\nu_c(\bm x_t)$, can be interpreted as the probability of being allocated to mixture component $c$, conditional on not being allocated to the previous $1,\dots,c-1$ mixture components \citep{rigonTractableBayesianDensity2020a}. The infinite mixture problem can hence be broken up into $c-1$ sequential binary choice problems. In the first step, a given observation is either allocated to the first component, $c=1$, with probability $\nu_1(\bm x_t)$ or any of the subsequent components, $c: h>1$, with probability, $1-\nu_1(\bm x_t)$. This procedure is then repeated for subsequent components and terminates once $G_t = 1$.

Second, one can now rewrite the continuation-ratio representation and introduce an assignment indicator $\zeta_{tc} = \mathbb{I} (G_t = c)$ given by

\begin{equation}
\zeta_{tc} = z_{tc}\prod_{l=1}^{c-1} (1-z_{tl})
\end{equation}

\noindent for each $t=1,\dots,T$. $z_{tc}\sim \text{Bern}(\nu_1(\bm x_t))$ is a Bernoulli random variable that encodes the decision of allocating a given observation to component $c$ or any of the subsequent ones. \cite{rigonTractableBayesianDensity2020a} introduce the logistic link into this sequential Bernoulli choice problem,  defining $\eta_c(\bm x_t) = logit(\nu_c(\bm x_t))=log(\nu_c(\bm x_t)/(1-\nu_c(\bm x_t)))$, and combining the result with Equation (\ref{sequentialproblem1}). This yields the logit stick-breaking prior on $w_c(\bm x_t)$

\begin{equation}\label{LSBP}
w_c(\bm x_t) = \frac{exp\left\{\eta_c(\bm x_t) \right\}}{1+exp\left\{\eta_c(\bm x_t) \right\}} \prod_{l=1}^{c-1}\left[ \frac{1}{1+exp\left\{\eta_c(\bm x_t) \right\}}\right],
\end{equation} 

\noindent where $\eta_c(\bm x_t)$ is the log-odds of the probability of observation $t$ being allocated to component $c$, given that it has not been allocated to any of the previous components. This approach enables the estimation of the parameters controlling the stick-breaking weights through a series of logistic regressions.

Importantly, the algorithms presented later in this paper for extracting the drivers of risk are not confined to this specific mixture model and can be universally applied alongside any flexible model that produces a forecast density. Nevertheless, this particular model formulation provides several advantages when studying the drivers of risk.

First, the stick-breaking formulation's sequential structure induces shrinkage. Redundant components unnecessary to characterize the data are automatically deleted, keeping the model parsimonious and computationally efficient. Specifically, because the above procedure terminates whenever $G_t = 1$, later components might never be reached and hence receive a zero probability. This avoids the need to estimate the number of mixture components. 

In addition, this formulation allows to leverage the results in \cite{polsonBayesianInferenceLogistic2013} that offer computational advantages in case of Bayesian logistic regression through Pólya-Gamma augmentation. Essentially, conditional on a Pólya-Gamma distributed variable $\omega_{t,h}\sim PG(1,\bm x_t'\bm \phi_c)$, the logistic regression collapses to a linear regression of the transformed assignment decision indicators $(z_{tc}-0.5)/ \omega_{t,h}$ onto the predictors $\bm x_t$ \citep{polsonBayesianInferenceLogistic2013,rigonTractableBayesianDensity2020a}. Compared to e.g. Metropolis–Hastings (MH) samplers, this approach is computationally fast, automatic, and requires no parameter tuning. These features make it suitable for large hierarchical models.

\subsection{Likelihood, Priors and Posterior}\label{sec:priors}

With the general framework in place, what remains is the specification of the likelihood function and priors. For the individual mixture components in \autoref{mixturemodel}, 

\begin{equation}
K_{\bm x}(y_{t+h};\bm \theta_c) = \sqrt{\tau_c}\phi\left(\sqrt{\tau_c}(y_{t+h}- \bm x_t \bm\beta_c)\right),
\end{equation} 

\noindent is assumed, where $\bm x_t =[x_{t,1},\dots,x_{t,n}]$ collects the constant, lags of $y$, and additional explanatory variables for $t=1,\dots,T$. $\phi$ denotes the Gaussian kernel, $\tau_c = \sigma_c^{-2}$ is the corresponding precision parameter, and $\bm\beta_c$ denotes the kernel specific regression coefficients. Correspondingly, \autoref{mixturemodel} is now given by

\begin{equation}
f_{\bm x}(y_{t+h}) = \sum^{\infty}_{c=1} w_c(\bm x)  \sqrt{\tau_c}\phi\left(\sqrt{\tau_c}(y_{t+h}- \bm x_t \bm\beta_c)\right).
\end{equation}

\noindent As a result, every mixture component is characterized by the same predictive relationship; however, the individual regression parameters are free to differ. The cluster allocation of each observation is governed by the categorical, $G_t$, which depends on the mixture probabilities, $w_c$, and hence the log-odds. Throughout the paper, it is assumed that $\eta_c(\bm x_t)$ depend linearly on $\bm x_t$. This yields
$\eta_c (\bm x_t) = \bm x_t'\bm \psi_c$.

In practice, the framework can accommodate different sets of predictors for the kernels and the log-odds. In forecasting applications for example it might be useful to let the mixture probabilities depend on a large set of covariates that broadly reflect the state of the economy, while keeping a parsimonious structure for the individual mixture components. However, akin to linear regression, logistic regression also suffers from the curse of dimensionality, if the number of covariates grows large. Likewise, it is beneficial to penalize the likelihood in out-of-sample forecasting applications to enforce parsimony. 

To address these issues, this paper extends the baseline model and imposes the horseshoe prior proposed in  \cite{carvalhoHorseshoeEstimatorSparse2010} on the parameters $\bm \psi_c$, where 

\begin{align}\label{horseshoe}
\bm\psi_c|\xi_c,\bar{\bm\tau }_{c} & \sim \prod_{j=1}^k N(0,\xi_c \bar{\tau}_{j,c}) \\
\xi_c & \sim C^+(0,1) \\
\bar{\tau}_{j,c} & \sim C^+(0,1), 
\end{align}

\noindent and $C^+$ denotes the half-Cauchy distribution. Under the horseshoe prior, the posterior remains almost unrestricted in small parameter spaces; however, as the parameter space grows large relative to the number of observations, increasing amounts of shrinkage are imposed. It is automatic and parameter tuning free, making it a default choice for many researchers  \citep{korobilisBayesianApproachesShrinkage2022}.  

Due to the conditionally normal form of the individual logistic regressions, the prior can be imposed without adjustments and sampling proceeds as in the standard regression case. The application of the horseshoe prior in the density regression context is novel. However, \cite{bhattacharyyaApplicationsBayesianShrinkage2022} and \cite{weiContractionPropertiesShrinkage2020} find that the favourable properties of the horseshoe prior carry over to Bayesian logistic regression. Because independent horseshoe priors are imposed for each mixture component, small clusters will be particularly regularized, alleviating small sample issues.

To complete the description of the priors, 
$\tau_c \sim IG(a_\tau, b_\tau)$, $\beta_c\sim N(0,1)$, and the LSBP in \autoref{LSBP} are imposed on the kernel specific precision parameter $\tau_c$, the kernel regression coefficients and the mixing probabilities $w_c(z_t)$, respectively. Finally, in application, the algorithm operates on a truncated version of the probability measure $p_x$ that serves as an approximation to the infinite mixture model. For a truncation point $C$, only the first $C-1$ components are modeled and $\nu_C=1$ for all $\bm x \in \mathcal{X}$, such that $\sum_{c=1}^Cw_{c}(\bm x) =1$. \cite{rigonTractableBayesianDensity2020a} show that this approximation is accurate, as long as the truncation point is set sufficiently large. The joint posterior of the model is hence given by 

\begin{align*}
p(\Theta|\bm y,\bm X) & \propto  \prod_{t=1}^{T} \left[ \prod_{c=1}^{C} \sqrt{\tau_c}\phi\left(\sqrt{\tau_c}({y}_{t+h}- \bm x_t \bm\beta_c)\right)^{\mathbb{I}(G_t=c)}  \prod_{c=1}^{C-1}   \nu_c(\bm x_t)^{\mathbb{I}(G_t=c)} (1-\nu_l(\bm x_t))^{\mathbb{I}(G_t>c)} \right]\\
&\phantom{=}  \cdot\prod_{c=1}^{C-1} p(\bm \psi_c | \xi_c,  \bar{\bm\tau}_c) p(\xi_c)\prod_{j=1}^k p(\bar{\tau}_{j,c})\cdot\prod_{c=1}^{C} p(\bm\beta_c) p(\tau_c),
\end{align*}

\noindent where $\nu_c(\bm x_t)= exp\left\{\bm x_t'\bm\psi_c \right\} /\left[ 1+exp\left\{\bm x_t'\bm\psi_c \right\} \right]$, $\bm X = [\bm x_1, \dots, \bm x_T]$, and $\Theta$ collects all model parameters. In all empirical exercises, $C=5$ is assumed. For readers interested in an automatic procedure, \autoref{app:mix_comp} in the Appendix details a simple algorithm for estimating $C$.

\subsection{Estimation Algorithm}\label{sec:estimation}
In this paper, the model is estimated with two complementary methods, MCMC and Variational Bayes (VB), extending the algorithms in  \cite{rigonTractableBayesianDensity2020a}. Whereas MCMC approximates the posterior distribution by sampling, VB solves an optimization problem, leading to significant gains in computational speed. However, VB provides an approximate solution, because this optimization problem is simplified by factorizing the posterior into independent variational families. In general, VB is thus particularly useful for big data applications, large scale forecasting exercises or real-time monitoring tasks.\footnote{For a general introduction to VB see \cite{bleiVariationalInferenceReview2017}.} 

In the context of this paper, VB offers an additional advantage. When applied to mixture models, MCMC algorithms can encounter label switching problems. If the interest is beyond the forecast distribution and in e.g. the mixture kernels' parameters, the posterior draws may require reordering. VB, however, converges on a single solution, eliminating the need for such post-processing.

In the remainder of the paper, results on the full sample are produced with MCMC, as computational demand is less of a concern. Conversely, VB is used in the forecasting exercises and to decompose the forecast distribution and risk measures. To provide evidence that the results are generally comparable, \autoref{app:comparison_MCMC_VB} in the Online Supplement shows inflation densities estimated with both algorithms. Notably, the results are very similar, indicating minimal sacrifice in accuracy when using VB for these applications. 

Both algorithms cycle through the following steps:

\begin{enumerate}
\item Sample the mixture indicator $G_t$ from a categorical distribution with probability $pr(G_t=c)$ and assign each observation to the corresponding mixture component. This step is implemented as in \cite{rigonTractableBayesianDensity2020a}.
\item Sample the parameters of the individual logistic regressions that govern the mixture component allocation, $\psi_c$. This step exploits the Pólya-gamma augmentation scheme for Bayesian logistic regression of \cite{polsonBayesianInferenceLogistic2013} as well as the continuation-ratio formulation introduced above. In particular, for each mixture component, $c=1,\dots,C-1$, subset the observations for which $G_t>c-1$ into two groups and define indicator variable $\bar{z}_{t,c}$, such that $\bar{z}_{t,c}=1$, if $G_t = c$ and $\bar{z}_{t,c} = 0$, if $G_t>c$. Conditional on a draw of the Pólya-gamma distributed variable $\omega_{t,c}\sim PG(1,\bm x_t'\bm \psi_c)$ the likelihood contribution of observation $t$ is proportional to a Gaussian kernel for $(\bar{z}_{tc}-0.5)/\omega_{t,c}$. The mixture parameters, $\psi_c$, can now be updated exploiting standard results for Bayesian linear regression.
\item Sample the parameters of the horseshoe prior based on the hierarchical parametrization of \cite{makalicSimpleSamplerHorseshoe2016}. This parametrization has the advantage that it leads to conjugate posterior distributions and hence facilitates the implementation of fast and simple samplers.
\item Sample the kernel specific regression and precision parameters, $\bm\beta_c$ and  $\tau_c$, exploiting the results for standard Bayesian linear regression.
\end{enumerate}

Overall, the algorithm collapses to a sequence of Bayesian linear regressions and has a simple and user-friendly structure. The full algorithms, which are extensions of the algorithms in \citep{rigonTractableBayesianDensity2020a} and additional derivations are provided in \autoref{app:EstimationAlgorithms}. Throughout the paper, MCMC results are based on 1,000,000 draws, after discarding the initial 100,000 draws of the chain. Out of the remaining draws, only every 100th is retained, as consecutive draws might be strongly correlated. 

\subsection{Decomposing the Forecast Density}\label{sec:dens_decomps}

\begin{figure}[h]
\centering
\caption{examples of 1-step ahead forecast distributions}\label{fig:example_densities}
\subcaptionbox{2022Q2\label{dist2022q2}}
{\includegraphics[width=.42\textwidth, trim={7cm 1cm 5cm 2.9cm}, clip]{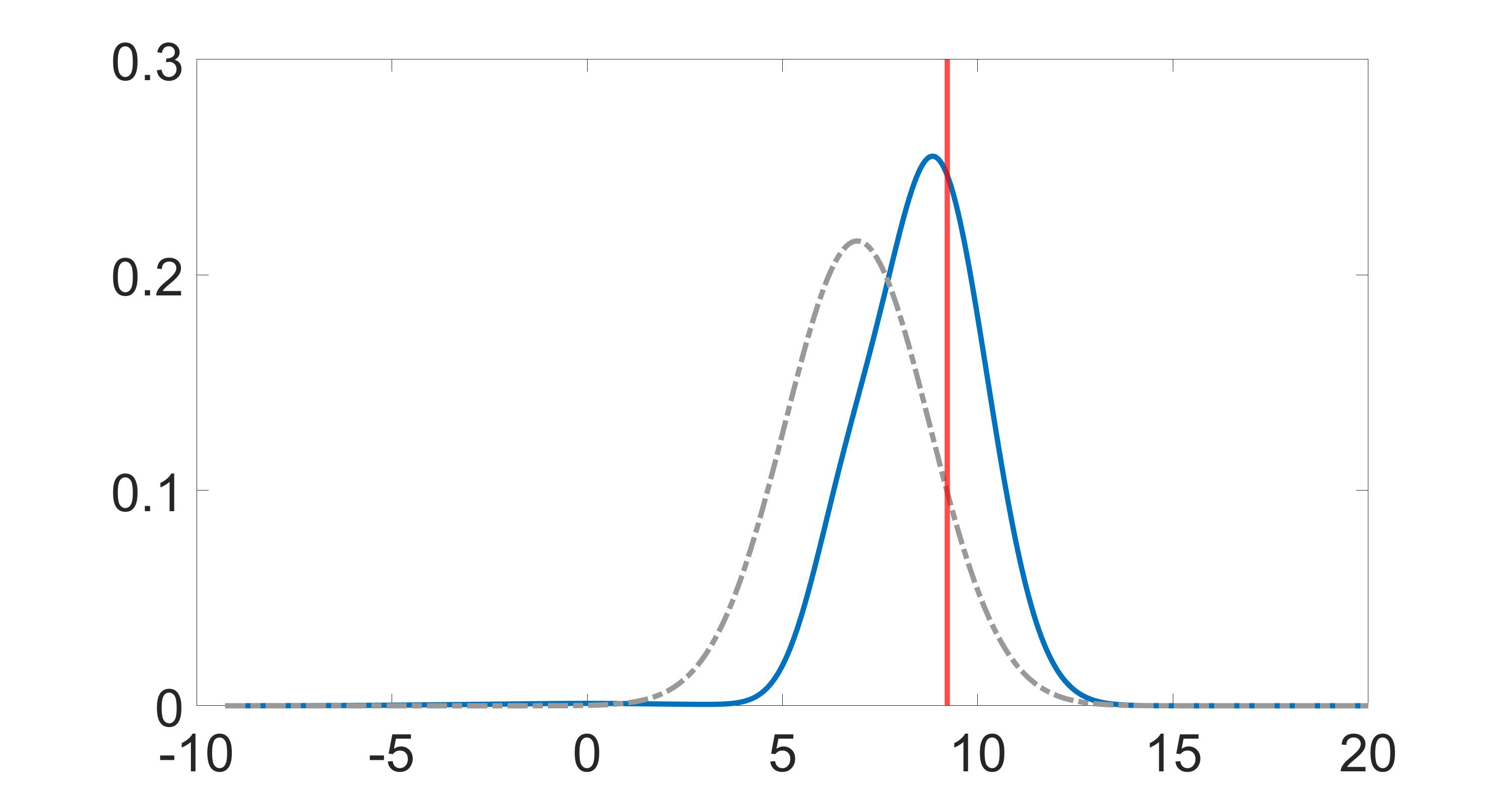}}
\subcaptionbox{2022Q3\label{dist2022q3}}
{\includegraphics[width=.42\textwidth, trim={7cm 1cm 5cm 2.9cm}, clip]{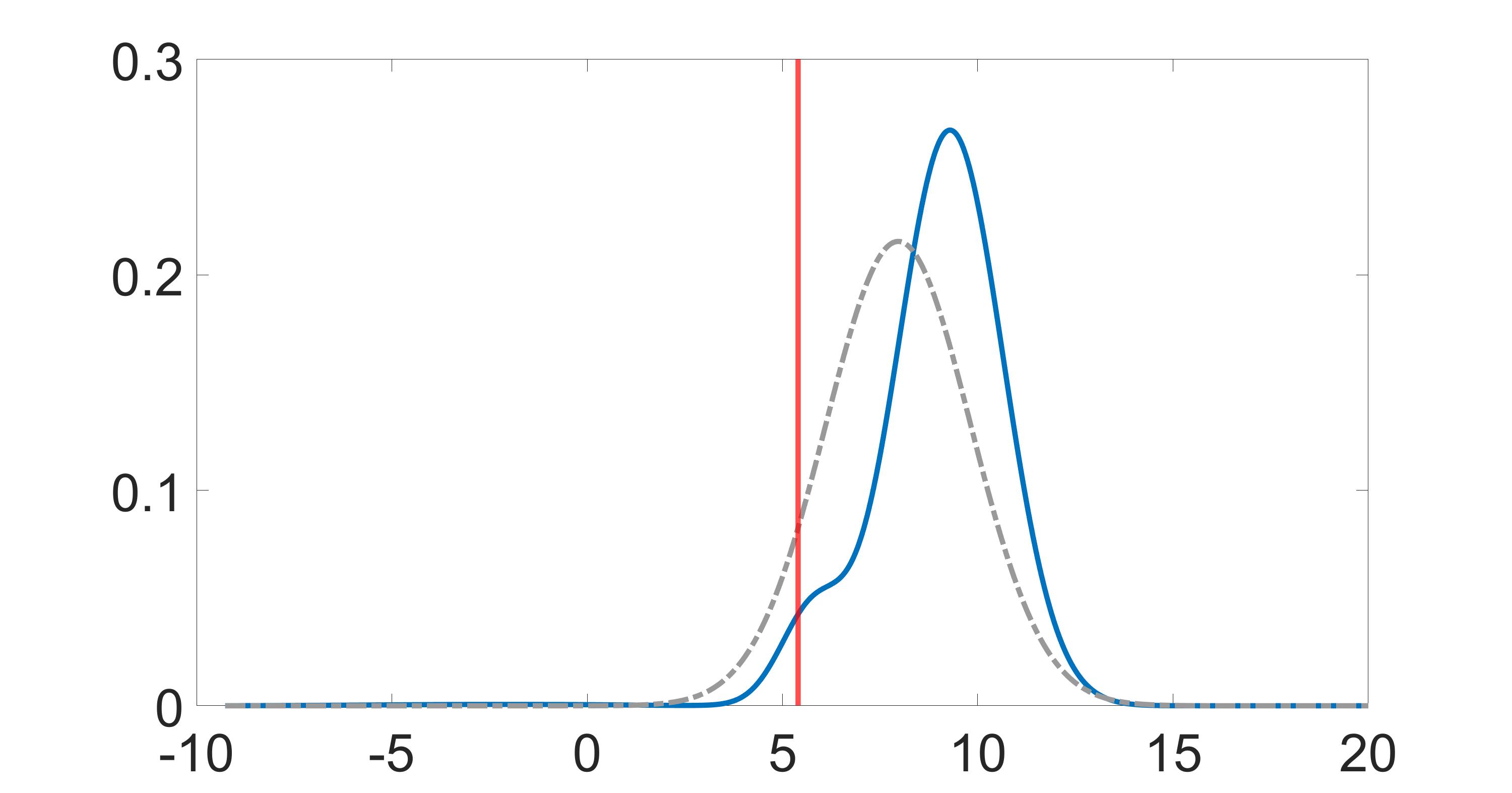}}
\begin{flushleft}
\footnotesize \singlespacing \textit{Notes: The panels depict the 1 quarter ahead forecast densities for 2022Q2, and 2022Q3 in blue, estimated with the density regression model introduced below, with data available at 2022Q1 and 2022Q2. The gray density is estimated with an AR model. Both models are estimated as specified in \autoref{sec:data_and_validation}. The red line indicates the realized value.}
\end{flushleft}
\end{figure}

Before introducing the remainder of the framework, it is illustrative to look at forecast densities generated by the baseline model, based on the specification that is later outlined in \autoref{sec:data_and_validation}. \autoref{fig:example_densities} displays one-step ahead out-of-sample forecast densities of the density regression (DR) model (blue) for the second and the third quarter of 2022, as examples. The gray density represents the corresponding forecast density of the AR model to make asymmetries in the DR forecast densities more apparent, while the red line indicates the realized value. 
Compared to the AR model, the forecast density of the DR model differs distinctly in both panels. In the second quarter of 2022, it is centered around the realized value, but is asymmetric with a pronounced left skew. In the third quarter of 2022, the left-skew morphs into a hump, indicating a higher probability of more moderate inflation realizations. While the density itself remains centered around high inflation, this hump is centered around the realized value.


A primary contribution of this paper is a framework that identifies the drivers of risk. This allows to link interesting features and changes in the shape of the forecast distribution, such as the emerging hump illustrated in \autoref{fig:example_densities}, to economic predictors. In addition, later sections show how all this information can be summarized with appropriate risk measures. 
Emergent upside and downside risk can thus be mapped directly to the economic environment. The drivers of risk hence become observables, providing more detailed insights into the economic outlook. Form a policy perspective, this allows to evaluate whether assumptions about the risk outlook are supported by the data.

Conventional methods lack the ability to provide such information. For instance, in QRs, variables affect specific conditional percentiles of the forecast distribution, making it difficult to understand their impact on the entire distribution. Additionally, QRs are usually estimated independently for different percentile levels, which may yield divergent results across percentiles that are hard to generalize across the distribution. Conversely, in regression models such as the GARCH models used in the risk management literature, variables only affect the conditional mean. Intuitively, one could discretize the forecast distribution and then attribute changes in mass in each probability bin to the variables based on their contribution to the conditional mean. However, because the dynamics of these models are not rich enough, the individual variables' relative contribution would remain constant across bins, lacking interpretabilty and empirical value. 

On the other hand, economically meaningful representations of local changes in probability mass are available with models that provide richer dynamics and generate more flexible forecast distributions, including the density regression model discussed in this paper. Specifically, because the mixing probabilities and the mixing kernels are functions of the data, the relationship between local probability mass and predictors becomes dynamic, allowing contributions to vary across bins. In addition, this also implies interaction effects between variables. For example, an increase in variable $x_j$ might lead to a change in the contribution of variable $x_i$, solely because the mixture weights change and hence the composition of the overall density. Based on this intuition, this paper defines a variable's contribution to a region of the forecast distribution as

\begin{defiM}[Contributions to the Forecast Distribution]\label{def:driver of cross sec risk}
The contribution of variable $x_j$ to the forecast distribution at time $t$ for horizon $t+h$ in the interval ${\pi}_{i-1}^\star$ to ${\pi}_i^\star$, is 

\begin{equation}
\varphi_j^{\pi_i^\star} = \int_{{\pi}_{i-1}^\star}^{{\pi}_{i}^\star} f_x(\pi)_{t+h|t}d\pi -\int_{{\pi}_{i-1}^\star}^{{\pi}_{i}^\star} f_x(\pi)_{t+h}d\pi  - \sum_{l\in [1,\dots,n] : \neq j} \varphi_l^{\pi_i^\star}.
\end{equation}

Then, $x_j$ is a driver of the forecast distribution, if $\varphi_j^{\pi_i^\star}\neq 0$.
\end{defiM}

Hence, the sum of all contributions, $\sum_l^n \varphi_l^{\pi_i^\star}$, in probability bin $\pi_{i-1}^\star$ to $\pi_{i}^\star$, is precisely equal to the difference between the probability mass of the forecast distribution and the probability mass of the average across all historical forecast distributions in that bin. This definition has key practical advantages: Firstly, unless the forecast distribution is degenerate, it will always have mass around the central tendency, and the predictors will always contribute mass to various bins. Expressing drivers or contributions as deviations from the historical average provides a simple normalization. This will also ensure that constants have zero contributions. Secondly, because the contributions must add up exactly, this implies efficiency. Thirdly, even though the model can be highly non-linear, the decomposition remains linear. Therefore, the contribution of a group of variables can be computed as the sum of the individual contributions. This greatly facilitates interpretation. Finally, with this definition, the individual contributions to the forecast distribution satisfy Lemma \autoref{def:lemma}.

\begin{defiL}\label{def:lemma}
Under definition \ref{def:driver of cross sec risk} and given $g$ grid points $\pi^\star$, such that $\pi_{i-1}^\star<\pi_i^\star$, the contributions of the individual predictors preserve integrability, i.e. 
\begin{equation*}
\sum_{i=2}^g \left[ \sum_j^n\varphi_j^{\pi_i^\star} + \int_{{\pi}^\star_{i-1}}^{{\pi}^\star_i} f_x(\pi)_{t+h}d\pi \right]=\int_{-\infty}^{\infty}f_x(\pi)_{t+h|t} dx = 1.
\end{equation*}
The proof is given in appendix \ref{app:proof}.
\end{defiL}

They are hence directly comparable across forecast horizons, time periods, and probability bins - an exclusive characteristic of the methods proposed in this paper.

\subsection{Computing Decompositions}\label{sec:dens_decomps_compute}

The main challenge in calculating the contributions lies in the highly non-linear nature of the forecast densities. Each variable affects both the mixture components and mixture weights. Consequently, computing the contributions involves approximating how each variable influences the output of these non-linear objects.\footnote{In addition, the risk measures introduced later are non-linear functions of these forecast densities, adding further complexity.}

The explainable machine learning literature proposes general approaches to address such cases. \cite{strumbelj2010efficient} reframe a forecasting model as a cooperative game. In this analogy, the variables contribute to the model's prediction, akin to players contributing to the payoff of a game: Adding a variable (player) to a model (game) that contains little additional information (contributes little additional skill) changes the prediction (pay-off) only slightly. The opposite is true for variables (players) that add a lot of information (skill). Once rephrased in this way, game theory provides established solutions in the form of Shapley values, pioneered by \cite{shapley1953value}, to attribute the prediction fairly to the variables. 

Let $\bm x$ denote a set of variables. Following \cite{strumbelj2010efficient} the Shapley value of variable $x_j$ at time $t$ for a model prediction $\hat{f}$ can be written as

\begin{equation}\label{shap}
\varphi_j(\bm x_t;) = \sum_{\bm x' \subseteq \pazocal{C}(x) \setminus\left\{k \right\}} \frac{|\bm x'|!(n-|\bm x'|-1)!}{n!}\left[ \hat{f}(x_j|\bm x'\cup \left\{k \right\}  )-\hat{f}(x_j|\bm x')\right], 
\end{equation} 

In this equation, $\bm x'$ denotes a subset of $\bm x$, $|\bm x'|$ denotes its cardinality, $\pazocal{C}(x) \setminus\left\{k \right\}$ is the set of all possible variable combinations excluding the $j^{th}$ variable, and $\frac{|\bm x'|!(n-|\bm x'|-1)!}{n!}$ is a combinatorial weight. Essentially, \autoref{shap} describes the weighted sum of the marginal contributions of variable $x_j$ to model prediction $\hat{f}$ in all possible coalitions of variables. In addition to the properties of linearity and efficiency discussed previously, Shapley values embody additional useful axioms, formally grounding the decompositions introduced above in theory. The dummy axiom implies that constants do not receive contributions, while monotonicity dictates that Shapley values do not decrease if a variable's contribution to the prediction remains unchanged \citep{strumbelj2010efficient}. 

In application, computing the contributions for all possible collocations at every point in time is computationally demanding and often infeasible. Therefore, \citep{strumbelj2010efficient} propose an approximation based on Monte Carlo sampling that provides accurate approximations under predictor independence. Building on this approach, this paper proposes a sampling algorithm for computing the contribution of variable $x_j$ to the forecast distribution following definition \autoref{def:driver of cross sec risk}. The full algorithm is given by \autoref{alg:shap_dens} in  \autoref{app:drivers}. Given the importance of computing contributions to risk in applied policy work, the algorithm is tailored for real-time monitoring and out-of-sample forecasting applications. In addition, it is applicable beyond the model proposed in this paper. For instance, \autoref{alg:shap_dens} can be employed to decompose the forecast density generated by any other flexible density forecasting model, such as other semi-parametric mixture models, with trivial adjustments.

For one time period $t$ and one variable $j$, the algorithm evolves as follows:

\begin{enumerate}
\item At given time period $t$, sample a random collocation, $S$, from $\pazocal{C}(x)$ and a random time period, $t^*$, from $1,\dots,t-1$. 
\item Replace the values in $\bm x_t$ for the variables in $S$ with the values in $\bm x_{t^*}$.
\item Create two vectors of observations. The first one contains $x_j$ at time $t$, the other one also has $x_{j,t}$ replaced with $x_{j,t^*}$.
\item Compute the probability mass for probability bins of the forecast density for both vectors. The difference yields the marginal contribution of $x_j$.
\item Repeat the procedure for a number of MC samples, $M$. The average of these marginal contribution yields an estimate of $\varphi_j^{\pi^\star}$.
\end{enumerate}

Repeating the algorithm at different time periods and for different variables, $x_j$, then yields the full sequence of contributions for all variables. In theory, the computations can also be repeated at different MCMC draws or simulated posterior draws in the case of VB, to yield estimates of credible intervals and standard errors. 

Notably, because the sampling-based algorithm does not explore all possible collocation-time-period combinations, in practice the contributions are not guaranteed to add up to the difference between the predicted balance of risk and the historical average. The latter is denoted by the black dashed lines in  \autoref{fig:drivers_density} and \autoref{fig:drivers_density_h4} in later sections, providing a direct test for the accuracy of the algorithm. Overall, the algorithm works well and deviations are mostly negligible.

\section{Drivers of the Forecast Distribution }\label{sec:data_and_validation}

Recent research examining the drivers of inflation during the Covid-19 pandemic, including studies by \cite{ballUnderstandingInflationCOVID2022}, \cite{blanchard2023caused}, \cite{haInflationPandemicWhat2021}, \cite{hallDriversSpilloverEffects2023}, and \cite{labelle2022global}, associate inflationary pressures with supply chain disruptions, the post-lockdown recovery of global demand, and fluctuations in commodity prices.

To showcase the methods introduced in this paper, they are applied to revisit these findings, but from the perspective of inflation risk, with a particular focus on the recovery of the domestic U.S. business cycle, the global business cycle, and commodity prices. 

This section introduces the data used throughout all subsequent sections and presents a forecasting exercise to empirically support the modeling approach. The last subsection applies the methods introduced in \autoref{sec:dens_decomps} and \autoref{sec:dens_decomps_compute} to decompose the forecast density of inflation during the recent high inflation period.





  
 
\subsection{Data}



The baseline dataset consists of key U.S. and global variables and spans the period from 1974Q2 to 2022Q3. Throughout the analysis, inflation is measured by the annualized U.S. quarterly CPI inflation rate. While the potential predictors are numerous, only a small dataset is considered, to keep the exercises illustrative and tractable. Nevertheless, the dataset aligns well with the variables considered in \cite{hallDriversSpilloverEffects2023} and the Phillips curve and inflation forecasting literature, including \cite{koopForecastingInflationUsing2012a},  \cite{korobilisQuantileRegressionForecasts2017}, \cite{lopez-salidoInflationRisk2022}, and \cite{stockForecastingInflation1999,stockModelingInflationCrisis2010}. An overview is provided in \autoref{tab:data}. 

\begin{table}[H]
\caption{Data Overview\\}\label{tab:data}
\centering\resizebox{.9\textwidth}{!}{
\begin{threeparttable}
\begin{tabular}{llllc}\hline \hline
			\textsc{Macroeconomic Indicators}	&	 \textsc{Source} &	 \textsc{Mnemonics}	 &	 \textsc{Tcode}\tnote{1}	\\ \hline
Consumer price index for all urban consumers 	 & FRED & CPI & 4 \\
U.S. business cycle indicator		& FRED-QD\tnote{2} & BC$_d$ & 1 \\
Excess Bond Premium & \cite{gilchristCreditSpreadsBusiness2012a}\tnote{3} & EBP & 1\\ 
Federal Funds Rate & FRED & FFR & 1\\
OECD+6 industrial production & \cite{baumeisterStructuralInterpretationVector2019b}\tnote{4} & BC$_g$& 4\\
Energy price index & ``Pink Sheet'', The World Bank\tnote{5} & Energy & 4\\
Food price index & ``Pink Sheet'', The World Bank\tnote{5} & Food & 4\\
Metal price index & ``Pink Sheet'', The World Bank\tnote{5} & Metals & 4\\
 \hline\hline
\end{tabular}
\footnotesize\begin{tablenotes}
\item [1] Tcode identifies the time series transformation: level (1), log-differences (4).
\item [2] The data is taken form the FRED-QD database. The indicator is based on own computations.
\item [3] \url{https://sites.google.com/site/cjsbaumeister/datasets?authuser=0}.
\item [4] \url{https://www.worldbank.org/en/research/commodity-markets}.
\item [5] \url{https://www.federalreserve.gov/econres/notes/feds-notes/ebp_csv.csv}.
\end{tablenotes}
\end{threeparttable}
}
\end{table}

The domestic U.S. business cycle measure is derived from the FRED-QD database, a comprehensive source covering various economic indicators. Developed by \cite{mccrackenFREDQDQuarterlyDatabase2020c}\footnote{The database spans a large set of economic indicators that cover 14 broad groups - NIPA; industrial production; employment and unemployment; housing; inventories, orders, and sales; prices; earnings and productivity; interest rates; money and credit; household balance sheets; exchange rates; others (uncertainty and expectations); stock markets, non-household balance sheets; - and has found wide applications in economic research.}, this database spans indicators across 14 broad groups, facilitating extensive economic research. To capture future inflation trends effectively, the activity factor is extracted using partial least squares for one-step ahead inflation. Prior to extraction, broad aggregates are excluded as recommended by \cite{stockDisentanglingChannels200720092012a}, and the data is transformed to ensure stationarity.

In addition, the global industrial production indicator, developed by \cite{baumeisterStructuralInterpretationVector2019b}, is included. Encompassing the OECD and six other major economies, it represents about 75\% of the IMF World Economic Outlook estimate of global GDP \citep{baumeisterStructuralInterpretationVector2019b}, thus serving as a valuable proxy for global business cycle dynamics and global demand.

Commodity prices are represented by the indices for energy, food, and metal prices sourced from the World Bank's ``Pink Sheet'' dataset. Notably, the metal price index, also functions as an indicator for the cost of industrial raw materials, offering initial insights into producer price pressures.

In addition to these core indicators central to the study, a few additional controls and variables are included. The macroeconomic uncertainty literature often highlights the significance of financial conditions measures \citep{banerjeeInflationRiskAdvanced2020,korobilisTimeVaryingEvolutionInflation2021,lopez-salidoInflationRisk2022,tagliabracciAsymmetryConditionalDistribution2020}. To this end, the excess bond premium (EBP), constructed by \cite{gilchristCreditSpreadsBusiness2012a}, is included in the data. \cite{gilchristCreditSpreadsBusiness2012a} demonstrate that the EBP is particularly informative about future economic activity, making it a forward-looking indicator that captures macro-financial linkages. Notably, in comparison to broad financial condition indices such as the NFCI, the EPB is constructed entirely from financial market data and hence less likely to capture information already contained in the macroeconomic variables \citep{reichlinFinancialVariablesPredictors2020}.

The last indicator is the federal funds rate, which serves as a measure of interest rates and monetary policy. In forecasting exercises the FFR is included without modifications. However, due to the various factors influencing interest rates beyond central bank decisions, relying solely on the FFR often leads to price or output puzzles in applied research. Similarly, in this paper, understanding the impact of interest rate changes on inflation risk requires a more nuanced approach. To address this, a monetary policy proxy is constructed from high-frequency information by re-running the Proxy-SVAR in \cite{miranda-agrippinoUnsurprisingShocksInformation2016} on an updated sample. The resulting structural shocks then replace the FFR in corresponding exercises.


All predictors are standardized before estimation. Apart from these transformations, the data remains unaltered and is not adjusted for pandemic outliers.

\subsection{Forecasting Exercise}

For the reasons detailed in \autoref{sec:dens_decomps}, forecasting models that yield symmetric forecast distributions, including auto-regressive (AR) models with stochastic volatility (SV) and time-varying parameters (TVP), lack the necessary flexibility to capture the drivers of risk in empirically meaningful ways. Therefore, the flexibility provided by the density regression model is essential for addressing research questions concerning the drivers of risk. Since forecast performance is hence not crucial for the remainder of the paper, the goal of this forecasting exercise is to assess whether the non-standard forecast densities are as well calibrated as those of simple benchmarks, thereby offering additional validation of the modeling approach.

To serve as benchmark models, a suite of univariate models commonly employed in economic forecasting exercises is included: a simple AR, SV-AR, TVP-AR, TVPSV-AR, AR with t-distributed residuals (T-AR), and a QR model. Each model, along with the density regression model, incorporates a constant, four lags of the annualized quarterly inflation rate, and the entire set of predictor variables mentioned above. The information set hence remains consistent across all models.

For the reasons outlined previously, all models are estimated with VB.\footnote{The entire exercise runs within a few hours on a laptop with Intel Core i5-8265U CPU with 1.6GHz without parallelization.} The estimation algorithms can be obtained as special cases of the Variation Bayes algorithm for the density regression model, as outlined in \autoref{supp:Benchmark_models}. Derivations for TVP-AR and SV-AR models draw from \cite{koopBayesianDynamicVariable2023}, while the AR model with t-distributed residuals follows \cite{christmasRobustAutoregressionStudentt2011}. Additionally, computations for the QR model are analogous to \cite{Limetal2020}.
For all models, estimation begins with 50\% of the total sample, i.e. data from 1974Q2 to 1996Q4. For each vintage, direct forecasts are then generated for horizons ranging from 1 to 12 quarters ahead. Subsequently the sample is extended with one more observation at the end of the sample, the models are re-estimated, and a new set of forecasts is generated. For all horizons, the last forecast is computed for 2023Q3.

\begin{table}[h]
\caption{p-values for tests on PITs\\}\label{tab:dens_body}
\centering\resizebox{1\textwidth}{!}{
\begin{threeparttable}
\begin{tabular}{lccccccc|ccccccc}\hline \hline
& \multicolumn{3}{c}{Uniformity} & \multicolumn{2}{c}{Identical} & \multicolumn{2}{c}{Independent}& \multicolumn{3}{c}{Uniform} & \multicolumn{2}{c}{Identical} & \multicolumn{2}{c}{Independent} \\
 & KS & AD & DH &  1$^{st}$ &  2$^{nd}$ &  1$^{st}$ &  2$^{nd}$& KS & AD & DH &  1$^{st}$ &  2$^{nd}$ &  1$^{st}$ &  2$^{nd}$\\
 & \multicolumn{7}{c|}{$H=1$}  & \multicolumn{7}{c}{$H=2$} \\
DR	&	\textbf{0.751}	&	\textbf{0.703}	&	\textbf{0.208}	&	\textbf{1.000}	&	\textbf{0.628}	&	\textbf{0.978}	&	\textbf{0.752}	&	\textbf{0.432}	&	\textbf{0.366}	&	\textbf{0.154}	&	\textbf{0.239}	&	\textbf{0.119}	&	\textbf{0.175}	&	\textbf{0.951}	\\
AR	&	\textbf{0.556}	&	\textbf{0.375}	&	\textbf{0.086}	&	\textbf{0.562}	&	\textbf{0.114}	&	\textbf{0.730}	&	0.010	&	\textbf{0.246}	&	\textbf{0.397}	&	0.001	&	\textbf{0.847}	&	\textbf{0.422}	&	0.005	&	0.000	\\
TVP-AR	&	0.046	&	0.001	&	\textbf{0.683}	&	\textbf{0.155}	&	0.000	&	\textbf{0.467}	&	0.000	&	0.021	&	0.000	&	\textbf{0.874}	&	\textbf{0.818}	&	\textbf{0.383}	&	\textbf{0.178}	&	0.001	\\
SV-AR	&	\textbf{0.562}	&	\textbf{0.391}	&	\textbf{0.115}	&	\textbf{1.000}	&	\textbf{1.000}	&	\textbf{0.538}	&	\textbf{0.669}	&	\textbf{0.485}	&	0.007	&	0.003	&	\textbf{0.874}	&	\textbf{0.347}	&	0.008	&	\textbf{0.070}	\\
TVPSV-AR	&	0.000	&	0.000	&	0.000	&	\textbf{1.000}	&	\textbf{1.000}	&	\textbf{0.124}	&	\textbf{0.124}	&	0.000	&	0.000	&	0.000	&	\textbf{0.858}	&	\textbf{0.858}	&	\textbf{0.473}	&	\textbf{0.473}	\\
T-AR	&	0.025	&	0.014	&	0.000	&	\textbf{0.657}	&	\textbf{0.208}	&	\textbf{0.585}	&	\textbf{0.310}	&	0.008	&	0.011	&	0.011	&	\textbf{0.361}	&	\textbf{0.064}	&	0.000	&	0.036	\\
QR	&	0.000	&	0.000	&	0.000	&	\textbf{1.000}	&	\textbf{0.740}	&	\textbf{0.428}	&	0.019	&	0.000	&	0.000	&	0.000	&	\textbf{0.843}	&	\textbf{0.450}	&	0.001	&	0.000	\\
[1ex]
 & \multicolumn{7}{c|}{$H=3$}  & \multicolumn{7}{c}{$H=4$} \\
DR	&	\textbf{0.280}	&	\textbf{0.363}	&	\textbf{0.484}	&	\textbf{0.136}	&	\textbf{0.103}	&	0.035	&	\textbf{0.860}	&	\textbf{0.316}	&	\textbf{0.407}	&	\textbf{0.741}	&	\textbf{0.775}	&	\textbf{0.591}	&	\textbf{0.213}	&	\textbf{0.359}	\\
AR	&	\textbf{0.070}	&	0.009	&	0.000	&	\textbf{0.835}	&	\textbf{0.350}	&	0.001	&	0.005	&	\textbf{0.071}	&	0.017	&	0.021	&	\textbf{1.000}	&	\textbf{0.712}	&	0.001	&	0.012	\\
TVP-AR	&	0.001	&	0.000	&	\textbf{0.570}	&	\textbf{0.441}	&	\textbf{0.895}	&	0.014	&	0.015	&	0.000	&	0.000	&	\textbf{0.364}	&	\textbf{0.096}	&	\textbf{0.310}	&	0.016	&	\textbf{0.167}	\\
SV-AR	&	\textbf{0.367}	&	0.004	&	0.048	&	\textbf{0.554}	&	\textbf{0.080}	&	0.025	&	0.000	&	\textbf{0.196}	&	0.000	&	\textbf{0.269}	&	\textbf{0.864}	&	\textbf{0.349}	&	0.010	&	\textbf{0.064}	\\
TVPSV-AR	&	0.000	&	0.000	&	0.000	&	\textbf{0.117}	&	\textbf{0.117}	&	0.019	&	0.019	&	0.000	&	0.000	&	0.000	&	\textbf{0.104}	&	\textbf{0.104}	&	0.007	&	0.007	\\
T-AR	&	0.001	&	0.005	&	0.004	&	\textbf{0.735}	&	\textbf{0.290}	&	0.000	&	0.046	&	0.000	&	0.002	&	0.000	&	\textbf{0.550}	&	\textbf{0.137}	&	\textbf{0.053}	&	\textbf{0.235}	\\
QR	&	0.000	&	0.000	&	0.000	&	\textbf{0.173}	&	0.038	&	0.001	&	0.002	&	0.000	&	0.000	&	0.000	&	\textbf{0.429}	&	\textbf{0.133}	&	0.004	&	0.009	\\
[1ex]
 & \multicolumn{7}{c|}{$H=8$}  & \multicolumn{7}{c}{$H=12$} \\
DR	&	\textbf{0.196}	&	\textbf{0.058}	&	\textbf{0.901}	&	\textbf{0.685}	&	\textbf{0.811}	&	0.043	&	\textbf{0.601}	&	\textbf{0.497}	&	\textbf{0.408}	&	\textbf{0.298}	&	\textbf{0.112}	&	0.046	&	0.000	&	0.000	\\
AR	&	0.000	&	0.000	&	\textbf{0.067}	&	\textbf{0.083}	&	\textbf{0.344}	&	0.002	&	0.000	&	0.000	&	0.000	&	0.005	&	0.000	&	0.000	&	0.000	&	0.000	\\
TVP-AR	&	0.000	&	0.000	&	0.047	&	0.019	&	0.018	&	\textbf{0.051}	&	\textbf{0.780}	&	0.000	&	0.000	&	0.028	&	0.000	&	0.000	&	0.000	&	0.000	\\
SV-AR	&	\textbf{0.416}	&	0.003	&	\textbf{0.328}	&	\textbf{0.442}	&	\textbf{0.175}	&	0.017	&	0.010	&	\textbf{0.597}	&	0.001	&	\textbf{0.234}	&	\textbf{0.190}	&	\textbf{0.104}	&	0.015	&	0.001	\\
TVPSV-AR	&	0.000	&	0.000	&	0.000	&	\textbf{0.322}	&	\textbf{0.322}	&	\textbf{0.380}	&	\textbf{0.380}	&	0.000	&	0.000	&	0.000	&	0.000	&	0.000	&	\textbf{0.293}	&	\textbf{0.293}	\\
T-AR	&	0.000	&	0.001	&	0.001	&	\textbf{0.394}	&	0.026	&	0.003	&	0.001	&	0.000	&	0.000	&	0.001	&	0.014	&	0.016	&	0.000	&	0.000	\\
QR	&	0.000	&	0.000	&	0.000	&	\textbf{0.430}	&	\textbf{0.799}	&	0.008	&	0.017	&	0.000	&	0.000	&	0.000	&	\textbf{0.162}	&	\textbf{0.058}	&	0.000	&	0.000	\\
\hline\hline
\end{tabular}
\footnotesize
\begin{tablenotes}
\item The table contains the p-values for several test statistics: KS=Kolmogorov-Smirnov, AD=Anderson-Darling, DH=Doornik-Hansen, Andrews, Ljung-Box. For the Andrews and Ljung-Box test, 1$^{st}$ and 2$^{nd}$ denote the results for the first and second moment, respectively. $H$ indicates the forecast horizon. Bold values mark p-values greater than 5\%.
\end{tablenotes}
\end{threeparttable}
}
\end{table}

The accuracy of the density forecasts is evaluated using probability integral transforms (PITs). PITs represent the probability of observing the realized value given the model's predictive density, i.e. $\text{pit}_{t+h|t}= \int^{y_{t+h}}_{-\infty}f_x(\pi)_{t+h|t}d\pi$. When forecast densities are correctly specified, PITs follow an independent and identically uniform distribution. \cite{rossiEvaluatingPredictiveDensities2014} offer a formal framework for testing these properties, which is applied in this paper using their algorithms. \autoref{tab:dens_body} presents the results.\footnote{Results for the test in \cite{berkowitzTestingDensityForecasts2001} are omitted, because they remained inconclusive.}

To assess uniformity, three test statistics are computed: the Kolmogorov-Smirnov (KS), Anderson-Darling (AD) \citep{andersonAsymptoticTheoryCertain1952,andersonTestGoodnessFit1954}, and Doornik-Hansen (DH) \citep{doornikOmnibusTestUnivariate2008} test. Compared to the KS, the AD test puts more weight on the tails of the distribution and is more powerful in smaller samples. In contrast, the DH test specifically tests for the absence of skewness and kurtosis,  operating on the inverse normal transformation of the PITs. For the first forecast horizon, uniformity is not rejected for the DR, AR, and SV-AR model. However, for all other horizons, only the DR model does not have uniformity rejected under all tests.

The next test examines whether the PITs have an identical distribution, implying that their moments must remain stable over time. Following \cite{rossiEvaluatingPredictiveDensities2014}, this property is evaluated by testing for structural breaks in the mean and variance, following the procedures in \cite{andrewsTestsParameterInstability1993}. For the DR, AR, and TVPSV-AR stability is only rejected for $h=12$. The SV-AR model is the only model for which stability is not rejected across all forecast horizons, while it is rejected for the TVP-AR at $h=1,8,12$, the T-AR at $h=8,12$, and the QR at $h=3$.

Finally, independence is assessed using the Ljung-Box test, following \cite{rossiEvaluatingPredictiveDensities2014}. Across the first two forecast horizons, independence is rejected for all models except the DR and TVPSV-AR model. At $h=4$ independence cannot be rejected for the DR and T-AR model, whereas independence is rejected for all models at $h=3$. For both $h=8$ and $h=12$ the  TVPSV-AR is the only model for which independence cannot be rejected.

Overall, the density regression model produces well calibrated density forecasts especially for horizons up to one year ahead. Additional results such as continuous ranked probability scores (CRPS) are provided in the Online Supplement in \autoref{supp:add_tables}. The overall ranking of the models is more mixed; however, the density regression model continues to outperform the majority of benchmark models, particularly at longer forecast horizons. Additionally, \autoref{supp:point_fore} in the Online Supplements provides RMSE and point forecast evaluations for interested readers. Non-negligible performance gains are observed for the density regression model at all forecast horizons, suggesting that accounting for asymmetries in the forecast distribution also improves point forecast performance.

\subsection{Forecast Distribution Decompositions}\label{sec:forecast_dist_decomp}

Having established the model's satisfactory density forecast performance, this section present the drivers of the forecast distribution. The results are computed with \autoref{alg:shap_dens}, based on $M=1000$ random collocation-time-samples. Notably, all exercises are conducted in pseudo-real-time. Therefore, the decompositions are computed out-of-sample and the algorithm is re-run at every vintage. Importantly, all results should be interpreted through the lens of predictions and not be viewed as a fully structural exercise, which is left for future research. In this context, the focus is on the contributions of the variables, abstracting from the constant and lags of inflation throughout.


\begin{figure}[h]
\hspace{-4cm}\centering
\caption{Drivers of the Forecast Distribution for $h=1$}\label{fig:drivers_density}
\subcaptionbox{2021Q4}
{\includegraphics[width=.49\textwidth, trim={3.5cm 2cm 5cm 1.5cm},clip]{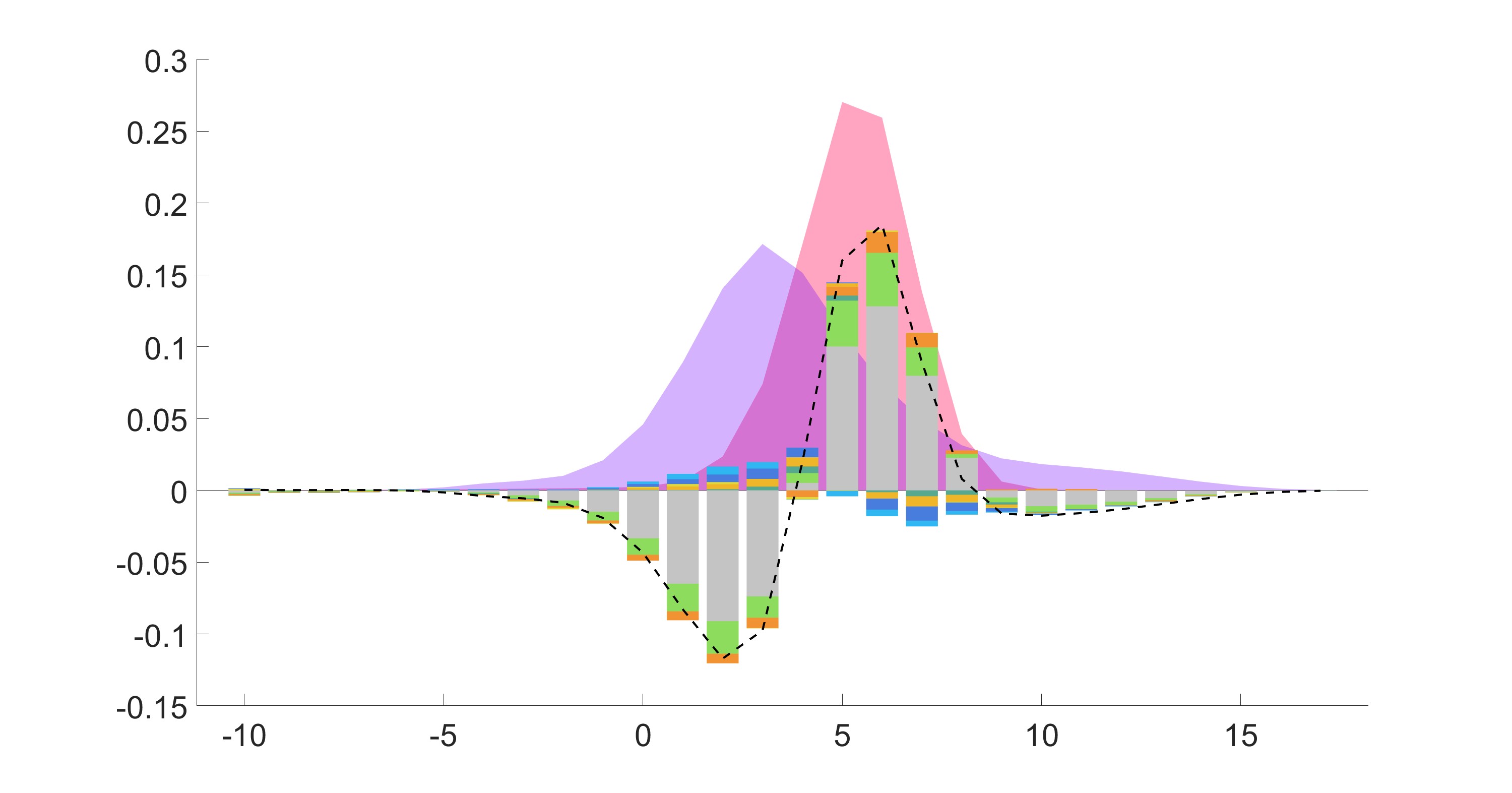}}
\subcaptionbox{2022Q1}
{\includegraphics[width=0.49\textwidth, trim={3.5cm 2cm 5cm 1.5cm},clip]{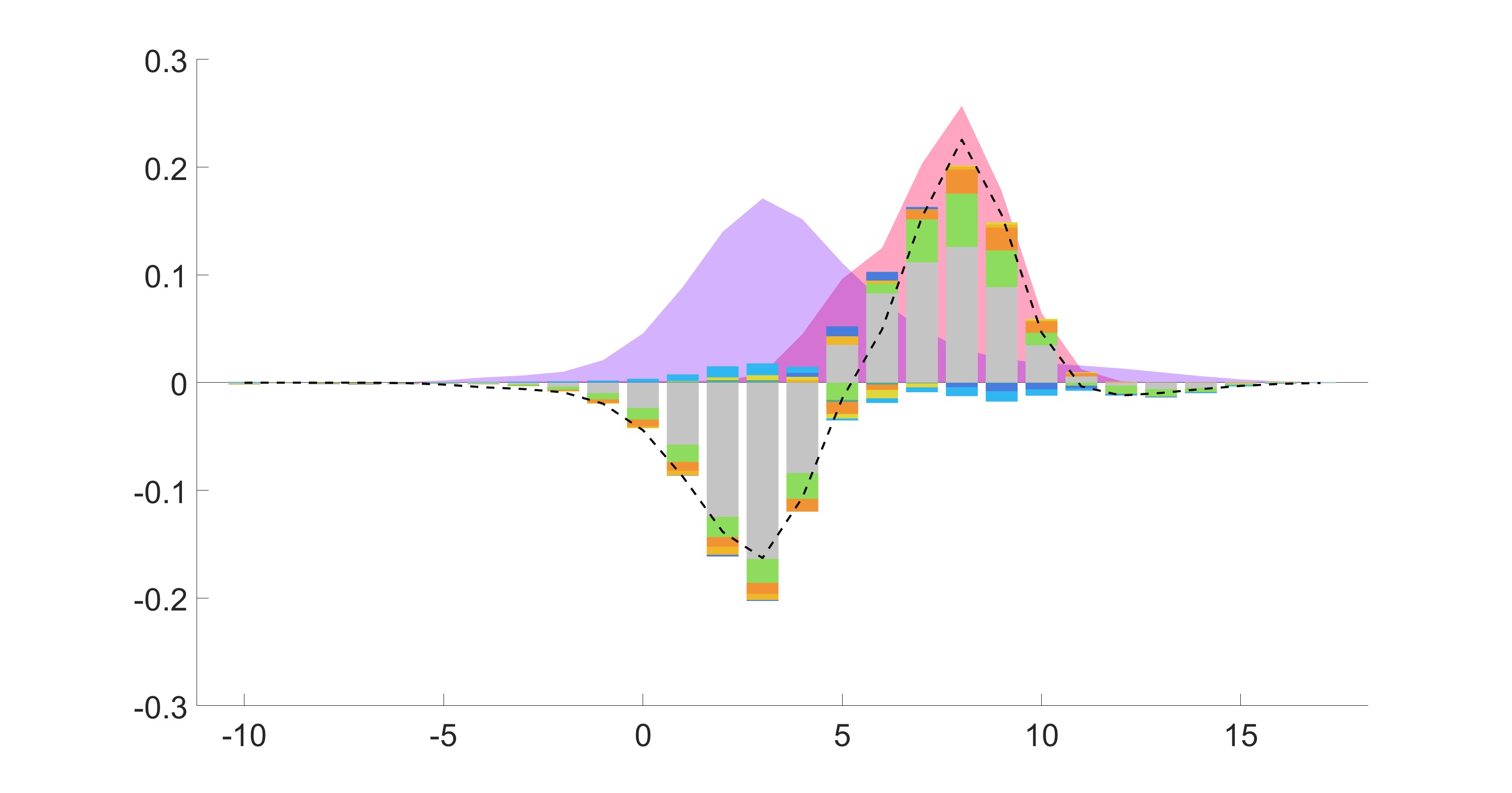}}
\subcaptionbox{2022Q2}
{\includegraphics[width=0.49\textwidth, trim={3.5cm 2cm 5cm 1.5cm},clip]{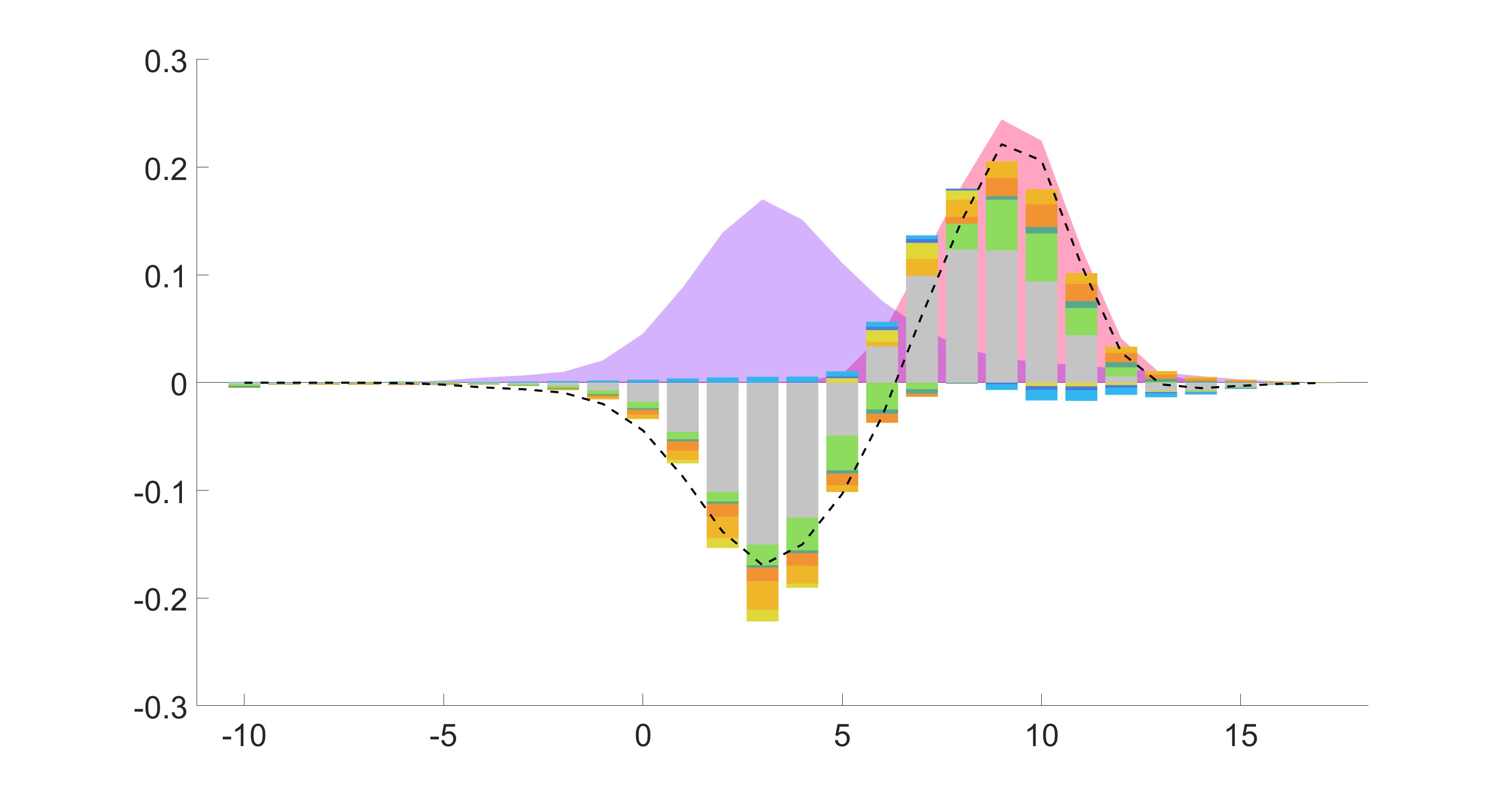}}
\subcaptionbox{2022Q3}
{\includegraphics[width=0.49\textwidth, trim={3.5cm 2cm 5cm 1.5cm},clip]{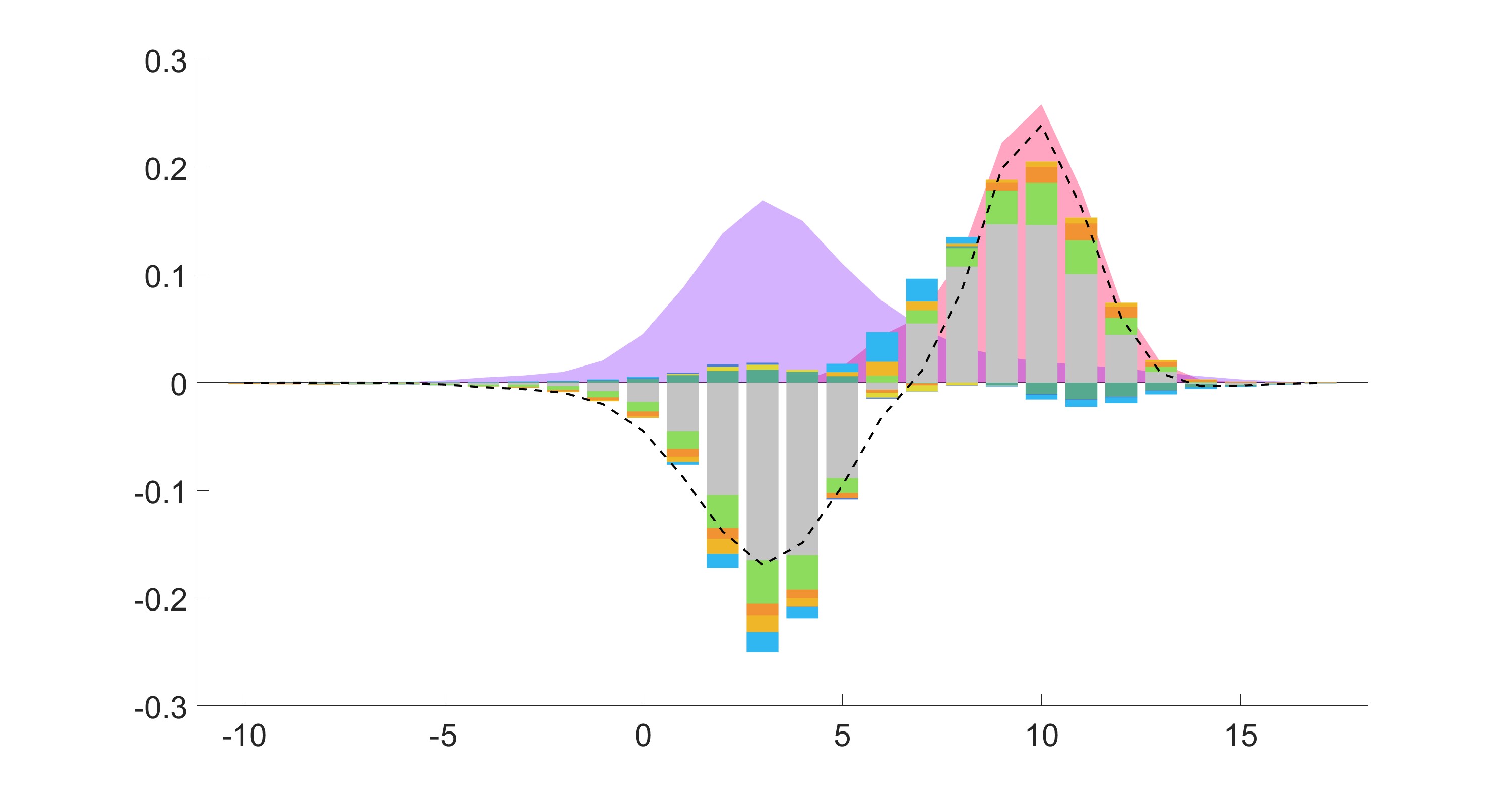}}
\vspace{-.5cm}
\begin{flushleft}
\footnotesize \singlespacing \textit{Notes: The purple density indicates the average historical density. The red density indicates the forecast density for 1-step ahead annualized quarterly inflation. The dashed black line indicates the difference between the predicted density and the historical average. The stacked bars contain the contributions, with: \mycirc[PastInflGrey] \footnotesize{past inflation}, \mycirc[DomBCGreen] \footnotesize{domestic business cycle}, \mycirc[GlobBCGreen] \footnotesize{global business cycle}, \mycirc[EnergyOrange] \footnotesize{energy prices}, \mycirc[FoodOrange] \footnotesize{food prices}, \mycirc[MetalYellow] \footnotesize{metal prices}, \mycirc[EBPBlue] \footnotesize{excess bond premium}, \mycirc[PolicyBlue] \footnotesize{policy rate}.}
\end{flushleft}
\end{figure}

Results for the one-step ahead forecast density are presented in \autoref{fig:drivers_density}. To keep things simple and illustrative, the focus is on the predictive out-of-sample densities for the last four quarters in the sample, i.e. 2021Q4 to 2022Q3. In each panel, the purple density denotes the historical average, while the red density indicates the predictive density for the respective quarter. Without loss of generality, both densities are discretized into 1 percentage point bins.\footnote{The densities can be discretized into arbitrarily fine grids, at the cost of increased computational demand} The variable contributions aggregate to the difference between the predicted density and the historical average, as indicated by the black dashed line. To avoid clutter, the constant and the 4 lags of quarterly inflation are grouped together under the label ``past inflation''. Leveraging the linearity of the variables' contributions, they can be treated directly as one group in the algorithm. In practice, rather than constructing $\bm x_t^{+j}$ and $\bm x_t^{-j}$ for a single variable $j$, the observations are replaced for the entire group, greatly reducing computational demand. In applied work, this allows to compute contributions even for very large data sets.   

Starting with target quarter 2021Q4 in the top left, the predicted density is shifted to the right compared to the historical average. Abstracting from past dynamics, energy prices and the domestic business cycle are the biggest forces that move probability mass towards higher inflation realizations. Food prices, monetary policy, credit spreads, and the global business cycle shift mass in the opposite direction. Overall, this gives rise to the slight negative skew. Because probability mass is shifted from a larger set of bins to a more concentrated area, this causes the forecast distribution to be less dispersed than the historical average.

In the next quarter, food prices emerge as an additional factor that shifts probability mass to the right. The overall effect is now stronger and mass is shifted to a wider interval compared to before, such that the forecast density becomes more dispersed and further shifts to the right. Credit spreads and the interest rate continue to shift mass back to realizations in line with the 2\% target. 

The next two panels are particularly interesting as they correspond to \autoref{dist2022q2} and \autoref{dist2022q3}. For 2022Q2, a slight negative skew can be observed. Energy, food, and metal prices increasingly shift probability mass towards higher inflation outcomes. Compared to energy and food prices, however, the effect of metal prices is more benign and shifts mass to less extreme realizations. This heterogeneity, along with the influence of monetary policy, is responsible for the observed negative skew. Furthermore, compared to the previous quarters, both, the domestic and global business cycle shift probability mass to the right, suggesting a delayed effect of global business cycle dynamics on the U.S. economy.

For 2022Q3, the effect of metal prices fades and the contribution of food and energy prices moderates slightly. The domestic business cycle continues to move probability mass from lower to higher realization, albeit with a smaller effect. In contrast, the global business cycle now redistributes mass from very extreme realisations in the right tail of the distribution to the left tail, countering inflation pressures. The contribution of monetary policy significantly contributes to a pronounced left skew and the emergence of the hump observed in \autoref{dist2022q3}.

\begin{figure}[h]
\hspace{-4cm}\centering
\caption{Drivers of the Forecast Distribution for $h=4$}\label{fig:drivers_density_h4}
\subcaptionbox{2023Q2}
{\includegraphics[width=.49\textwidth, trim={3.5cm 2cm 5cm 1.5cm},clip]{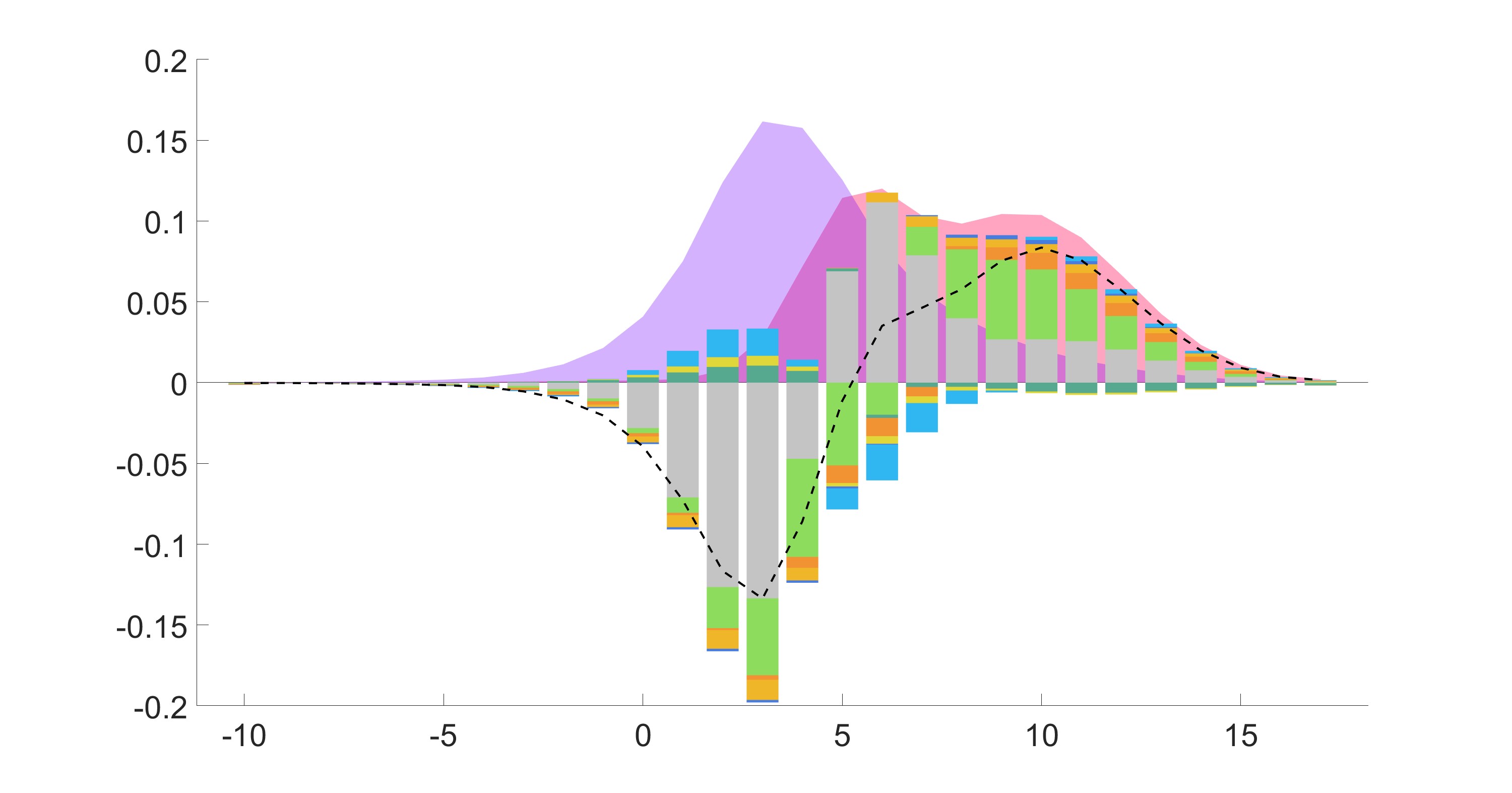}}
\subcaptionbox{2023Q3}
{\includegraphics[width=0.49\textwidth, trim={3.5cm 2cm 5cm 1.5cm},clip]{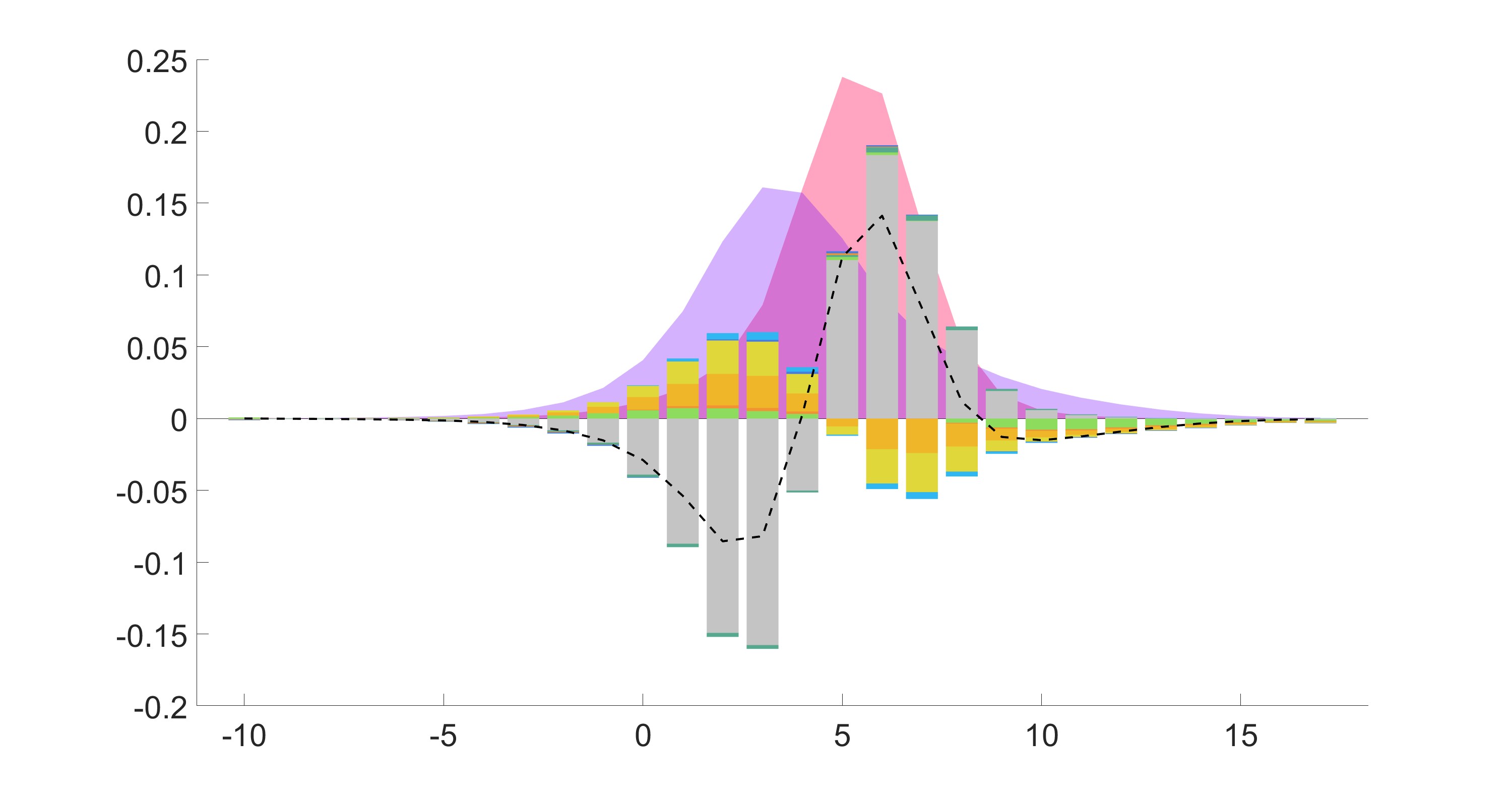}}
\vspace{-.5cm}
\begin{flushleft}
\footnotesize \singlespacing \textit{Notes: The purple density indicates the average historical density. The red density indicates the forecast density for 4-step ahead annualized quarterly inflation. The dashed black line indicates the difference between the predicted density and the historical average. The stacked bars contain the contributions, with: \mycirc[PastInflGrey] \footnotesize{past inflation}, \mycirc[DomBCGreen] \footnotesize{domestic business cycle}, \mycirc[GlobBCGreen] \footnotesize{global business cycle}, \mycirc[EnergyOrange] \footnotesize{energy prices}, \mycirc[FoodOrange] \footnotesize{food prices}, \mycirc[MetalYellow] \footnotesize{metal prices}, \mycirc[EBPBlue] \footnotesize{excess bond premium}, \mycirc[PolicyBlue] \footnotesize{policy rate}.}
\end{flushleft}
\end{figure}

Observing forecast densities centered around very high inflation realizations as in the second half of 2022 might raise concerns from a policy standpoint. However, these short-run densities alone do not offer insights into the persistence of price risks and their underlying drivers. To address this, four-step-ahead forecast densities and contributions for the target quarters 2023Q2 and 2023Q3, constructed using data available a year prior, are depicted in \autoref{fig:drivers_density_h4}. For 2023Q2, the forecast density has a second mode, centered roughly around 10\% inflation. This second mode is mostly driven by the domestic U.S. business cycle and contributions of commodity prices. In 2023Q3, however, the second mode vanishes and the entire distribution shifts towards more moderate inflation outcomes. This shift is primarily driven by declining food and energy prices, which redistribute probability mass back towards lower realizations. Additionally, the previously significant contributions of the U.S. business cycle have faded.

This suggests that, based on data available in 2022, inflation risk persists until late 2023, driven by the recovery of the U.S. domestic business cycle and commodity prices. Furthermore, inflation pressures are expected to ease towards the end of the year as commodity price pressures fade and the business cycle moderates.

 \section{Risk Measures and their Drivers }

In practical applications, such as real-time monitoring tasks and policy making, it is common practice to reduce the information contained in the forecast distribution to a single summary statistic or risk measure. This section explains how risk measures, aligned with central bank risk preferences, can be derived directly from the forecast density. Additionally, it introduces an algorithm to break down these risk measure into their underlying drivers, extending the algorithm in \autoref{sec:dens_decomps_compute}. Continuing the analysis in \autoref{sec:forecast_dist_decomp}, \autoref{sec:drivers_balance} presents the results for inflation risk and an additional validation exercise. 

\subsection{Measuring Macroeconomic Risk}

The macroeconomic uncertainty literature, including \cite{adrianVulnerableGrowth2019a} and \cite{lopez-salidoInflationRisk2022}, constructs risk measures directly from the predicted quantiles of economic variables. These measures have limitations. Firstly, central banks concerned with price stability seek to evade deflation as well as excessive inflation risks, resulting in a policy trade-off \citep{bernanke2003constrained}. However, predicted quantiles only reflect one tail of the forecast distribution, providing only limited insights. Moreover, the choice of a specific percentile may seem arbitrary, as it is unclear why, for instance, the 5th percentile should be a better risk measure than the 6th. Secondly, central banks have clear preferences over risks; for example, they may view a 10 percentage point overshoot of inflation as more costly than a 1 percentage point overshoot. Beyond the statistical characteristics of the forecast distribution, a comprehensive risk measure should hence also reflect the pricing of risk of the central bank \citep{kilianQuantifyingRiskDeflation2007,kilianCentralBankerRisk2008}.

To take central bank preferences into account, \cite{kilianQuantifyingRiskDeflation2007,kilianCentralBankerRisk2008} propose risk measures, drawing from the risk management and finance literature, including \cite{machinaRisk1987}. In this framework, upside risks are defined as inflation, denoted by $\pi$, overshooting an upper threshold, $\bar{\pi}$, and downside risks as $\pi$ falling short of a lower bound, $\underline{\pi}$, where $\underline{\pi}\leq \bar{\pi}$. For point targets, such as the FED's inflation target, $\underline{\pi}=\bar{\pi}=\pi^*$. Introducing the preference parameters $\alpha$ and $\beta$, this results in




\begin{align}
DR_\alpha &\equiv - \int_{-\infty}^{\underline{\pi}} (\underline{\pi}-\pi)^\alpha dF_\pi(\pi), \text{ with } \alpha\geq 0 \label{DR}\\
EIR_\beta &\equiv \int_{\bar{\pi}}^{\infty} (\pi-\bar{\pi})^\beta dF_\pi(\pi), \text{ with } \beta\geq 0,\label{EIR}
\end{align}  

\noindent where $DR_\alpha$ and $EIR_\beta$ denote deflation risk and excess inflation risk. Intuitively, these risk measures are hence preference and probability weighted target deviations.

 
In the case of $\alpha=\beta=0$, the risk measures collapse to the marginal distribution of inflation , i.e. the probability of target overshoots in either direction. Here, the central bank does not consider the size of the target deviation. For $\alpha=\beta=1$, the risk measure becomes a weighted measure of expected shortfall, while for $\alpha=\beta=2$, the risk measures resemble a weighted measure of the target semi-variance. Both incorporate the size of the target deviation, but $\alpha=\beta=2$ penalizes larger target deviations more strongly than smaller ones. For a risk averse central bank, this specification provides a useful benchmark case \citep{blinderDistinguishedLectureEconomics1997,kilianQuantifyingRiskDeflation2007,svenssonInflationForecastTargeting1997,svenssonInflationTargetingShould2002}. 

The probability weighting is introduced via $F_\pi(\pi)$; the distribution of inflation. In \cite{kilianQuantifyingRiskDeflation2007,kilianCentralBankerRisk2008}, this distribution is estimated with simple GARCH models. As the previous section illustrates, however, a symmetric forecast distribution does not accurately reflect the risk outlook. Tail risks, particularly important for policy considerations, might hence be misrepresented. On the other hand, quantile regressions do not provide a complete forecast distribution, and approximating it from a grid of independent quantile regressions is susceptible to approximation errors \citep{koenkerRegressionQuantiles1978, chernozhukovQuantileProbabilityCurves2010}. The model proposed in this paper effectively bridges these two literatures. The mixture representation accurately reflects the risk outlook while also providing the full forecast distribution.

To see this, let $\Theta^m$ denote a set of posterior draws and $\pi$ denote a point of interest on the support of the distribution. At time $t$ and for draw $m$ the full predictive PDF at $\pi$ is then given by

\begin{equation}
f_{x}(\pi)_{t+h|t} = \sum_{c=1}^C \left[ \nu_c^m(\bm x_t) \prod_{l=1}^{c-1}\left\{ 1-\nu_l^m(\bm x_t)\right\}  \right] \cdot \sqrt{\tau_c^m}\phi\left(\sqrt{\tau_c^m}(\pi- \bm x_t \beta_c^m)\right).
\end{equation}

\noindent Repeating the computation for a set of $\pi$ values that span the support of $f_{\bm x}(\pi_{t+h})$ yields the entire distribution function at one posterior draw, at time $t$ for forecast horizon $t+h$. Averaging across draws yields the final PDF estimate and credible intervals are estimated with quantiles.\footnote{Notably, this procedure applies to both estimation algorithms. For MCMC, the draws are obtained directly from the sampler, whereas for VB draws can be obtained by sampling from the posteriors at the VB estimate. Alternatively, the VB estimates can be used directly, collapsing $\Theta^m$ to a singleton. The unconditional PDF can be estimated by averaging across time. Computations for the CDF are analogous.} Plugging the results in, yields the risk measures


\begin{align}
DR_{\alpha,t+h|t} &\equiv - \int_{-\infty}^{\underline{\pi}} (\underline{\pi}-\pi)^\alpha f_{x}(\pi)_{t+h|t} d\pi, \text{ with } \alpha\geq 0, \label{DR2}\\
EIR_{\beta,t+h|t} &\equiv \int_{\bar{\pi}}^{\infty} (\pi-\bar{\pi})^\beta f_\pi(\pi)_{t+h|t} d\pi, \text{ with } \beta\geq 0.\label{EIR2}
\end{align}  

Given the density regression model, they fully account for the non-standard distribution of future inflation. Gaussian quadrature methods are employed to evaluate the integrals in \autoref{DR2} and \autoref{EIR2} across the paper, resulting in a fast and simple numerical procedure. 

Besides their flexible yet simple form, these risk measures have another advantage. Under mild conditions, \cite{kilianCentralBankerRisk2008} establish equivalence between a risk-managing central bank and an expected utility maximizing central bank for suitably chosen loss functions. A natural selection of such a loss function, satisfying these conditions, is 

$$
L = w\cdot\mathbb{I}(\pi<\underline{\pi})(\underline{\pi}-\pi)^{\alpha} + (1-w)\cdot\mathbb{I}(\pi>\bar{\pi})(\pi-\bar{\pi})^{\beta},
$$

\noindent where $w$ is a weight parameter with $0\leq w \leq1$. Taking expectations yields the balance of risk, $BR_{\alpha,\beta}$, or expected loss

\begin{equation}
BR_{\alpha,\beta} \equiv E(L) = -wDR_\alpha+(1-w)EIR_\beta.
\end{equation}

This risk modeling framework is hence fully compatible with standard approaches to modeling the decision problem faced by central banks and remains consistent for a broad range of preference functions. This includes non-integer values for $\alpha$ and $\beta$, different combinations of both parameters or policy-regime-specific parameter settings. While beyond this text, this formal connection to the monetary policy and central banking literature allows the computation of purely empirical risk measures under the same policy preference assumptions as in structural models. Consequently, it allows for direct comparisons between the outputs of the two models and facilitates backtesting exercises, useful for policy applications.

\subsection{Historical Inflation Risk Measures}\label{sec:in_sample}
 
Various assumptions on $\alpha$, $\beta$, and $w$ can be considered plausible. For instance, as the Covid-19 pandemic neared its end, policymakers exercised caution in raising rates too quickly to avoid stifling the nascent recovery. In such a scenario, the central bank's preferences might lean towards accepting higher inflation rates, suggesting $\beta<\alpha$ or $w<0.5$. Conversely, during periods when policy rates are constrained by the effective lower bound, central banks might be particularly concerned about deflationary pressures. This can be accommodated by setting $w>0.5$. Moreover, preferences are likely to vary over time or among different central bank governors.

However, the paper's primary focus is to present methods for decomposing the forecast distribution and risk measures into their drivers within a unified framework. Therefore, it abstains from specifying or estimating complex policy functions and instead concentrates on illustrative benchmark cases. Specifically, to align with the FED's policy objective of 2\% annual inflation, the bounds and weights are set to $\underline{\pi}=\bar{\pi}=2$ and $w=0.5$.

\begin{figure}[h]
\hspace{-4cm}\centering
\caption{The Balance of Risk}\label{fig:risk_measures1}
\subcaptionbox{BR for $\alpha=\beta=0$ and $h=1$}
{\resizebox*{!}{0.25\textwidth}{\includegraphics[width=0.5\textwidth, trim={6cm 2cm 5cm 2cm},clip]{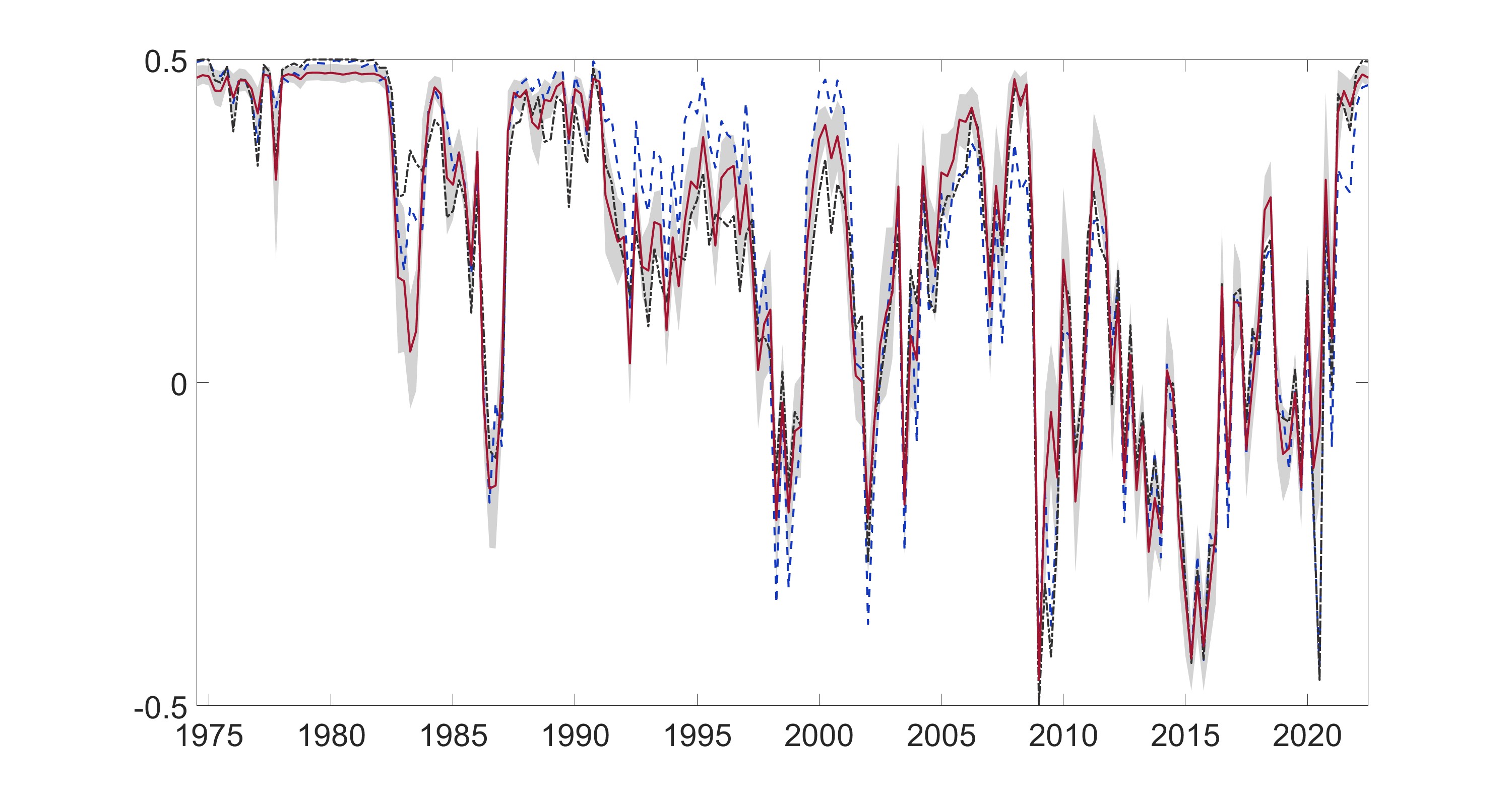}}}
\subcaptionbox{BR for $\alpha=\beta=2$ and $h=1$}
{\resizebox*{!}{0.25\textwidth}{\includegraphics[width=0.5\textwidth, trim={6cm 2cm 5cm 2cm},clip]{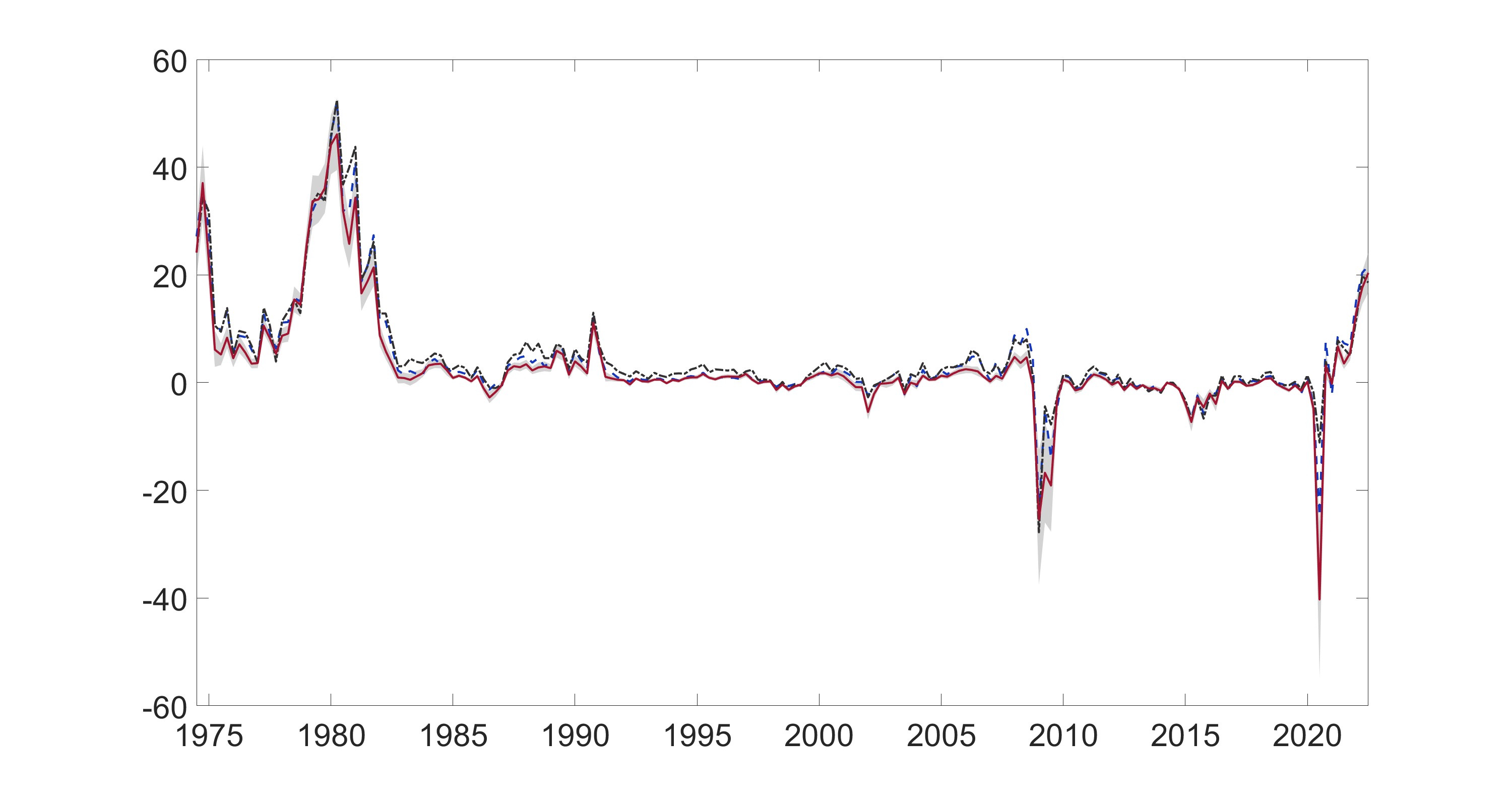}}}
\subcaptionbox{BR for $\alpha=\beta=0$ and $h=4$}
{\resizebox*{!}{0.25\textwidth}{\includegraphics[width=0.5\textwidth, trim={6cm 2cm 5cm 2cm},clip]{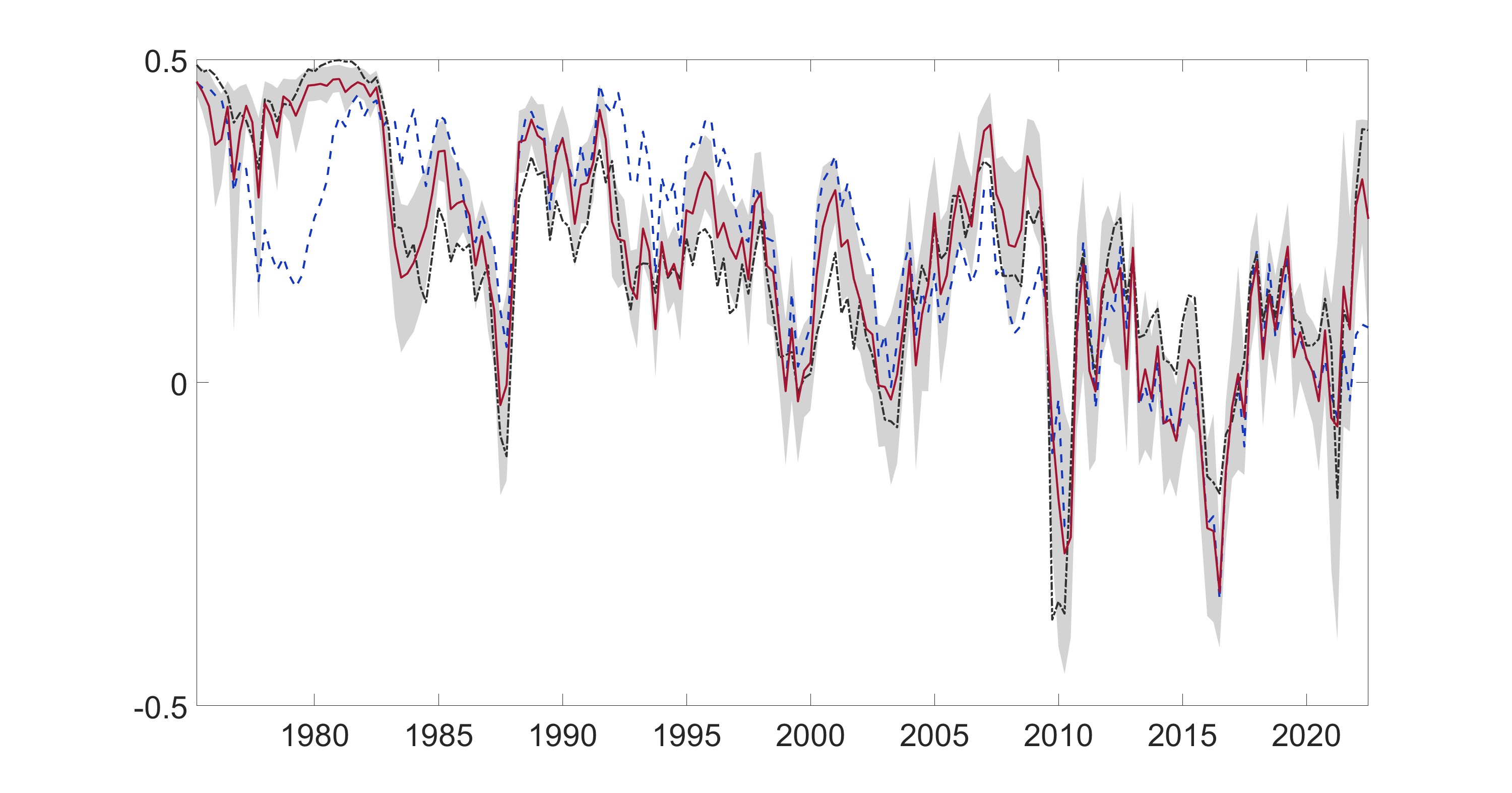}}}
\subcaptionbox{BR for $\alpha=\beta=2$ and $h=4$}
{\resizebox*{!}{0.25\textwidth}{\includegraphics[width=0.5\textwidth, trim={6cm 2cm 5cm 2cm},clip]{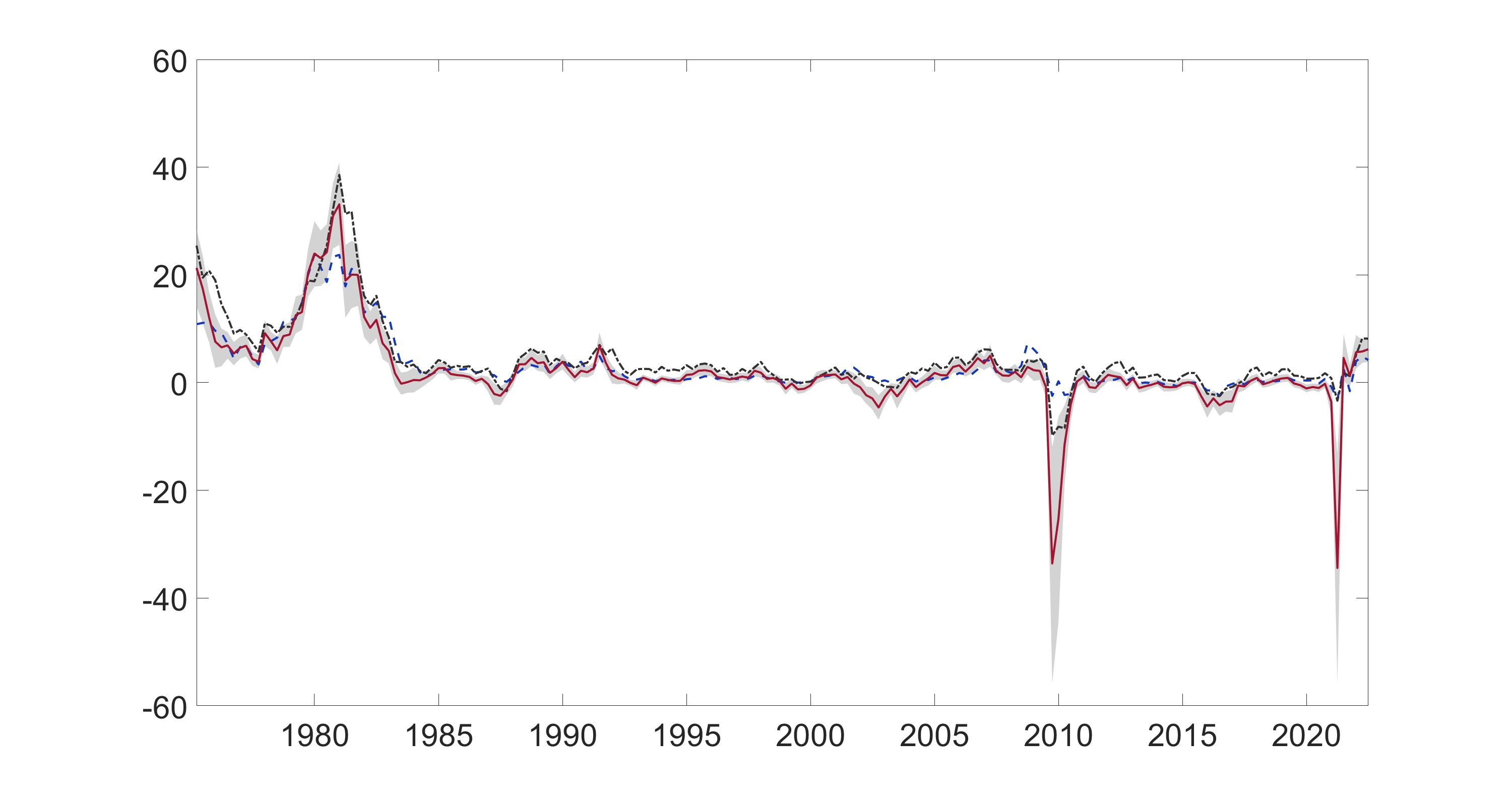}}}
\begin{flushleft}
\vspace{-4mm}
\footnotesize \singlespacing \textit{Notes: The solid red line indicates the MCMC estimate over the full sample obtained with the density regression model. The shaded area is the corresponding 86\% credible interval. The dashed blue and dot-dashed black line show the corresponding estimates computed with a model with stochastic volatility and constant volatility, respectively.}
\end{flushleft}
\end{figure}

\autoref{fig:risk_measures1} displays the estimated balance of risk for one quarter ahead inflation in the first row and four quarter ahead inflation in the second row. The shaded area indicates 86\% credible intervals. As a benchmark, corresponding estimates derived from univariate forecasting models with stochastic and constant volatility are depicted by the dashed blue and dot-dashed black lines, respectively. All models are estimated using MCMC over the entire sample of available data. Positive (negative) values in all risk measures signify excess inflation (deflation) risk.

The left panel shows the results for $\alpha=\beta=0$. Here, the risk measure reflects a purely statistical perspective, representing a central bank that solely considers the probability of inflation overshooting or undershooting the target, without regard for the size of the target deviation. Consequently, the risk measure is bounded between -0.5 and 0.5, indicating inflation risk exclusively on the downside or upside, respectively. 

Under this parameter setting, inflation risk was strictly positive until the early 80s for the first forecast horizon, $h=1$. It then gradually decreased, with a brief spike in 1984, falling to a temporary low in 1986. Towards the end of the Volcker era inflation risk was hence tilted to the downside. Subsequently, inflation risk jumped again to pre-Volcker levels before stabilizing around lower levels from roughly 1991, i.e. the end of the fist Gulf war, to the financial crisis. This period is interrupted by two marked troughs in 1997 and 2002 that coincide with the Asian financial crisis and the burst of the dot-com bubble and subsequent recession. The financial crisis marks a turning point. Inflation risk decreased to historic lows, with risks being almost exclusively on the downside. Up until 2021 risks then remained subdued and tilted to the downside on average with peaks around 2011 and 2018 and a sharp decline from 2014 to 2016. Both peaks roughly align with periods in which energy prices increased, whereas the trough aligns with the oil and gas price plunge that lasted from 2014 to 2016. Finally, risk again fell during the pandemic before quickly increasing to historical highs in 2022. For $h=4$ the results correspond, but the dynamics are slightly less volatile.

When comparing the risk measure to those computed based on the benchmark models, at $h=1$ all three measures appear fairly similar. However, notable differences emerge during crucial time periods. Specifically, the stochastic volatility model tends to overstate inflation risks during the 1990s and understate inflation risks during recent high inflation period compared to the density regression model. Conversely, the constant volatility model suggest a balance of risk that is more on the downside from the mid-80s up until the financial crisis. These differences are even more pronounced for $h=4$. The SV model indicates substantially less inflation risk during the Volcker-era and the recent high inflation period. Additionally, it suggests substantially higher inflation risks for most of the great moderation compared the density regression model. Conversely, the constant parameter model indicates more extreme inflation risks to the upside and downside for most of the sample.  
 

However, in practice most central bank communication rules out $\alpha=\beta=0$. This is illustrated with an example in \cite{kilianCentralBankerRisk2008}. Assume that a central banker is faced with two scenarios: (a) 2.001\% inflation with certainty, or (b) 10\% inflation with 20\% probability and inflation less than 2\% with 80\% probability. If $\alpha=\beta=0$, then (a) is considered worse, because $|BR_{0,0}^{(a)}|=|0.5|>|BR_{0,0}^{(b)}|=|-0.3|$. Conversely, since in practice central bankers would likely favour scenario (a) over (b), this implies $\alpha>0$ and $\beta>0$. Considering these insights, as a second benchmark, the right column of figure \ref{fig:risk_measures1} shows the estimates for $\alpha=\beta=2$. Under these preferences, the central bank considers the size of the target deviation and assigns greater weight to larger deviations. This specification aligns with the concept of ``low and stable inflation'' and serves as a useful benchmark for risk averse central banks \citep{kilianCentralBankerRisk2008}. 

With these preference parameters, the BR seems less erratic and changes in inflation risk more persistent.\footnote{With the target deviation included in the calculations, the measure is no longer bounded between -0.5 and 0.5. To facilitate the interpretation, it can be rescaled with the historical standard deviation in practice.} For $h=1$, the BR is strongly positive during the early to mid 70s,  before falling to lower levels between 1975 to 1978. In the left panel, inflation risk was elevated throughout the entire period. This suggests that the probability of missing the target was high throughout the 70s, but that the size of the expected target deviation decreased during the later half of this sub-sample. Until 1980 inflation risk reverted to its previous levels before decreasing to very low levels during the early 80s. Again, this period aligns well with the Volcker-era. Throughout the great moderation, inflation risk remained generally subdued, however, a distinct peak coincides with the first Gulf war and a marked dip in inflation risk with the dot-com recession of 2002. The financial crisis marks a turning point. After the crisis, inflation risk was strongly on the downside before gradually normalizing. Nonetheless, deflation risk dominated slightly, coinciding with the zero lower bound period. Contributing to this, the oil and gas price plunge from 2014 to 2016 leaves a deflationary footprint. Finally, during the pandemic, the balance of risk tilts strongly to the downside, jumps back up to pre-financial crisis levels and then gradually increases during the high inflation period towards the end of the sample, tilting increasingly to the upside. The stark contrasts to the left panels underscore the empirical significance of central bank preferences. Importantly, information crucial for policy considerations can be obscured when relying solely on statistical risk measures. As before, moving to $h=4$ leaves the overall dynamics almost unaffected. 

Comparing the results to the benchmark models overall confirms the previous picture. The models produce similar results, but diverge significantly during times of more volatile inflation. Both the stochastic volatility model and, to a greater extent, the constant parameter model suggest heightened inflation risks before the financial crisis and diminished deflationary pressures during subsequent recessions, including the one following the Covid-19 pandemic. Conversely, they imply lower inflation risks for parts of the recent high inflation period. These differences are even more pronounced for $h=4$. The stochastic volatility model indicates lower inflation risk during the Volcker-era and lower deflation risks during recessions. Similarly, during the recent high inflation period, inflationary pressures are understated compared to the density regression framework. On the other hand, the constant parameter model tends to suggest higher inflation risks throughout the sample, albeit lower deflation risks during recessions.

For readers interested in further exploration, \autoref{supp:asymmetric} in the Online Supplement revisits this analysis with asymmetric policy preferences, specifically $\alpha=3$ and $\beta=2$, and $\alpha=2$ and $\beta=3$. Under these settings, the risk measures exhibit qualitative similarities to the $\alpha=\beta=2$ case. However, the risk measures are distorted in the direction of the larger preference parameter.

In addition, recognizing that true inflation risk is inherently unobserved, \autoref{supp:shadow} in the Online Supplement provides a validation exercise by constructing an ``inflation risk shadow rate''. This rate represents the implicit interest rate that balances upside and downside inflation risks implied by the model. The estimated interest rate and the observed federal funds rate generally exhibit a high degree of correlation (0.71) and track similar dynamics. Particularly during the Volcker era and the great moderation, the rate aligns closely with the actual FFR, but suggests more aggressive rate adjustments during recessions and the recent high inflation period.


\subsection{Analysing the Drivers of Risk}\label{sec:drivers_theory}

While decomposing the forecast distribution provides a cross sectional view, decompositions of summary statistics such as the risk measures proposed above are equally significant. Because the inflation risk measures are constructed directly from the model's densities, there is a direct mapping from the drivers of the forecast distribution to the drivers of the risk measures. Analogously to \autoref{sec:dens_decomps}, they can be defined as

\begin{defiM}[Contributions to Risk Measures]\label{def:driver of risk}
A driver of downside risk at time $t$ for horizon $t+h$ is a predictor, $x_j$, for which

\begin{equation*}
\varphi_{t,j}^{DR_\alpha} = DR_{\alpha,t+h|t}-\mathbb{E}\left( DR_{\alpha,t+h} \right) - \sum_{i\in [1,\dots,n] : \neq j} \varphi_i^{DR_\alpha} < 0.
\end{equation*}

\noindent a driver of upside risk is a variable, $j$, for which

\begin{equation*}
\varphi_{t,j}^{EIR_\beta} = EIR_{\beta,t+h|t}-\mathbb{E}\left( EIR_{\beta,t+h} \right) - \sum_{i\in [1,\dots,n] : \neq j} \varphi_i^{EIR_\beta} > 0.
\end{equation*}

\noindent If $|\varphi_j|>|\varphi_i|, \text{ for } i\neq j$, then variable $x_j$ is a bigger risk factor at time $t$ than variable $x_i$. The definition for the balance of risks, BR, is analogous.
\end{defiM}

Therefore, if $|\varphi_j|>|\varphi_i|, \text{ for } i\neq j$, then variable $x_j$ is a bigger risk factor at time $t$ than variable $x_i$. A driver of downside (upside) risk is hence a variable that contributes negatively (positively) to deviations of the corresponding risk measure from its historical average. As before, these definitions ensure efficiency and linearity for the decompositions of risk measures.

However, it is important to note that the preference parameters $\alpha$ and $\beta$ also play a role in defining the drivers of risk. Consider a simple example: If a central bank is only concerned with deflation risks ($w=1$), but an increase in commodity prices shifts probability mass within the upper tail of the distribution, increasing the likelihood of very high inflation events, then commodity prices should not emerge as a driver of risks. Furthermore, the relative size of the contributions will vary based on the preference parameters and on how the individual predictors shift probability mass. For instance, if a variable shifts mass towards extreme inflation realizations, the relative contribution will be larger for higher levels of risk aversion. Alternatively, under the same preference parameters, a predictor that shifts mass further away from the target will have a larger contribution compared to one that shifts the same mass closer to the target. Therefore, analogous to the risk measures themselves, the drivers of risk cannot be interpreted without taking into account risk preferences. The requirement that any risk measure must be linked to the preferences of the economic agent as stated in \cite{machinaRisk1987}, hence generalizes to the drivers of risk.

In practice, computing the contributions remains complicated, because the risk measures are non-linear transformations of the mixture model. However, the same principles as outlined in \autoref{sec:dens_decomps_compute} to solve this problem apply and motivate \autoref{alg:shap_risk}, which is provided in full in \autoref{app:drivers}. The individual steps are analogous to \autoref{alg:shap_dens}, replacing the density with the individual risk measures:

\begin{enumerate}
\item At given time period $t$, sample a random collocation, $S$, from $\pazocal{C}(x)$ and a random time period, $t^*$, from $1,\dots,t-1$. 
\item Replace the values in $\bm x_t$ for the variables in $S$ with the values in $\bm x_{t^*}$.
\item Create two vectors of observations. The first one contains $x_j$ at time $t$, the other one also has $x_{j,t}$ replaced with $x_{j,t^*}$.
\item Compute the risk measure for both vectors. The difference yields the marginal contribution of $x_j$.
\item Repeat the procedure for a number of MC samples, $M$. The average of these marginal contribution yields an estimate of $\varphi_{t,j}^{DR_\alpha}$ and $\varphi_{t,j}^{EIR_\beta}$.
\end{enumerate}

%
%
%
%
%
%
%
%
%

\subsection{Drivers of the Balance of Risk} \label{sec:drivers_balance}

Drawing on insights from preceding sections, this subsection addresses the initial research question and investigates the drivers of inflation risk in the U.S. during the recent high inflation period, placing particular emphasis on the contributions of commodity prices and the U.S. and global business cycle. The exercise is extends the analysis in \autoref{sec:forecast_dist_decomp}. Additionally, to validate that these contributions capture economically meaningful information, the last exercise in this section compares selected contributions to structural economic shocks. Results for the individual upside and downside risk are shown in the Online Supplement in \autoref{supp:add_figures}.

\begin{figure}[h]
\hspace{-4cm}\centering
\caption{Drivers of the Balance of Risk}\label{fig:drivers_risk_measures1}
\subcaptionbox{Risk Drivers for $\alpha=\beta=0$, $h=1$}
{\resizebox*{!}{0.25\textwidth}{\includegraphics[width=\textwidth, trim={6cm 2cm 5cm 2cm},clip]{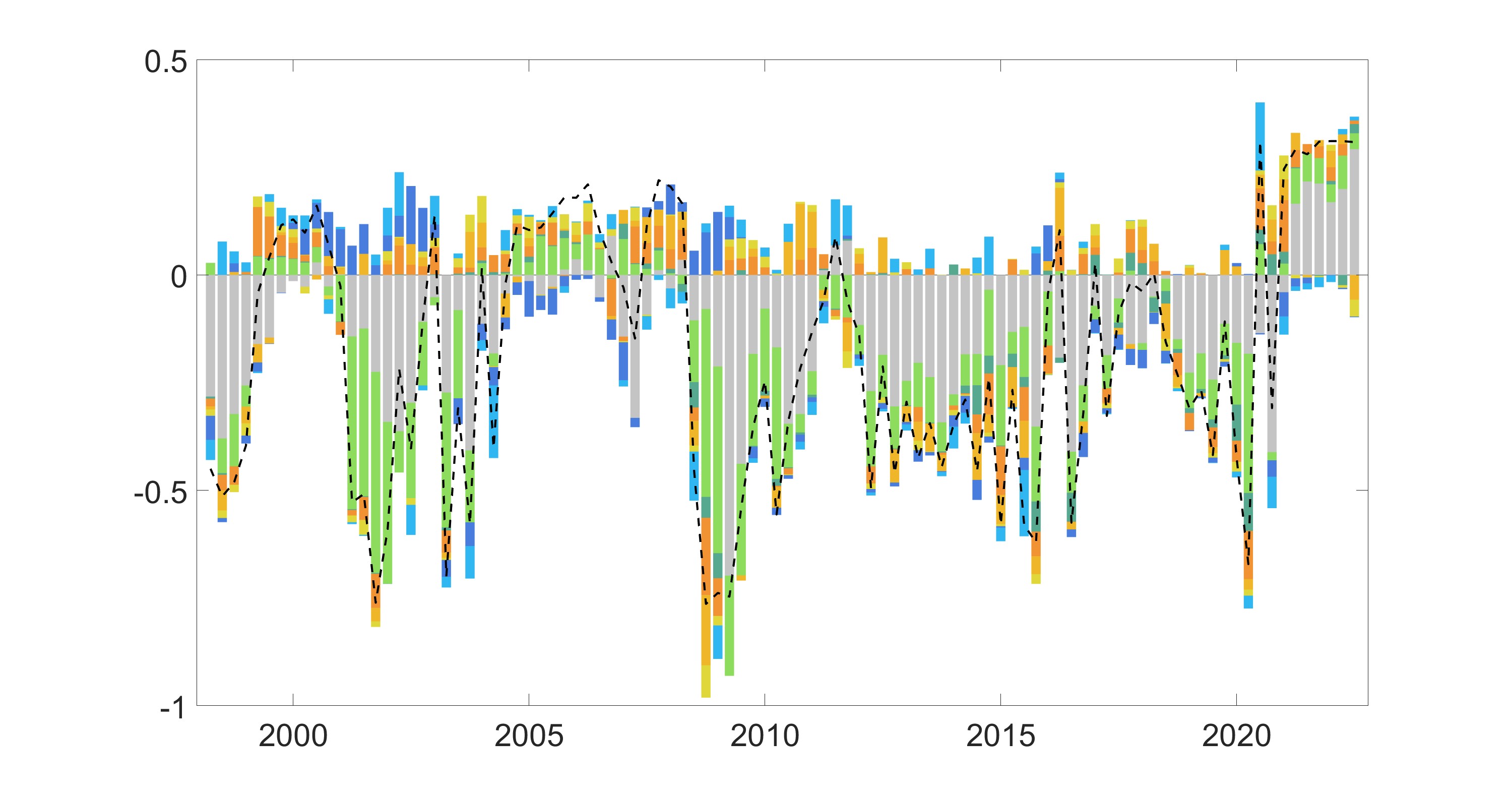}}}
\subcaptionbox{Risk Drivers for $\alpha=\beta=2$, $h=1$}
{\resizebox*{!}{0.25\textwidth}{\includegraphics[width=\textwidth, trim={6cm 2cm 5cm 2cm},clip]{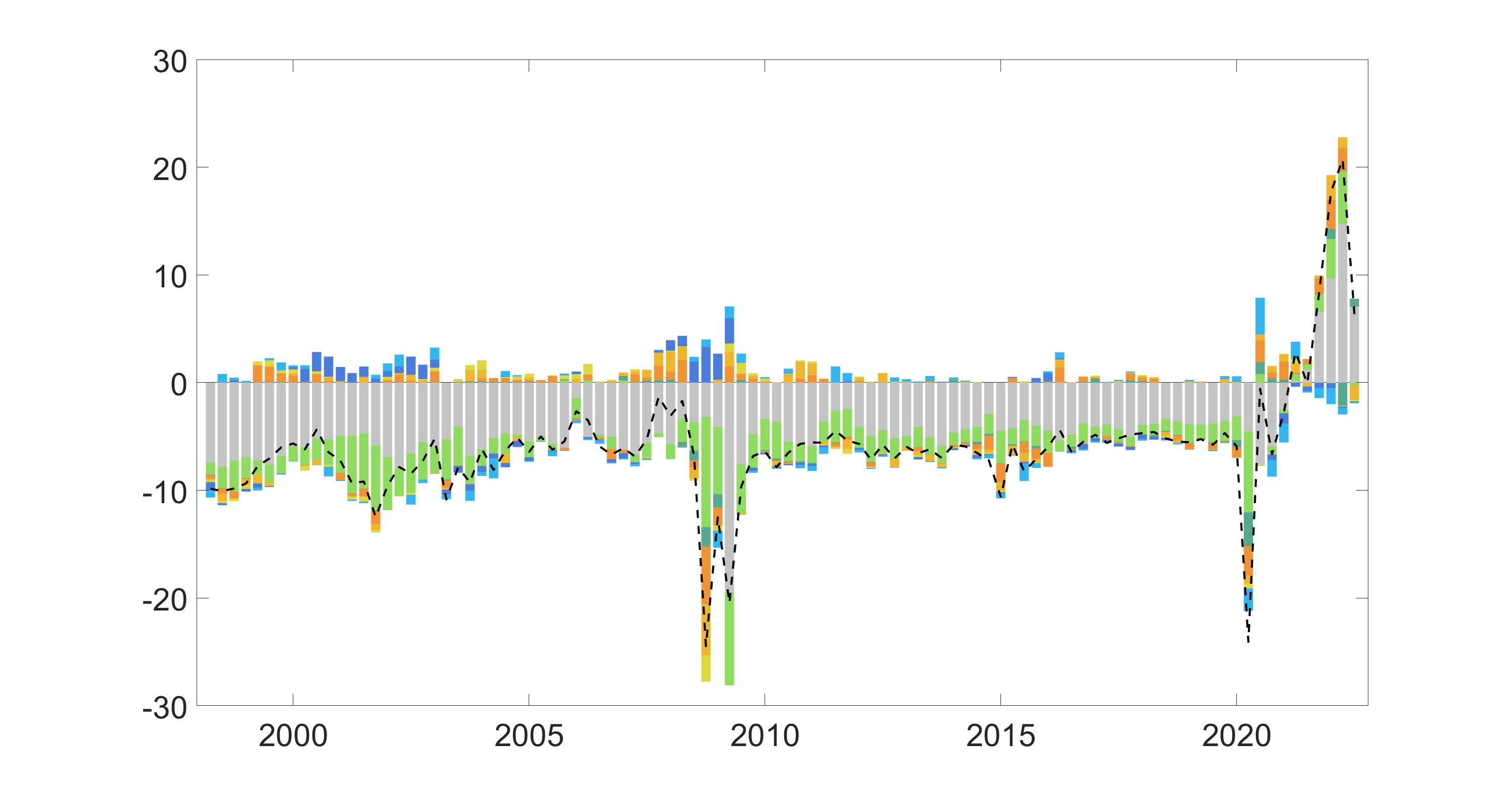}}}
\subcaptionbox{Risk Drivers for $\alpha=\beta=0$, $h=4$}
{\resizebox*{!}{0.25\textwidth}{\includegraphics[width=\textwidth, trim={6cm 2cm 5cm 2cm},clip]{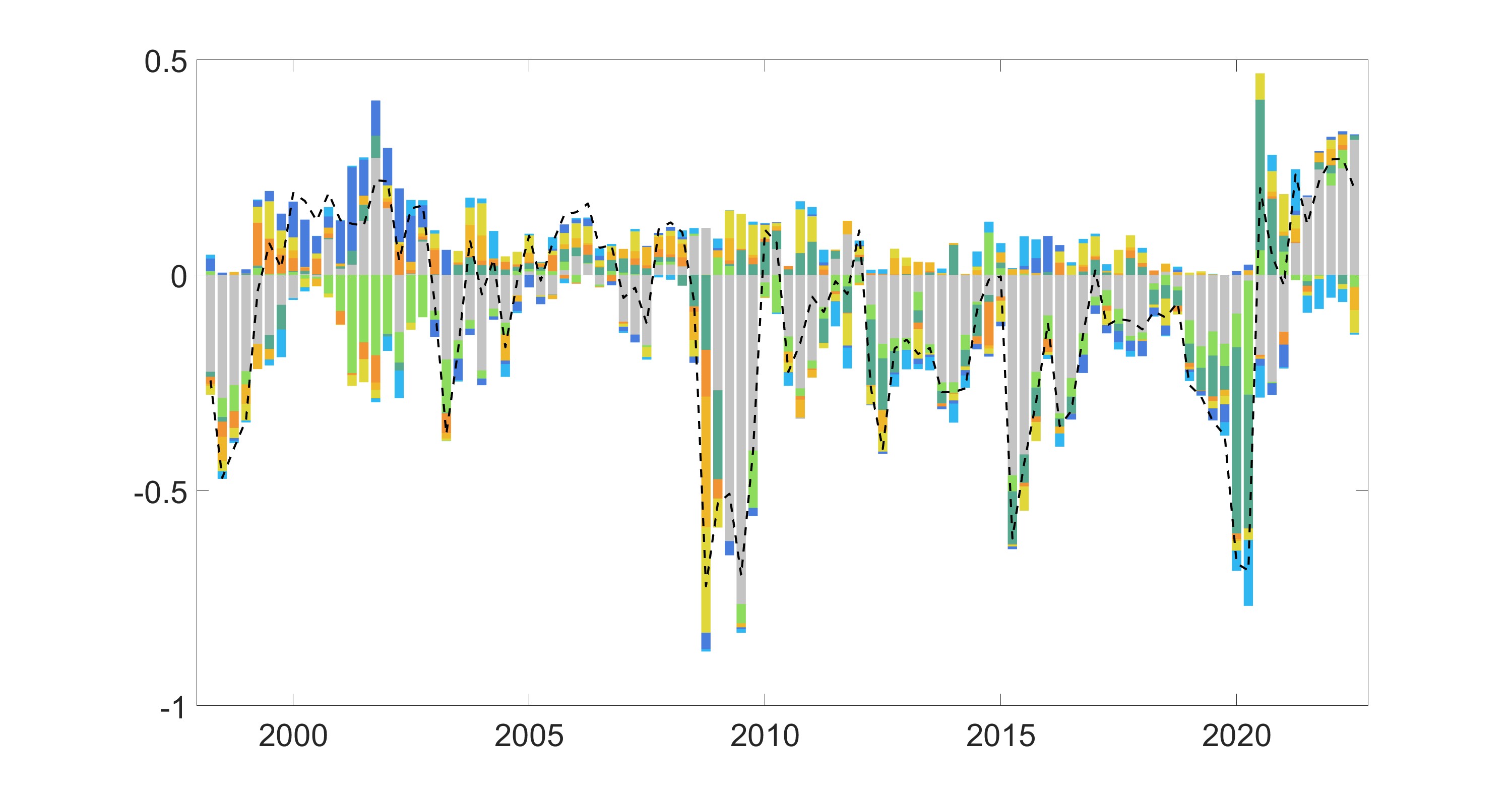}}}
\subcaptionbox{Risk Drivers for $\alpha=\beta=2$, $h=4$}
{\resizebox*{!}{0.25\textwidth}{\includegraphics[width=\textwidth, trim={6cm 2cm 5cm 2cm},clip]{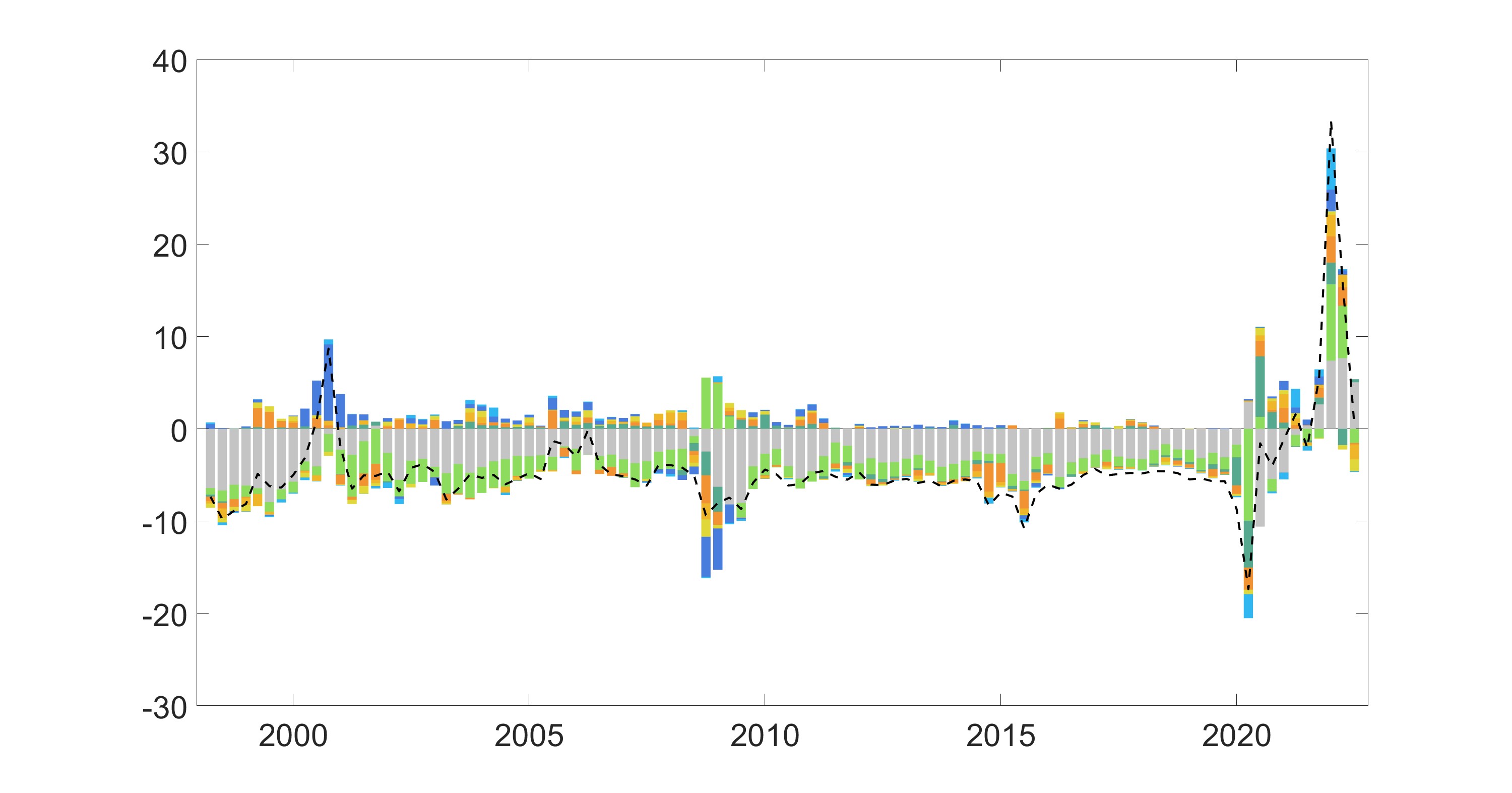}}}
\begin{flushleft}
\vspace{-5mm}
\footnotesize \singlespacing \textit{Notes: The dashed line indicates the observed value for the balance of risk minus the sample average. The stacked bars contain the variables' contributions, with: \mycirc[PastInflGrey] \footnotesize{past inflation}, \mycirc[DomBCGreen] \footnotesize{domestic business cycle}, \mycirc[GlobBCGreen] \footnotesize{global business cycle}, \mycirc[EnergyOrange] \footnotesize{energy prices}, \mycirc[FoodOrange] \footnotesize{food prices}, \mycirc[MetalYellow] \footnotesize{metal prices}, \mycirc[EBPBlue] \footnotesize{excess bond premium}, \mycirc[PolicyBlue] \footnotesize{policy rate}.}
\end{flushleft}
\end{figure}

The decompositions of the balance of risk are presented in \autoref{fig:drivers_risk_measures1}. The first and second row contain the decomposition of the one-quarter-ahead and four-quarter-ahead balance of risk, respectively. Within each row, the left panel provides results for $\alpha=\beta=0$, while the right panel shows the corresponding results for $\alpha=\beta=2$. The time axis represents the data vintage used to construct the balance of risk and decompositions.  

Starting with the left column, the 1997 decline in predicted inflation risk is mostly attributed to the domestic business cycle and falling commodity prices. Monetary policy and credit spreads contribute positively, counteracting the deflationary pressures in case of $h=1$. Subsequently, inflation risk increases, largely due to a rebound in the domestic business cycle and rising commodity prices. The collapse of inflation risk following the burst of the dot-com bubble is predominantly driven by the recession and associated slowdown of the domestic business cycle. Monetary policy and credit conditions exert significant counteractive pressure, evident from their positive contributions to the balance of risks, both, one quarter ahead and four quarters ahead. This coincides well with the FED easing cycle that began in late 2000. 

From 2004 onward the domestic business cycle and increasing commodity prices tilt the balance of risk to the upside. Financial conditions contribute downward pressure, in line with FED tightening starting in late 2004, for $h=1$. In comparison, the contribution of the business cycle and credit spreads to the four quarter ahead balance of risk is more muted. However, the energy price shocks observed during 2007 and 2008 leave a marked footprint for both forecast horizons, driving the balance of risks upwards right before the financial crisis. At the beginning of the financial crisis, deflation risks are driven by mostly the domestic business cycle for $h=1$. Falling energy prices, monetary policy, and a contraction of global output provide additional downward pressure, whereas credit conditions stabilize the balance of risks. Conversely, for $h=4$ commodity prices and the global business cycle contribute most strongly to downward pressures. Subsequently, the balance of risk recovers, supported by increasing commodity prices and monetary policy. Although the model attributes most weight to food prices, the increase in energy prices correlates with the oil price shock of 2011. Four quarters ahead, the recovery of the global business cycle contributes significantly to inflation risks, while monetary policy shows only little effect.

The results for the low inflation period that lasted up until the pandemic are more mixed. Overall, the domestic and global business cycle contribute mostly deflationary pressures. Energy prices exert mostly downward pressure with the exception of the period around 2017 and 2018 that coincides with historic oil price increases. These features are more apparent for $h=4$. With the start of the pandemic and the introduction of containment measures, all predictors contribute deflationary pressures. The biggest drivers of downside risks, however, are the domestic and global business cycle, where the domestic (global) business cycle contributes a larger share for $h=1$ ($h=4$). One quarter ahead, the subsequent recovery of inflation risk is largely driven by monetary policy, commodity prices, and to a lesser extend the business cycle, while four quarters ahead, most of the recovery pertains to the global business cycle. Towards the end of the sample, inflation risk can mostly be attributed to the domestic business cycle as well as increasing commodity prices. Monetary policy provides deflationary pressures, during the recent inflation surge. Mirroring the results of the previous section, moderating commodity prices contribute negatively to the balance of risk for the final data vintage. 

In the right panels, $\alpha=\beta=2$ introduce risk averse central bank preferences. Notably, ``past inflation'' now almost exclusively contributes deflationary pressure over the entire sample. One reason is that the balance of risk was very high at the beginning of the sample and then decreased to lower levels up until the pandemic. In consequence, the average past balance of risk is decreasing as the sample expands, but almost always higher than the predicted value, leading to a negative difference between the two. Because all contributions aggregate to this difference, they must compensate for the excess in historical inflation risk.\footnote{Generally, the balance of risk and the contributions can be re-based, however, to keep things straightforward this paper refrains from such corrections.} 

Compared to the left panels, the results across both forecast horizons are more homogenous. In particular, the contributions of the domestic business cycle to the four quarter ahead balance of risk are more pronounced. 
Additional differences emerge for specific sample periods. For $h=1$, the decrease in the balance of risk at the start of the financial crisis is now roughly shared between the contraction of the domestic and global business cycle and a decline in commodity prices. For the four quarter ahead balance of risk, the global business cycle, commodity prices, and spreads contribute the most to deflation risks. Additionally, while the balance of risk again decreases mostly due to a contraction in the domestic and global business cycle at the start of the pandemic, compared to before, the following recovery in the balance of risks is largely driven by monetary policy for $h=1$ and the global business cycle for $h=4$. Finally, during the recent high inflation period, monetary policy exerts increasing downward pressure one quarter ahead and the domestic business cycle contributes more strongly to inflation risks, compared to the left panel. For $h=4$, the recovery of the domestic business cycle and commodity prices contribute significantly to inflation risks, while the contribution of the global business cycle is reduced.



Differences in the relative contributions across forecast horizons and preferences are the empirical manifestation of the theoretical results stated in \autoref{sec:drivers_theory}. The relative size of the contributions will depend on where incoming data shifts probability mass. In particular, once preferences are introduced, a predictor that shifts mass to more extreme inflation realisations will contribute more than a predictor that shifts mass to lower realizations. Considering preferences when constructing drivers of risk, is hence also empirically relevant.

Irrespective of the preference parameter setting, however, commodity prices and the recovery of the domestic U.S. business cycle are the main drivers of inflation risks during the recent high inflation period. Contributions from global business cycle dynamics are overall small, but non-negligible for $h=4$.

\subsection{Additional Validation}

\begin{figure}[h]
\hspace{-4cm}\centering
\caption{Comparison with Structural Shocks}\label{fig:struct_shocks}
\subcaptionbox{Comparison with Oil Supply Shocks}
{\resizebox*{!}{0.25\textwidth}{\includegraphics[width=\textwidth, trim={6cm 3.2cm 4.5cm 2cm},clip]{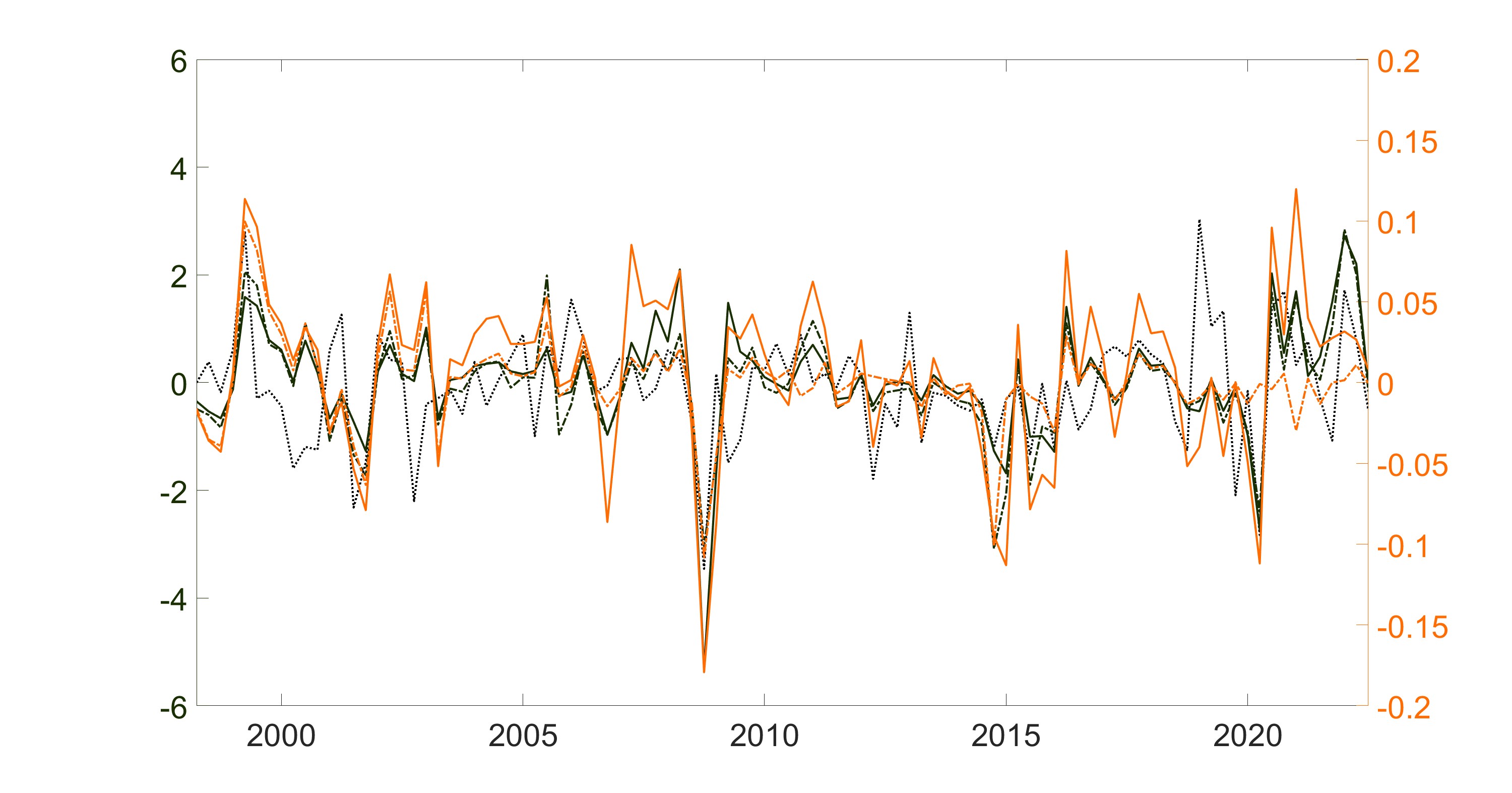}}}
\subcaptionbox{Comparison with Global Activity Shocks}
{\resizebox*{!}{0.25\textwidth}{\includegraphics[width=\textwidth, trim={6cm 3.2cm 4.5cm 2cm},clip]{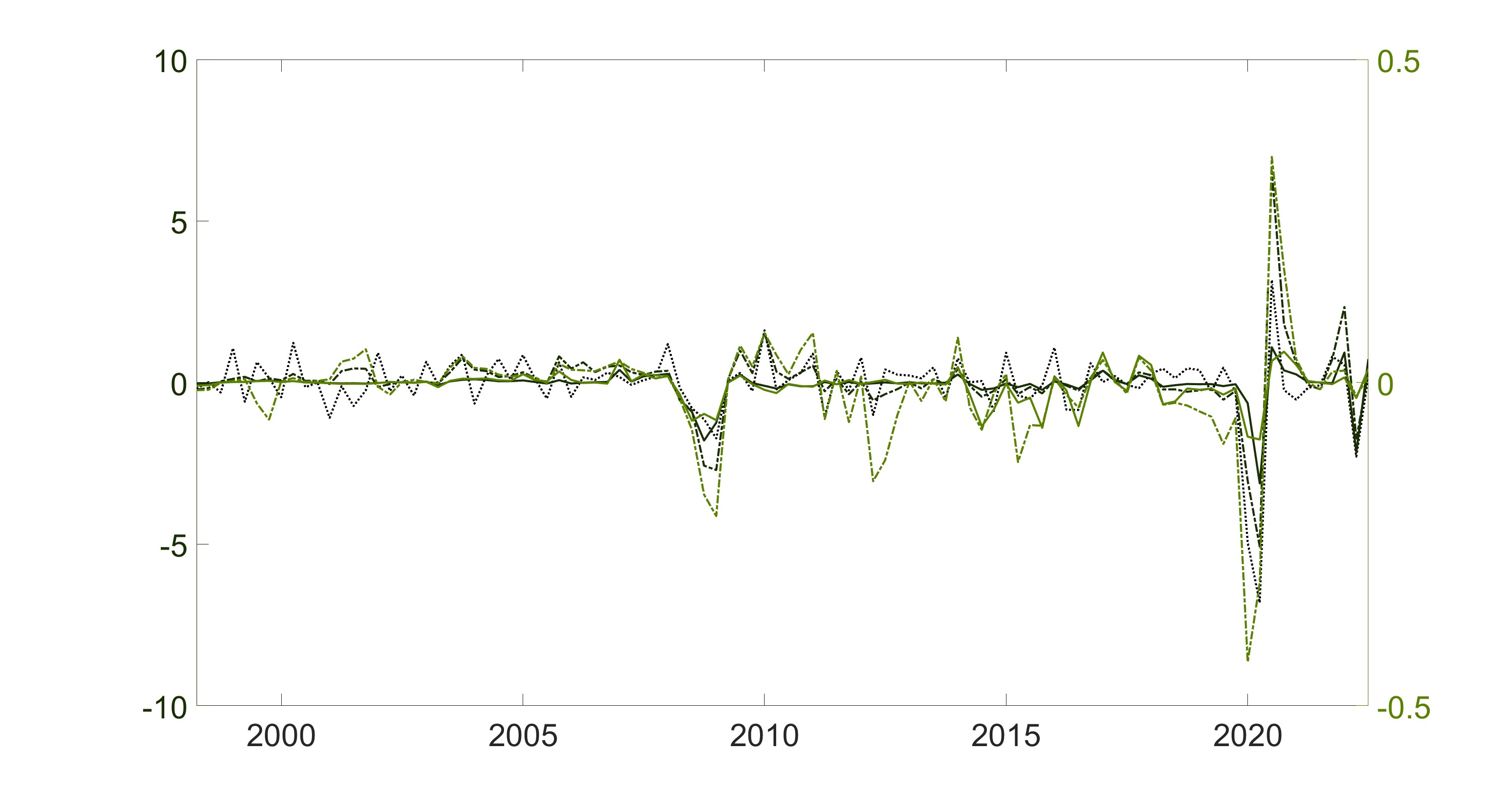}}}
\begin{flushleft}
\vspace{-5mm}
\footnotesize \singlespacing \textit{Notes: The left (right) panel compares the structural oil supply shock (economic activity shock) in \cite{baumeisterStructuralInterpretationVector2019b} (dotted black lines), to the contributions of energy (global business cycle). The solid lines in dark denote $h=1$ and $h=4$ for $\alpha=\beta=2$ on the left axis and dashed lines in orange green denote $h=1$ and $h=4$ for $\alpha=\beta=0$ on the right axes. The oil supply shock is inverted to account for the inverse relationship between oil prices and oil supply.}
\end{flushleft}
\end{figure}

The results in the previous subsection are intuitive as contributions from energy prices to inflation risk, for example, should correlate with historical oil supply shocks. Similarly, the contributions of the global business cycle, should correlate with shocks to global economic activity. These relationships are more formally investigated in \autoref{fig:struct_shocks}. The left panel shows the contributions of energy prices for $h=1$ (solid lines) and $h=4$ (dashed lines) as well as $\alpha=\beta=2$ (dark gray) and $\alpha=\beta=0$ (orange) to inflation risk together with a structural oil supply shock (dotted black). Correspondingly, the right panel shows the contributions of the global business cycle (shades of green) together with a structural global activity shock. These shocks are sourced directly from the structural VAR in \cite{baumeisterStructuralInterpretationVector2019b} and provided by the authors. To enhance visibility, the oil supply shock is inverted and both shocks are aggregated to quarterly frequency. Apart from this, the shocks and contributions are shown without additional transformations.

The left panel confirms that the contributions of energy prices and the oil supply shock track similar dynamics, especially during pivotal events such as the financial crisis, the oil price decline from 2014 to 2016, and the recent inflation surge. Correlation coefficients ranging from 0.34 to 0.46 support this observation. Even stronger comovements are observed for the global business cycle contributions and the global activity shock in the right panel, with correlation coefficients ranging from 0.58 to 0.78. In particular, the series align well around the financial crisis, the Covid-19 pandemic, the recovery of demand post-lockdowns, and the recent high inflation period.

Overall, even though the exercises presented in this section are fundamentally not structural, these findings suggest that the estimated contributions indeed relate to the underlying economic variables in a meaningful way, lending additional support to the proposed risk measurement framework.

\section{Conclusion}
This paper develops a unified framework to assess and monitor macroeconomic risk, without imposing restrictions on the shape and nature of the forecast distribution. By providing a full density estimate, the model facilitates the construction of summary statistics, in the form of risk measures, that fully incorporate the risk preferences of central banks. A key contribution of this paper lies in the development of two algorithms that enable the computation of the drivers of risk. Specifically, the first algorithm attributes changes in the local probability mass of the forecast distribution to economic variables, while the second algorithm decomposes the risk measures, significantly improving the interpretability of the economic outlook. Moreover, these tools are not confined to the forecasting model in this paper, making them applicable in conjunction with other density forecasting models.

To illustrate the proposed framework, it is applied to analyze the recent surge of U.S. inflation. In this context, the recovery of the U.S. business cycle and increasing commodity prices emerge as the main drivers of inflation risks. Additionally, monetary policy provides alleviating pressure, contributing negatively to inflation risks.

Finally, the proposed framework can trivially be extended to other economic settings and model classes. One potential avenue for future research is the extension to a multivariate mixture model to dynamically estimate the joint distribution of key economic aggregates, as well as their drivers.

\newpage
\clearpage
\bibliographystyle{Bib_files/bst/ecta}
\addcontentsline{toc}{section}{\refname}
\bibliography{Bib_files/bib/DensityRegression}
 
\clearpage
\setcounter{page}{1}
\appendix


\section{Technical Appendix: Estimation Algorithms}\label{app:EstimationAlgorithms}

\subsection{MCMC Sampler}

This section presents the full algorithm for the MCMC sampler outlined in methodological section of the paper. 

\RestyleAlgo{ruled} 
\begin{algorithm}[h!]
\scriptsize
\caption{MCMC algorithm}\label{alg:MCMC}
$\textbf{[0]}$ Initialize the model parameters.\\
\Begin
{
$\textbf{[1]}$ Assign each observation $t=1,\dots,T$ to a mixture component $c=1,\dots,C$: \\
\For {$t=1:T$}
{Sample $G_t\in {1,\dots,C}$ from a categorical distribution with probabilities\\
\begin{equation*}
\text{pr}(G_t=c|-) = \frac{\left[ \nu_c(\bm x_t) \prod_{l=1}^{c-1}\left\{ 1-\nu_l(\bm x_t)\right\}  \right] \sqrt{\tau_c}\left[ \sqrt{\tau_c} \left\{ y_t -\bm x_t' \bm\beta_c\right\}\right]}{\sum_{q=1}^{C}\left[ \nu_q(\bm x_t) \prod_{l=1}^{q-1}\left\{ 1-\nu_l(\bm x_t)\right\}  \right] \sqrt{\tau_q}\left[ \sqrt{\tau_q} \left\{ y_t -\bm x_t' \bm\beta_q\right\}\right]},
\end{equation*}
for every $c=1,\dots,C$.
}
$\textbf{[2]}$ Sample the parameters $\bm\psi_c$ and the horseshoe prior parameter $\bm\upsilon_c$, $\tau_c$, $\bm\eta_c$, and $\xi_c$ for each $c=1,\dots,C-1$:\\
\For {$c=1,\dots,C-1$}
{
\vspace{0.2cm}
$\textbf{[2a]}$ Sample the parameters $\bm\psi_c$ for $c=1,\dots,C-1$ exploiting the continuation-ratio parametrization and the results for Bayesian logistic regression in \cite{polsonBayesianInferenceLogistic2013}:\\
\For {$t \text{ such that } G_t>c-1$}
{
Sample the Pólya-gamma distributed data $\omega_{t,c}$ from $(\omega_{t,c}|-)\sim PG(1,\bm z_t'\bm\psi_c)$.
}
Update $\psi_c$ from $(\psi_c|-)\sim N(\bm\mu_{\psi,c},\bm\Sigma_{\psi,c})$ by Pólya-gamma augmentation, where \\
$\bm\mu_{\psi,c}=\bm\Sigma_{\psi,c}\left[ \bm X' \bm\kappa_c \right]$, $\bm\Sigma_{\psi,c}=\left[ \bm X' diag(\omega_{1,c},\dots,\omega_{\bar{T}_c,c}) \bm X + \bm \tilde{\Upsilon}_c \right]^{-1}$, with \\
$\bm \kappa_c = (\bar{z}_{1,c}-0.5,\dots,\bar{z}_{\bar{T}_c,c}-0.5)'$, $\bar{z}_{t,c}=1$ if $G_t =c$ and $\bar{z}_{t,c}=0$ if $G_t>c$, and $\bm\tilde{\Upsilon}_c = diag(\bm \bar{\tau}^{-2}\bm\upsilon_c^{-2})$. \\
\vspace{0.2cm}
$\textbf{[2b]}$ Update the Horseshoe prior parameters following \cite{makalicSimpleSamplerHorseshoe2016} from: \\
$(\upsilon_{j,c}^2|-) \sim IG\left(1, \frac{\psi_{j,c}^2}{2\bar{\tau}^2_c} + \frac{1}{\eta_{j,c}}\right)$, $(\eta_{j,c}|-) \sim IG\left(1, 1+ \frac{1}{\lambda_{j,c}^2}  \right)$ \\
$(\bar{\tau}_{c}^2|-) \sim IG\left(\frac{K+1}{2}, \frac{1}{\xi_c} +  \sum_{j=1}^k \frac{\psi_{j,c}^2}{2\upsilon_{j,c}^2} \right)$, 
$(\xi_{c}|-) \sim IG\left(1, 1+ \frac{1}{\bar{\tau}_{c}^2}  \right)$.
}
$\textbf{[3]}$ Sample the parameters $\bm\beta_c$ from $(\bm \beta_c|-)\sim N(\bm\mu_{\beta_c},\Sigma_{\bm\beta_c})$, where\\ 
$\bm\mu_{\beta_c} =\bm \Sigma_{\beta_c} \left( \tau_c \bm X_c'\bm y_c + \underline{\bm\Sigma}^{-1}_\beta \underline{\bm\mu}_\beta\right)$, $\bm\Sigma_{\beta_c} = \left(\tau_c \bm X_c' \bm X_c +\underline{\bm\Sigma}^{-1}_\beta \right )^{-1}$. $\bm X_c$ and $\bm y_c$ correspond to the observations for which $G_t=c$. \\
$\textbf{[4]}$ Draw the precision parameter $\tau_c$ for $c=1,\dots,C$:\\
\For {$c=1,\dots,C$}
{
Sample $\tau_c$ from $(\tau|-)\sim G\left( a_\tau + \frac{1}{2} \sum_{t=1}^T  \mathbb{I} (G_t =c), b_\tau + \frac{1}{2} \sum_{t:G_i=c}\left\{ y_t-\bm x_t' \bm \beta_c \right\}^2 \right)$
}
}
\end{algorithm}

\subsection{Variational Bayes Inference}

Variational Bayes (VB) and MCMC both provide approximations to a given posterior. However; while MCMC relies on sampling, VB approximates the posterior through solving an optimization problem. For a general introduction to VB see \cite{bleiVariationalInferenceReview2017}. Because this optimization problem requires fewer iterations than sampling from an MCMC chain, VB's main advantage lies in its computational speed. As a caveat, whereas MCMC guarantees exact draws from the posterior, VB yields a solution close to the posterior. As such, VB is particularly useful for tasks where computational time is critical and accurate parameter estimates are less of a concern, such as, big-data problems, models with highly intractable posteriors, real-time monitoring tasks, or forecasting exercises. The algorithm proposed in this paper extends the VB algorithm in \cite{rigonTractableBayesianDensity2020a} and introduces shrinkage into the sequential logistic regressions. \\

The algorithm evolves as follows:
\begin{enumerate}
\item Update the variational density of the mixture component assignment indicators $\zeta_{tc}$. Here the model is augmented using the binary indicators $z$ instead of the mixture component membership indicators $G_t$. This step differs from the MCMC sampler and follows \cite{rigonTractableBayesianDensity2020a}
\item Update the variational densities corresponding to the regression coefficients in the logistic regressions, $\bm\psi_c$, for all mixture components. 
\item Update the variational density of the horseshoe prior coefficients. This step differs slightly from the MCMC sampler in that the horseshoe prior is parametrized differently.  
\item Update the variational density of the Pólya-Gamma indicators $\omega_tc$.
\item Update the variational density of the regression coefficients, $\bm \beta_c$, and the precision parameters, $\tau_c$, of the individual mixture components. 
\end{enumerate}

Throughout lower case letters denote scalars, bold lower case letters denote vectors and bold capital letters denote matrices. To ease notation, whenever indices referring to the mixture assignment indicators, time periods, or variables are suppressed, they are collected in a single vectors. Collect all model parameters in $\bm \Theta = (\bm \beta, \bm \tau, \bm z, \bm \psi, \bm \omega, \bm \nu, \bm \bar{\bm \tau}^2, \bm \eta, \bm \xi)$. For a family of tractable densities $q(\bm\Theta)$, we aim to find a density $q^\star$ that best approximates the posterior $p(\bm \Theta| \bm x)$ by minimizing

\begin{equation}
q_x^\star(\bm \Theta ) = \underset{q\in \mathcal{Q}}{\text{argmin}} \: \mathbb{D}_{KL} \left( q(\bm \Theta || p(\bm \Theta |\bm x)\right).
\end{equation}

This is equivalent to maximizing

\begin{equation}
ELBO = \mathbb{E}_{q_x(\bm \Theta )}\left[ \text{log} \: p(\bm x)\right] + \mathbb{E}_{q_x(\bm \Theta )}\left[ \text{log} \: p(\bm \Theta)\right] - \mathbb{E}_{q_x(\bm \Theta )}\left[ \text{log} \: q_x(\bm \Theta)\right],
\end{equation}

where $KL$ denotes the Kullback-Leibler divergence. Importantly, the solution requires optimizing over a family of distribution functions and hence the application of variational calculus. A key part of VB is to simplify the variational posterior by factorization, such that the resulting optimization problem becomes tractable. On the one hand, the simpler the resulting posterior, the easier the optimization problem. On the other hand, this implies independence assumptions between the individual variational components and hence leads to approximation error. This gives rise to a trade-off \citep{ormerodVariationalBayesApproach2017}. In the proposed estimator, the so-called mean-field factorization is applied. This yields

\begin{align}\label{VBpostApp}
q_x(\Theta) = &\prod_{c=1}^{C-1}  q_x(\bm\psi_c) \prod_{c=1}^{C-1} q_x(\xi_c) \prod_{c=1}^{C-1}\prod_{j=1}^{n} q_x(\nu_{j,c}^2) \prod_{c=1}^{C-1}\prod_{j=1}^{n} q_x(\bar{\tau}_{j,c}^2)\prod_{c=1}^{C-1}\prod_{j=1}^{n}q_x(\eta_{j,c}^2) \nonumber\\
& \cdot \prod_{c=1}^C  q_x(\bm \beta_c) \prod_{c=1}^C q_x(\tau_c) \prod_{c=1}^{C-1} \prod_{t=1}^T q_{x_t}(z_{tc}) \prod_{c=1}^{C-1}\prod_{t=1}^T q_{x_t}(\omega_{tc}). 
\end{align}

This implies independence between the regression coefficients in the logistic regressions and kernel regressions, the cluster assignment indicators, the parameters of the horseshoe prior, and the Pólya-Gamma indicators. Additionally, conditional independence between the priors is assumed.

%

Given these assumptions, the solution to the optimization problem can be obtained by sequentially iterating over the densities

\begin{equation}\label{VBdensAPP}
q_x(\bm \Theta_l ) \propto \text{exp} \:\: \mathbb{E}_{q_x(\bm \Theta_{(-l)} )} \left( \text{log} \:\: p(\bm \Theta_l | \bm \Theta_{(-l)}, \bm x) \right)
\end{equation}
where $\bm\Theta_{(-l)}$ denotes all elements of $\bm\Theta$, excluding those in the $l^{th}$ group, $l=1,\cdots,L$.

As a result, the variational posterior can be obtained by calculating the variational expectation of the conditional posterior. As described above, the accuracy of the variational approximation hinges on how well the partitioning matches the independence structure of the parameters in the target posterior. For a general discussion of this issue see e.g. \cite{ormerodVariationalBayesApproach2017}. To arrive at the final variational densities, insert \ref{VBpostApp} and the priors into \ref{VBdensAPP}. The final algorithm is given in algorithm \ref{alg:VB}.



\RestyleAlgo{ruled} 
\begin{algorithm}[H]
\scriptsize
\caption{Variational Bayes algorithm}\label{alg:VB}
$\textbf{[0]}$ Initialize the model parameters. Let $q^m(.)$ denote a generic variational distribution at iteration m. \\
\Begin
{
$\textbf{[1]}$ Compute the variational density $q^*_{\bm x_t}(z_{ic})$ for each $t=1,\dots,T$ and $c=1,\dots,C-1$: \\
\For {$t=1:T$}
{
\For {$c=1:C$}
{
Following \cite{rigonTractableBayesianDensity2020a}, $q^*_{x_t}(z_{ih})$ coincides with the probability mass function of $Bern(\rho_{ih})$, where
\begin{equation*}
\rho_{ih}=\frac{1}{1 + exp\left(-\left[\bm x_t'\mathbb{E}(\phi_c)+ \sum_{l=c}^C \zeta_{tl}^c \left(  0.5 \cdot\mathbb{E}(\text{log } \tau_l) - 0.5 \cdot\mathbb{E}(\text{log } \tau_l)\cdot\mathbb{E}(y_t-\bm x_t' \bm\beta_l)^2 \right) \right]\right)}.
\end{equation*}
with $\zeta_{tl}^c= \prod_{r=1}^{l-1}(1-\rho_{tr})$ for $l=c$ and $\zeta_{tl}^c= -\rho_{tl}\prod_{r=1,r\neq c}^{l-1}(1-\rho_{tr})$, and $\rho_{tC} =1$.\\
\textbf{Assign:} $\mathbb{E}(\zeta_{tc}) = \mathbb{E}\left(z_{tc}\prod_{l=1}^{c-1}(1-z_{tl})\right) = \rho_{tc}\prod_{l=1}^{c-1}(1-\rho_{tl})$, and $\mathbb{E}(z_{tc})=\rho_{tc}$.
}
}
$\textbf{[2]}$ Update $q_x^*(\bm\psi_c)$ and $q_x^*(\bm\upsilon_c^2)$, $q_x^*(\bar{\bm \tau}_c^2)$, $q_x^*(\bm\eta_c)$, and $q_x^*(\xi_c)$ for $c=1,\dots,C-1$:\\
\For {$c=1,\dots,C-1$}
{
\vspace{0.2cm}
$\textbf{[2a]}$ Update $q_x^*(\bm\psi_c)$ from $N(\bm\mu_{\psi,c},\bm\Sigma_{\psi,c})$, with\\
$\bm\Sigma_{\psi,c}=\left( \bm X'\bm V_c \bm X +  \bm \Lambda^{-1}\right)^{-1}$, $\bm\mu_{\psi,c} = \bm\Sigma_{\psi,c}\left(\bm X'\bm\kappa_c \right)$, with $\bm \kappa_c = (\mathbb{E}({z}_{1,c})-0.5,\dots,\mathbb{E}({z}_{T,c})-0.5)'$, \\$\bm V_c=diag\left(\mathbb{E}(\omega_{1c}),\dots,\mathbb{E}(\omega_{Tc}) \right)$, and $\bm\Lambda^{-1}_{c} = diag\left( \mathbb{E}(\bm \upsilon_c^{-2})\mathbb{E}(\bm\tau^{-2}_c)\right)$.\\
\textbf{Assign:} $\mathbb{E}(\bm \psi_c) = \bm\mu_{\psi,c}$ and  $\mathbb{E}(\bm \psi_c^2) = \bm\mu_{\psi,c}^2+diag(\bm\Sigma_{\psi,c})$.\\
\vspace{0.2cm}

$\textbf{[2b]}$ Update $q_x^*(\bm\upsilon_c^2)$, $q_x^*(\bar{\bm \tau}_c^2)$, $q_x^*(\bm\eta_c)$, and $q_x^*(\xi_c)$ for $c=1,\dots,C-1$:\\

$q^*(\bm \upsilon_{j,c}^2) = IG( a_{\upsilon,c}, b_{\upsilon,c}) = IG\left(1,\frac{\mathbb{E} (\psi_{j,c}^2)}{2 } + \mathbb{E}\left(\frac{1}{\eta_{j,c}}\right)\right)$, 

$q^*({\eta}_{j,c}) = IG( a_{\eta,c}, b_{\eta,c}) = IG\left(1,\mathbb{E}\left(\frac{1}{\upsilon_{j,c}^2}\right) + b_\psi^{-2} \mathbb{E}\left( \frac{1}{\bar{\tau}^2_{j,c}}\right)\right)$,

$q^*(\bar{\tau}_{j,c}^2)  = IG( a_{\bar{\tau},c}, b_{\bar{\tau},c}) = IG\left(1,  b_\psi^{-2}  \mathbb{E}\left( \frac{1}{\eta_{j,c}}\right) + \mathbb{E}\left( \frac{1}{\xi_{c}}\right) \right)$,

$q^*({\xi}_{c})  = IG( a_{\xi,c}, b_{\xi,c}) = IG\left(\frac{n+1}{2}, 1 + \sum_{j=1}^{n} \mathbb{E}\left(\frac{1}{\bar{\tau}_{j,c}^2} \right)\right)$, \\
for $j = 1,\cdots,n$. 

\textbf{Assign:} $\mathbb{E}(\upsilon_{j,c}^2) =   \frac{ a_{\upsilon,c}}{ b_{\upsilon,c}}  $, $\mathbb{E}( \eta_{j,c}) =  \frac{ a_{\eta,c}}{ b_{\eta,c}} $, $\mathbb{E}(\bar{\tau}_{j,c}) = \frac{ a_{\tau,c}}{ b_{\tau,c}}  $, and $\mathbb{E}(\xi_c)=  \frac{ a_{\xi,c}}{ b_{\xi,c}}  $.
}

$\textbf{[3]}$ Update $q^*_x(\omega_{tc})$ for $t=1,\dots,T$ and $c=1,\dots,C$.\\
\For {$t=1:T$}
{
\For {$c=1:C$}
{
Update $q^*_x(\omega_{tc})\sim PG(1,\delta_{tc})$, with $\delta_{tc}^2=\bm x_t'\mathbb{E}(\bm \psi_c^2) \bm x_t$.\\
\textbf{Assign:} $\mathbb{E}(\omega_{tc})=0.5\cdot\delta_{tc}^{-1}tanh(0.5\cdot\delta_{tc})$.
}
}

$\textbf{[4]}$ Update $q^*_x(\bm\beta_c)$ and $q^*_x(\tau_c)$ for $c=1,\dots,C$:\\
\For {$c=1:C$}
{
 
 Update $q^*_x(\bm\beta_c)\sim N(\bm\mu_{\beta_c},\bm\Sigma_{\beta_c})$ and $q^*_x(\tau_c)\sim IG(a_{\tau_c},b_{\tau_c})$, with $\bm\Sigma_{\beta_c} = \left( \bm X'\bm \Gamma_c \bm X +\underline{\bm\Sigma}^{-1}_\beta \right)^{-1}$, \\ $\bm\mu_{\beta_c} = \bm\Sigma_{\beta_c} \left(\bm X\bm \Gamma_c \bm y+ \underline{\bm\Sigma}^{-1}_\beta \underline{\bm\mu}_\beta\right)$, $a_{\tau_c}=a_\tau + 0.5\sum_{t=1}^T \mathbb{E}(\zeta_{tc})$, and \\$b_{\tau_c}= b_\tau +0.5 \sum_{t=1}^T \mathbb{E}(\zeta_{tc}) \mathbb{E}\left( (y_t-\bm x_t'\bm \beta_c)^2 \right)$, with $\Gamma_c = \mathbb{E}(\tau_c)diag(\mathbb{E}(\zeta_{1c}),\dots,\mathbb{E}(\zeta_{Tc}))$.\\
 \textbf{Assign:} $\mathbb{E}(\bm \beta_c)=\bm \mu_{\beta_c}$, $\mathbb{E}(\bm \beta_c^2)=\bm \mu_{\beta_c}^2 + diag \left( \bm\Sigma_{\beta_c}\right)$, and $\mathbb{E}(\tau_c)=\frac{a_{\tau_c}}{b_{\tau_c}}$.
 }

}
\end{algorithm}

Finally, note that a different parametrization of the horseshoe prior is used: 

\begin{align}
\bm\psi_{c}| \left\{\nu_{j,c}^2, \eta_{j,c}, \bar{\tau}^2_{j,c}  \right \}_{j=1}^{n}, \xi_{c}  & \sim  N(\bm 0, \bm\Lambda_{c}), \\
\bar{\lambda}^{2}_{j,c} | \upsilon_{j,c} & \sim G^{-1}\left(\frac{1}{2},\frac{1}{\upsilon_{j,c}} \right), \:\:\: \text{for } j = 1,\cdots,n,\nonumber\\
\upsilon_{j,c}|\bar{\tau}^2_{j,c} & \sim G^{-1}\left(\frac{1}{2},\frac{1}{b_\psi^2  \tau^2_{j,c}  }\right), \:\:\: \text{for } j = 1,\cdots,n,\nonumber\\
\bar{\tau}^2_{j,c} | \xi_{c} & \sim G^{-1}\left(\frac{1}{2},\frac{1}{\xi_{c}}\right),\nonumber\\
\xi_{c} & \sim G^{-1}\left(\frac{1}{2},1\right),\nonumber
\end{align}

where $b_\psi$ is a hyperparameters, which is set to $0.0001$ in application.

\section{Algorithms for Computing the Drivers of Risk}\label{app:drivers}


%

\RestyleAlgo{ruled} 
\begin{algorithm}[h!]
\scriptsize
\caption{Forecast Density Decompositions}\label{alg:shap_dens}
$\textbf{[0]}$ Specify the number of samples $M$, and the number of grid points, $g$. Generate a grid for $\hat{\pi}$, with $\hat{\pi}_i \in [\lfloor \pi_{1:T} \rfloor,...,\lceil \pi_{1:T} \rceil]$ for $i=1,\dots,g$. \\
$\textbf{[1]}$ Compute the contribution for variable $j$ to the forecast distribution for $t+h$ at time $t$. \\
\Begin
{

$\textbf{[a]}$ Iterate over random samples. The sampling population is the set of all time-period/variable permutation pairs. \\
\For {$m=1:M$}
{
$\textbf{[i]}$Sample a random time period, $t^* \in_R [1:t-1]$, from the the preceding sample of observations and set $\bm z_t = \bm x_{t^*}$.\\
$\textbf{[ii]}$ Choose a random permutation of variables, $S$, i.e. sample $S \subseteq_R [1,\dots,j-1,j+1,\dots,n]$, where $|S|\leq n-1$. \\
$\textbf{[iii]}$ Create two new data vectors. For the first vector, replace the values in $\bm x_t$ with the values in $\bm z_t$ for the indices in $S$. For the second vector, also replace $x_{t,j}$ with $z_{t,j}$. 
As an example, let $S=[1,j+1,\dots,\j+n]$. This yields: 
\begin{align*}
\bm x_t^{+j} & = [z_{t,1},x_{t,2}, \dots, x_{t,j-1},x_{t,j},z_{t,j+1}, \dots  ,z_{t,n}]\\
\bm x_t^{-j} & = [z_{t,1},x_{t,2}, \dots, x_{t,j-1}, z_{t,j},z_{t,j+1}, \dots  ,z_{t,n}].
\end{align*} 
$\textbf{[iv]}$ Compute the contribution for $j$ between the grid points $\hat{\pi}_i$ and $\hat{\pi}_{i-1}$ at draw $m$\\

\begin{equation*}
V_{m,j}^i = \int_{\hat{\pi}_{i-1}}^{\hat{\pi}_i} f_{x}(\pi)_{t+h|t}^{+j}d\pi - \int_{\hat{\pi}_{i-1}}^{\hat{\pi}_i} f_{x}(\pi)_{t+h|t}^{-j}d\pi \text{\phantom{aaa}   for every   \phantom{aaa}} i=2,\dots,g.
\end{equation*}
}
$\textbf{[b]}$ Compute the contribution of $j$ between grid points $\hat{\pi}_i$ and $\hat{\pi}_{i-1}$, $\hat{\varphi}_j^{i}$, by taking the average over all samples, $M$. 
\begin{equation*}
\hat{\varphi}_j^{i}=\frac{1}{M}\sum_m^M V_{m,j}^i.
\end{equation*}
}
Notes: $\subseteq_R$ and $\in_R$ denote a random subset and random element, respectively. The integrals can be computed with quadrature methods.
\end{algorithm}

This section gives the full algorithms for computing the drivers of risk introduced in the main body of the text. Specifically, \autoref{alg:shap_dens} allows to decompose the forecast density into its drivers, while \autoref{alg:shap_risk} decomposes the risk measures.  

 \RestyleAlgo{ruled} 
\begin{algorithm}[h!]
\scriptsize
\caption{Risk Measure Decompositions}\label{alg:shap_risk}
$\textbf{[0]}$ Specify the number of samples $M$, and set $\alpha$, $\beta$, $\bar{\pi}$, $\underline{\pi}$. \\
$\textbf{[1]}$ Compute the contribution for variable $j$ to the risk measures for $t+h$ at time $t$. \\
\Begin
{

$\textbf{[a]}$ Iterate over random samples. The sampling population is the set of all time-period/variable permutation pairs. \\
\For {$m=1:M$}
{
$\textbf{[i]}$Sample a random time period, $t^* \in_R [1:t-1]$, from the the preceding sample of observations and set $\bm z_t = \bm x_{t^*}$.\\
$\textbf{[ii]}$ Choose a random permutation of variables, $S$, i.e. sample $S \subseteq_R [1,\dots,j-1,j+1,\dots,n]$, where $|S|\leq n-1$. \\
$\textbf{[iii]}$ Create two new data vectors. For the first vector, replace the values in $\bm x_t$ with the values in $\bm z_t$ for the indices in $S$. For the second vector, also replace $x_{t,j}$ with $z_{t,j}$. 
As an example, let $S=[1,j+1,\dots,\j+n]$. This yields: 
\begin{align*}
\bm x_t^{+j} & = [z_{t,1},x_{t,2}, \dots, x_{t,j-1},x_{t,j},z_{t,j+1}, \dots  ,z_{t,n}]\\
\bm x_t^{-j} & = [z_{t,1},x_{t,2}, \dots, x_{t,j-1}, z_{t,j},z_{t,j+1}, \dots  ,z_{t,n}].
\end{align*} 
$\textbf{[iv]}$ Compute the contribution for $j$ to $DR_\alpha$ and $EIR_\beta$ at draw $m$.\\

\begin{align*}
V_{m,j}^{DR_\alpha} &= -\left(\int_{-\infty}^{\underline{\pi}} (\underline{\pi}-\pi)^\alpha f_{x}(\pi)_{t+h|t}^{+j}d\pi -\int_{-\infty}^{\underline{\pi}}(\pi-\bar{\pi})^\alpha f_{x}(\pi)_{t+h|t}^{-j}d\pi\right)\\
V_{m,j}^{EIR_\beta} &= \int_{\bar{\pi}}^{\infty} (\underline{\pi}-\pi)^\beta f_{x}(\pi)_{t+h|t}^{+j}d\pi - \int_{\bar{\pi}}^{\infty}(\pi-\bar{\pi})^\beta f_{x}(\pi)_{t+h|t}^{-j}d\pi
\end{align*}
}
$\textbf{[b]}$ Compute the contribution of $j$ to $DR_\alpha$ and $EIR_\beta$, $\hat{\varphi}_j^{DR_\alpha}$ and $\hat{\varphi}_j^{EIR_\beta}$, by taking the average over all samples, $M$. 
\begin{equation*}
\hat{\varphi}_j^{DR_\alpha}=\frac{1}{M}\sum_m^M V_{m,j}^{DR_\alpha} \text{     and     } \hat{\varphi}_j^{EIR_\beta}=\frac{1}{M}\sum_m^M V_{m,j}^{EIR_\beta}.
\end{equation*}
}
Notes: $\subseteq_R$ and $\in_R$ denote a random subset and random element, respectively. The integrals can be computed with quadrature methods.
\end{algorithm}


\section{Additional Algorithms} 
\subsection{Computing Quantiles}\label{app:quantiles}
\noindent By the properties of the normal distribution, the sum of multiple normals is again normal. The CDF and PDF are hence available as the sum of the appropriately reweighted individual mixture components, as shown above. However, this convenient property does not hold for the quantile function. An analytical solution is hence not available, however, quantiles can be computed using simple root finding procedure, where

\begin{equation}\label{quantile_function}
\Phi_x({\pi}_{t+h})\big\rvert_{\hat{Q}_x^\tau({\pi}_{t+h})} - \tau = 0,
\end{equation} 

\noindent where $\tau$ denotes a percentile of interest and $\Phi_x({\pi}_{t+h})$ denotes the CDF of the mixture model that is evaluated at a candidate solution $\hat{Q}_x^\tau({\pi}_{t+h})$. The desired percentile, $Q_x^\tau({\pi}_{t+h})$, is the solution to Equation (\ref{quantile_function}). Note that because the CDF is proper and we estimate the entire distribution simultaneously, the resulting conditional quantiles do not cross. This is an advantage compared to univariate quantile regressions models. \\ 
 
\subsection{Computing the Optimal Number of Mixture Components}\label{app:mix_comp}
As mentioned above, the performance of the mixture model also hinges on the truncation point of the infinite sum of mixture components. Especially in forecasting exercises, it might hence be useful to select this truncation point automatically. In the Bayesian context, the marginal data likelihood allows to solve this model selection problem; however the computation of the marginal likelihood is usually computationally cumbersome. In the context of Variational Bayes, the ELBO, i.e. the convergence criterion of the optimization problem, provides an accurate approximation of the marginal likelihood. Because the algorithm is computationally fast, solving the model selection problem with VB is hence an ideal alternative. This gives rise to the following algorithm: 

\begin{enumerate}
\item Set an upper bar on the number of mixture components $\bar{C}$.
\item For $C = 1,\dots,\bar{C}$, re-estimate the model with VB an store the ELBO.
\item Set $$C^*=\argmax_C ELBO_C.$$
\item Re-estimate/store the estimates for the model estimated with $C^*$.
\end{enumerate}

\noindent Because the algorithm is fast, this optimization can even be repeated in real-time at every new data vintage, provided $\bar{C}$ is not too large. For macroeconomic data, $\bar{C}=5$ or $\bar{C}=10$ are useful starting points.

\section{Proof}\label{app:proof}

\subsection{Proof of Lemma \ref{def:lemma}}

\begin{proof}
Let $f_x(\pi)_{t+h|t}$ denote the forecast distribution of inflation at time $t$ for horizon $t+h$. Further, let $\pi^\star$ denote a set of $g$ ordered grid points across the entire support of $\pi$. ${\pi}^\star_i$ and ${\pi}^\star_{i-1}$ hence denote two neighbouring grid points such that ${\pi}^\star_{i-1}<{\pi}^\star_i$. Further, let $\pi_1^\star = -\infty$ and $\pi_g^\star=\infty$. Then

\begin{align*}
\int_{-\infty}^{\infty}f_x(\pi)_{t+h|t} dx= \sum_{i=2}^g \int_{{\pi}^\star_{i-1}}^{{\pi}^\star_i}f_x(\pi)_{t+h|t} dx = 1
\end{align*}

Analogously,

\begin{align*}
\int_{-\infty}^{\infty}f_x(\pi)_{t+h} dx= \sum_{i=2}^g \int_{{\pi}^\star_{i-1}}^{{\pi}^\star_i}f_x(\pi)_{t+h} dx = 1.
\end{align*} 

Given the definition of the predictor contributions
 
\begin{equation*}
\sum_j^n\varphi_j^{\pi_i^\star} + \int_{{\pi}^\star_{i-1}}^{{\pi}^\star_i} f_x(\pi)_{t+h}d\pi =\int_{{\pi}^\star_{i-1}}^{{\pi}^\star_i}f_x(\pi)_{t+h|t} dx.
\end{equation*} 

As a result

\begin{equation*}
\int_{-\infty}^{\infty}f_x(\pi)_{t+h|t} dx= \sum_{i=2}^g \int_{{\pi}^\star_{i-1}}^{{\pi}^\star_i}f_x(\pi)_{t+h|t} dx = \sum_{i=2}^g \left[ \sum_j^n\varphi_j^{\pi_i^\star} + \int_{{\pi}^\star_{i-1}}^{{\pi}^\star_i} f_x(\pi)_{t+h}d\pi \right]=1.
\end{equation*}
 
\end{proof}

\newpage
\onehalfspace
\begin{center}
{\Large Online Supplement to ``Mixing it up: Inflation at risk''}\\
Maximilian Schr\"{o}der
\end{center}

\setcounter{page}{1}
\setcounter{footnote}{0}
\renewcommand{\thesection}{\arabic{section}} \setcounter{section}{0}
\renewcommand{\theequation}{\arabic{equation}} \setcounter{equation}{0} %
\renewcommand{\thetable}{\arabic{table}} \setcounter{table}{0}
\renewcommand{\thefigure}{\arabic{figure}} \setcounter{figure}{0}

\section{Benchmark Models}\label{supp:Benchmark_models}
This appendix provides details on the estimation of the benchmark models used in the forecasting exercise in the main body of the paper. All benchmarks can be obtained as special cases of the Variational Bayes algorithm for the density regression model in \autoref{alg:VB}.

\begin{enumerate}
\item[AR:] The AR model can be obtained as a special case of step [4] in \autoref{alg:VB}, by setting $\mathbb{E}(\zeta_{tc}),\dots,\mathbb{E}(\zeta_{Tc})=1$.  The horseshoe prior can be implemented by modifying step [2b] of \autoref{alg:VB} , by setting $\psi_c = \beta_c$, and by replacing $\underline{\Sigma}_\beta^{-1}$ in step [4] by $\Lambda^{-1}$ from step [2a].
\item[TVP-AR:] The TVP-AR can be implemented as a special case of the AR model. In particular, the TVP regression can be rewritten as a static regression model with $T\cdot n$ parameters as described in \cite{korobilisBayesianApproachesShrinkage2022}. In particular, the regression model is then given by 

\begin{equation*}
\begin{bmatrix}
y_1\\
y_2\\
\vdots\\
y_T
\end{bmatrix} = 
\begin{bmatrix}
\bm x_1 & 0   &  \dots & \dots & 0 \\
\bm x_2 & \bm x_2   &  \ddots & \dots  & 0 \\
\vdots	&  \vdots   &  \ddots & \ddots   & \vdots \\
\bm x_T & \bm x_T   &  \dots & \dots & \bm x_T \\
\end{bmatrix}
\begin{bmatrix}
\bm\beta_1\\
\Delta\bm\beta_2\\
\vdots\\
\Delta\bm\beta_T
\end{bmatrix} +\varepsilon_t,
\end{equation*} 
where $\bm x_t = [x_{t,1}, x_{t,2}, \dots, x_{t,n}]$ and $\Delta\bm \beta_t = [\Delta\beta_{t,1}, \Delta\beta_{t,2}, \dots, \Delta \beta_{T,1} , \Delta \beta_{t,n}]'$. Estimates of $\bm \beta_t$ are then obtained by computing the cumulative sum of the parameter vector across time for all variables. Apart from these transformations, the TVP-AR model can then be estimated analogously to the AR model. Applying e.g. the horseshoe prior on the parameters of the regression model mitigates  overparameterization issues. 
\item[SV-AR:] To allow for stochastic volatility, the algorithm in  \cite{koopBayesianDynamicVariable2023} can be used to modify the algorithm for the AR model. In particular, the variational expectation of time-varying precision parameter $\tau_t\sim Gamma(\delta a_{t-1},\delta a_{t-1})$, is $\mathbb{E}(\tau_t)=a_t/b_t$, where $a_t = 0.5 + \delta a_{t-1}$, $b_t = 0.5\cdot\varepsilon_t^2 + \delta b_{t-1}$, and $\delta$ is a smoothing parameter set to 0.8, following \cite{koopBayesianDynamicVariable2023}, and $\varepsilon_t$ is the regression residual. Smoothed precision parameters, $\tau_t^s$, can be obtained by iterating computing $\tau_t^s = (1-\delta)\tau_t + \delta \tau_{t+1}$. backwards through time. 
\item[TVPSV-AR:] The TVPSV-AR can be computed by combining the modifications for the TVP-AR and SV-AR model.
\item[T-AR:] The AR with t-distributed residuals can be obtained as another special case of the AR algorithm, following the steps in \cite{christmasRobustAutoregressionStudentt2011}. In particular, in step [4] of \autoref{alg:VB}, $\mathbb{E}(\zeta_{Tc})$ in $\Gamma_c$ is replaced with $\mathbb{E}(z_t)$, where $z_t\sim Gamma(a_z,b_z)$ is a latent variable that modifies the precision. The coefficient updates are $a_z = (\mathbb{E}(d)+1)/2$ and $b_z=\mathbb{E}(d)/2 +\mathbb{E}(\tau)/2 \cdot \mathbb{E} (\varepsilon^2)$. The degrees of freedom parameter $d\sim Gamma(a_d,b_d)$ has parameter updates $a_d = \bar{a}_d + T/2$ and $ b_d = \bar{b}_d - 0.5\left( T + \sum_t^T  [\mathbb{E}(log(z_t))-\mathbb{E}(z_t)]   \right)$, where $\bar{a_d}=0.04$ and $\bar{b_d}=0.01$ are prior parameters. Additionally, $\mathbb{E}(z_t) = a_z/b_z$, $\mathbb{E}(log(z_t))=\psi(a_z) - log(b_z)$ , where $\psi(\cdot)$ is the digamma function, and $\mathbb{E}(d) = a_d/b_d$.
\item[QR:] In the QR model, the residuals have an Asymmetric Laplace distributions. However, similar to the T-AR, the QR model can be estimated as a special case of the AR model by rewriting the Asymmetric Laplace as a normal-exponential mixture, as in \cite{Limetal2020} and others. For quantile level $q$, define $\kappa_1 = (1-2q)/(q(1-q))$ and $\kappa_2 = 2/(q(1-q))$. The modified regression equation reads $y_t = \bm\beta(q)' \bm x_t + \kappa_1 z_t(q) + \kappa_2 \sqrt{\sigma(q) z_t(q)} \varepsilon_t$. In step [4] of \autoref{alg:VB}, $\Gamma_c=1/\kappa_2^2\mathbb{E}(1/\sigma)diag(\mathbb{E}(1/z_1(q)),\dots,\mathbb{E}(1/z_T(q)))$. In addition, $\mu(q)_{\beta_c} = \Sigma_\beta(q)[\bm x' \Gamma_c \bm y -\kappa_1/\kappa_2^2 \mathbb{E}(1/\sigma(q))\sum_{t=1}^T \bm x_t]$. $z_t$ is given by the inverse Gaussian distribution with $z_t(q)\sim InG(0.5,a^z(q),b_t^z(q))$, where $a^z(q)= \mathbb{E}(1/\sigma(q)) (2+\kappa_1/\kappa_2^2)$, $b_t^z(q) =\mathbb{E}(1/\sigma(q))[1/\kappa_2^2((\bm y_t - \bm x_t \beta(q))^2 + \bm x_t diag(\Sigma_\beta(q)) \bm x_t')] $. The expectations $\mathbb{E}(z_t)= (\sqrt{b_t^z(q)}K_{3/2}(\sqrt{a_t^z(q)b_t^z(q)}))/(\sqrt{a_t^z(q)}K_{1/2}(\sqrt{a_t^z(q)b_t^z(q)}))$ and $\mathbb{E}(1/z_t)= (\sqrt{a_t^z(q)}K_{3/2}(\sqrt{a_t^z(q)b_t^z(q)}))/(\sqrt{b_t^z(q)}K_{1/2}(\sqrt{a_t^z(q)b_t^z(q)}))-1/b^z_t(q)$, where $K_p(\bullet)$ is the Bessel function of order $p$. Finally, for $\sigma\sim IG(a_\sigma(q),b_\sigma(q))$ the updates are $a_\sigma(q)= \bar{a}_\sigma(q)+3T$ and $b_\sigma(b)= \bar{b}_\sigma(q)  + \sum_{t=1}^T [\mathbb{E}(1/z_t(q)) ((y_t - \mathbb{E}(\bm\beta(q))'\bm x_t)^2 + \bm x_t \Sigma_\beta(q)\bm x_t')/(2\kappa_2^2) - \kappa_1(y_t - \mathbb{E}(\bm\beta(q))'\bm x_t)/(\kappa_2^2)+(1+ \kappa_1^2/2\kappa_2^2)\mathbb{E}(z_t(q))]$, where $\bar{a}_\sigma(q)$ and $\bar{b}_\sigma(q)$ are the priors. Expectations are computed analogously to the other cases.

\end{enumerate}

\section{Additional Results}\label{Additional_Results}

\subsection{Point Forecast Performance}\label{supp:point_fore}

\begin{table}[H]
\caption{Relative RMSE vis-à-vis AR\\}\label{tab:RMSE}
\centering\resizebox{.7\textwidth}{!}{
\begin{threeparttable}
\begin{tabular}{lcccccc}\hline \hline
$h$  & DR$^{\phantom{***}}$     & TVP-AR$^{\phantom{***}}$  &  SV-AR$^{\phantom{***}}$  &  TVPSV-AR$^{\phantom{***}}$  &  T-AR$^{\phantom{***}}$ &  QR$^{\phantom{***}}$ \\\hline
1	&	0.881$^{**\phantom{*}}$	&	1.235$^{\phantom{***}}$	&	0.994$^{\phantom{***}}$	&	1.465$^{\phantom{***}}$	&	0.991$^{\phantom{***}}$	&	0.996$^{\phantom{***}}$	\\
2	&	0.826$^{\phantom{***}}$	&	1.809$^{\phantom{***}}$	&	0.925$^{\phantom{***}}$	&	1.666$^{\phantom{***}}$	&	0.961$^{\phantom{***}}$	&	0.922$^{\phantom{***}}$	\\
3	&	0.853$^{\phantom{***}}$	&	1.006$^{\phantom{***}}$	&	0.871$^{\phantom{***}}$	&	1.980$^{\phantom{***}}$	&	0.950$^{\phantom{***}}$	&	0.907$^{\phantom{***}}$	\\
4	&	0.923$^{\phantom{***}}$	&	1.874$^{\phantom{***}}$	&	0.885$^{\phantom{***}}$	&	2.630$^{*\phantom{**}}$	&	0.969$^{\phantom{***}}$	&	0.894$^{***}$	\\
8	&	0.836$^{***}$	        &	1.151$^{\phantom{***}}$	&	0.796$^{***}$	        &	2.132$^{*\phantom{**}}$	&	0.882$^{***}$	        &	0.836$^{***}$	\\
12	&	0.860$^{**\phantom{*}}$	&	1.501$^{\phantom{***}}$	&	0.769$^{***}$	        &	    1.627$^{***}$	        &	0.819$^{***}$	        &	0.784$^{***}$	\\
\hline\hline
\end{tabular}
\footnotesize
\begin{tablenotes}
\item The table contains the RMSE relative to the AR model. Statistical significance of the Diebold-Mariano test is indicated by the $*$, where $\{***\}=1\%$, $\{**\}=5\%$, $\{*\}=10\%$. For the QR model, RMSE are computed based on forecasts for the median.
\end{tablenotes}
\end{threeparttable}
}
\end{table}

RMSEs relative to the AR models for the forecast horizons $h={1,2,3,4,8,12}$ are shown in \autoref{tab:RMSE}, with the absolute RMSEs available in \autoref{supp:add_tables}. Statistical significance of the Diebold-Mariano test vis-à-vis the AR model is indicated by the asterisks. Notably, only the DR, SV-AR, T-AR, and QR model consistently outperform the AR model, with significant improvements observed mostly for longer forecast horizons. The DR model performs particularly well for forecast horizons of less than one year, while the DR, SV-AR, T-AR, and QR models perform similarly for longer horizons. The findings suggest that the DR model yields accurate point forecasts and that accounting for asymmetries in the forecast distribution also facilitates point forecasts performance gains.

\subsection{Asymmetric Preferences}\label{supp:asymmetric}
 
 \begin{figure}[h]
\hspace{-4cm}\centering
\caption{The Balance of Risk}\label{fig:risk_measures2}
\subcaptionbox{BR for $\alpha=2, \beta=3$ and $h=1$}
{\resizebox*{!}{0.25\textwidth}{\includegraphics[width=0.5\textwidth, trim={6cm 2cm 5cm 2cm},clip]{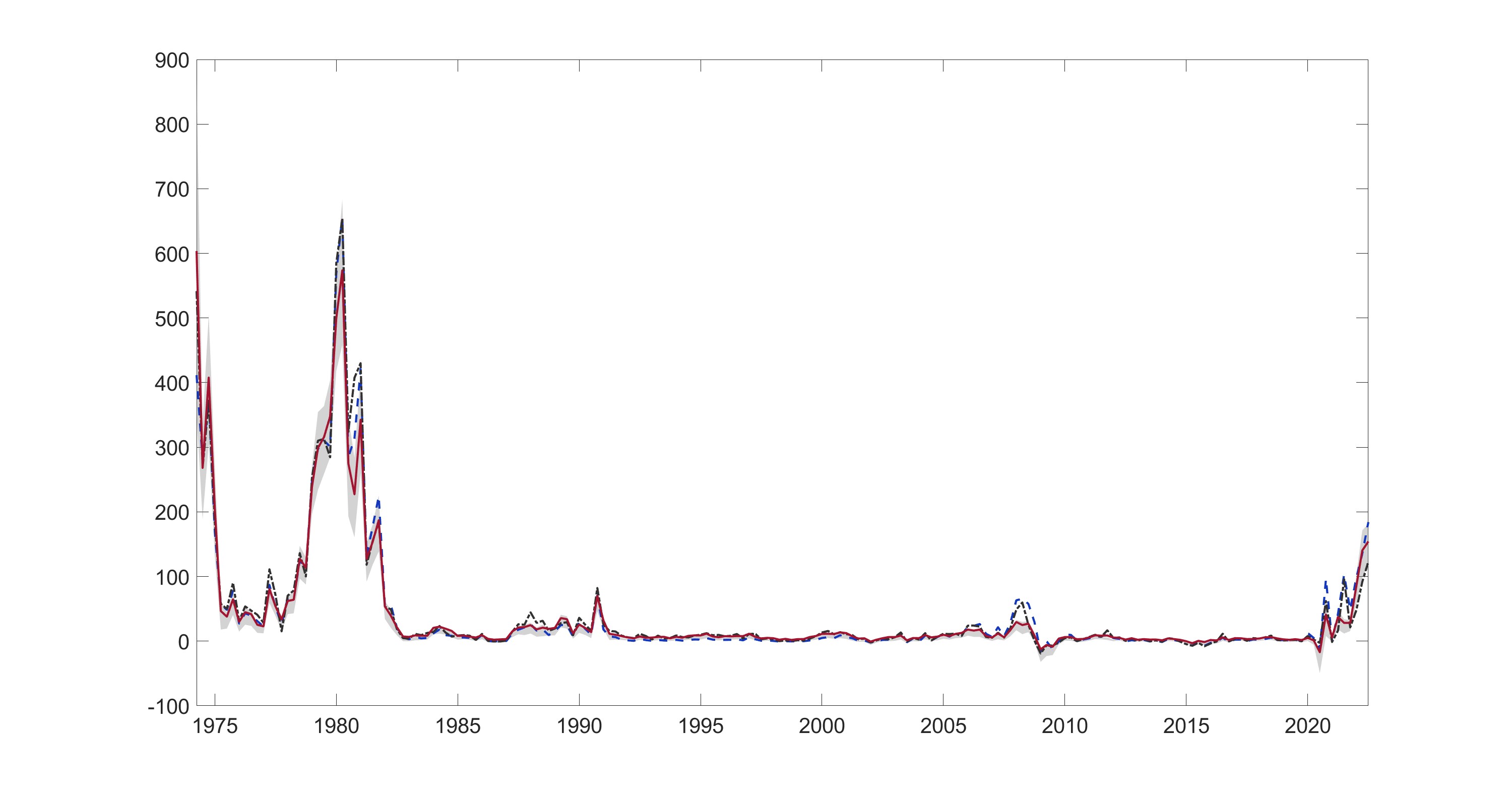}}}
\subcaptionbox{BR for $\alpha=3, \beta=2$ and $h=1$}
{\resizebox*{!}{0.25\textwidth}{\includegraphics[width=0.5\textwidth, trim={6cm 2cm 5cm 2cm},clip]{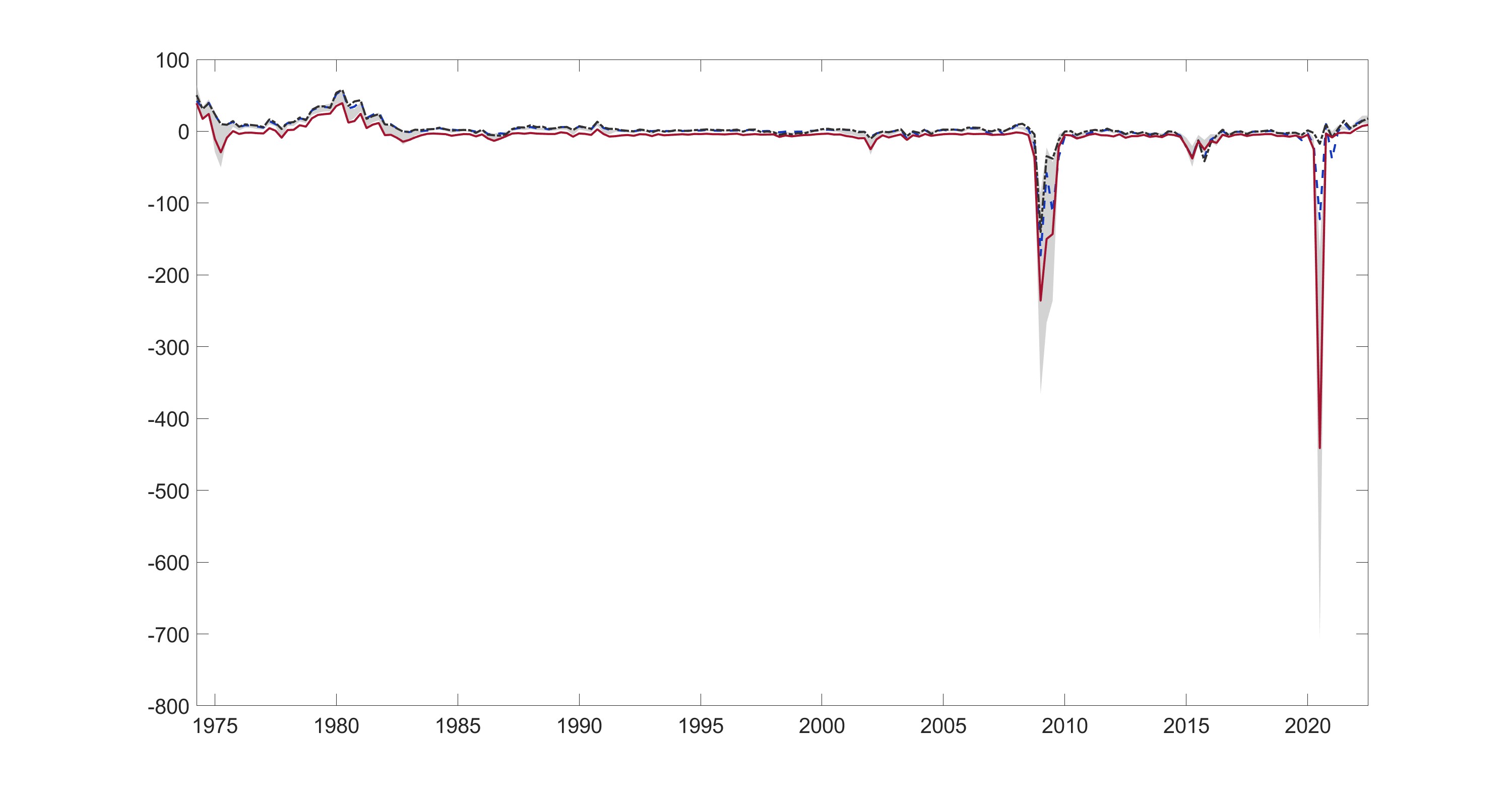}}}
\subcaptionbox{BR for $\alpha=2, \beta=3$ and $h=4$}
{\resizebox*{!}{0.25\textwidth}{\includegraphics[width=0.5\textwidth, trim={6cm 2cm 5cm 2cm},clip]{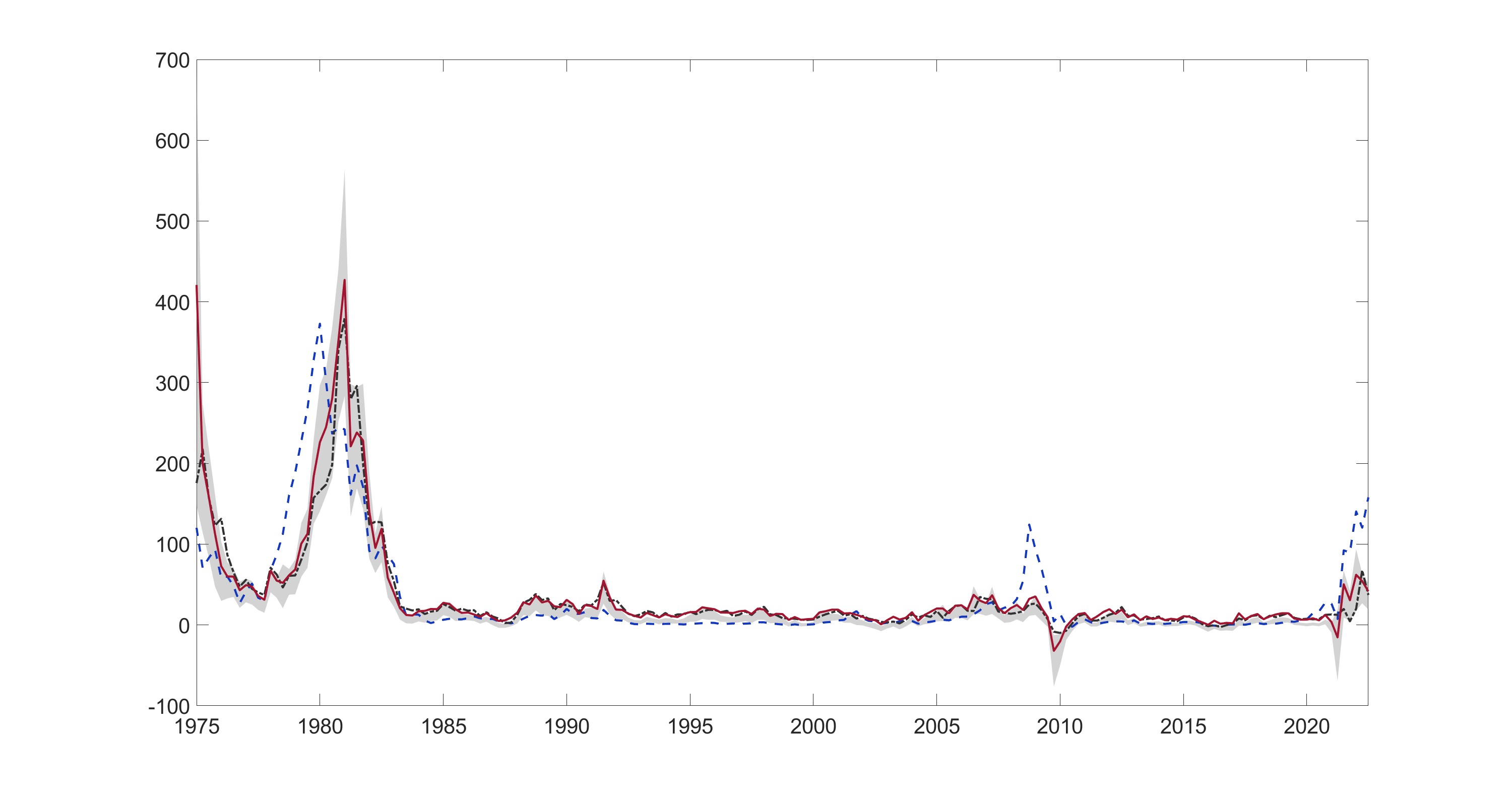}}}
\subcaptionbox{BR for $\alpha=3, \beta=2$ and $h=4$}
{\resizebox*{!}{0.25\textwidth}{\includegraphics[width=0.5\textwidth, trim={6cm 2cm 5cm 2cm},clip]{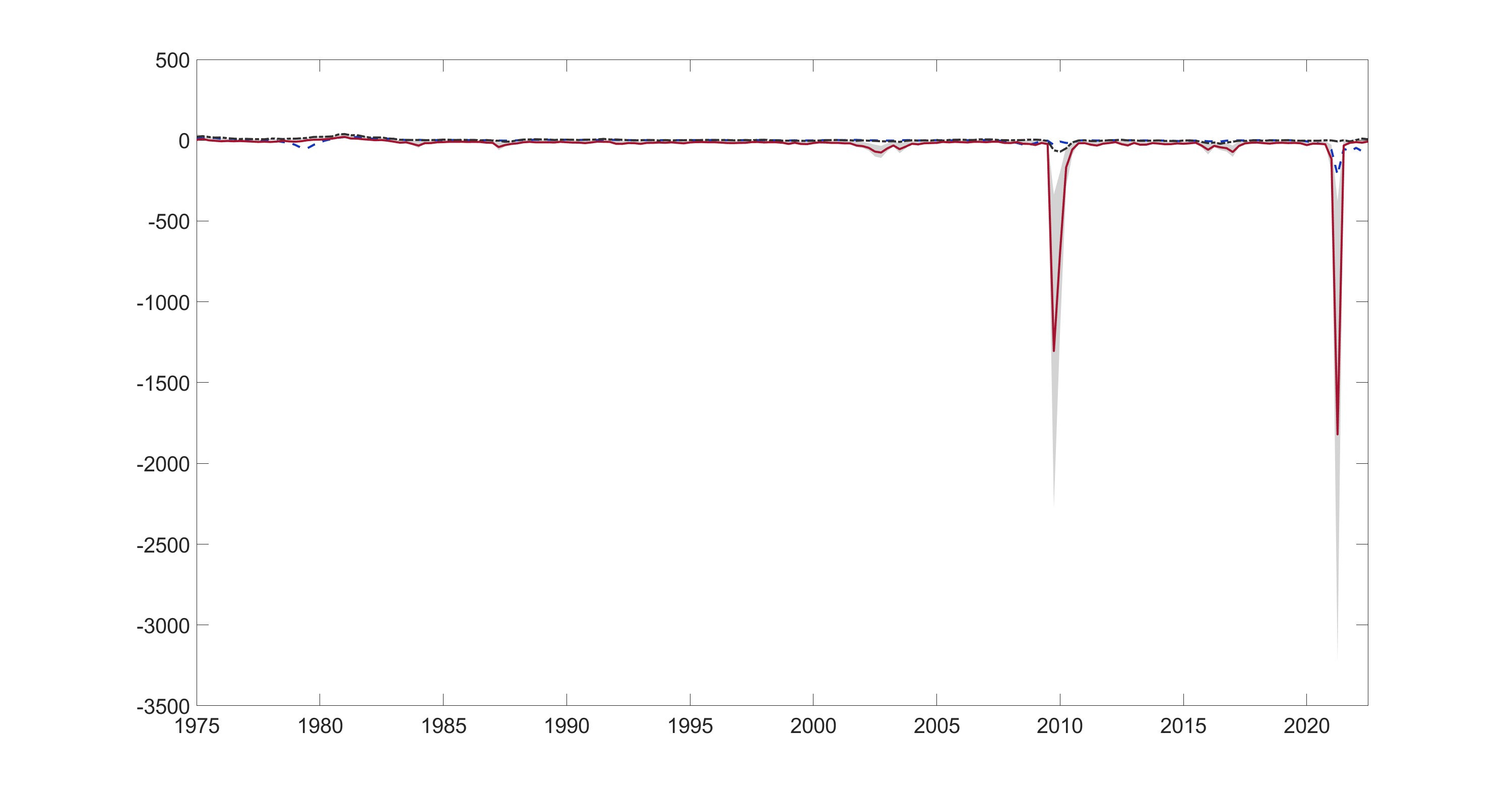}}}
\begin{flushleft}
\footnotesize \singlespacing \textit{Notes: The solid red line indicates the MCMC estimate over the full sample obtained with the density regression model. The shaded area is the corresponding 86\% credible interval. The dashed blue (dot-dashed black) line show the corresponding estimates computed with a model with stochastic volatility and constant volatility, respectively.}
\end{flushleft}
\end{figure}

This section repeats the analysis in the main body of the paper for asymmetric risk measures, that weight deflation risks more strongly than inflation risks and vice versa. These cases are illustrative, because central banks might e.g. be particularly cautious regarding deflationary pressure when interest rates are constrained by the effective lower bound. Similarly, central banks might be more concerned about further inflationary pressures, when inflation is already high and deflation risks seem generally muted. To illustrate such cases, Figure \ref{fig:risk_measures2} contains the results for $\alpha=2$ and $\beta=3$ in the left column and the results for $\alpha=3$ and $\beta=2$ in the right column. 

For preferences tilted towards inflation risks, the dynamics of the resulting risk measure are similar to $\alpha=\beta=2$, albeit at an inflated scale and with deflation risks subdued at both forecast horizons. For $h=1$ differences to the benchmark models emerge before the financial crisis and towards the end of the sample. At $h=4$, these differences are again exaggerated. In particular, the stochastic volatility model suggest inflation risks to pick up earlier in the late 70s and more extreme inflation risk before the financial crisis and at the end of the sample.  

In the case of preferences tilted towards deflationary pressures, the general image is inverted. Inflationary pressures become subdued and deflationary pressures emphasized. For $h=1$ this gives rise to lower inflation risks during the Volcker-era and at the end of the sample and more extreme deflation risks during recessions. Compared to the model proposed in this paper, the benchmark models generally indicate higher inflation risks and lower deflation risks during recessions. In case of $h=4$, inflationary pressures become even more subdued, leaving the overall balance of risk dominated by deflation risk. In contrast to before, the constant parameter model now indicates no deflation risk during the recession that followed the pandemic and the stochastic volatility model implausibly suggests deflation risks during the post-Covid high inflation period. This likely results from the assumed symmetry of the inflation distribution. Unlike the density regression model, which allows for skewness, an increase in volatility increases probability mass equally to the left and right of the conditional mean. In extremely uncertain environments, the AR-SV model hence rarely predicts inflation/deflation risks exclusively.

\subsection{An Inflation Risk Shadow Rate}\label{supp:shadow}

True inflation risk is inherently unobserved, which poses challenges for empirically validating the estimated risk measures. However, to offer some illustrative evidence that the estimated dynamics indeed capture useful information, this section constructs an ``inflation risk shadow rate''. Assuming that the infinite horizon problem can be broken up into period-by-period problems, the implied policy rate is given by

\begin{equation}
\underset{i_t}{\text{argmin}} \:\: \mathbb{E}(L_t),\text{ for } t=1,2,\dots.
\end{equation}

To keep the exercise in line with the FED's policy objective, the loss function is set to $\mathbb{E}(L_t)=\sum_{h=1}^4 \delta^{h-1} BR_{t+h}$ for $\delta=(0,1)$, implying that the central bank minimizes the quarterly balance of risk accumulated over the next year, discounted with discount factor $\delta$. Crucially, in case of $\alpha=\beta=2$, $w=0.5$, and $\underline{\pi}=\bar{\pi}$, $BR_t=0.5(\pi-\bar{\pi})^2$, minimizing the expected loss entails minimizing the variance of inflation around its target \citep{kilianQuantifyingRiskDeflation2007,svenssonInflationForecastTargeting1997}.

The density regression model indirectly connects the forecast distribution to the interest rate through the mixture kernels and weights. Moreover, the model's output directly provides the variance of the forecast distribution, facilitating a straightforward minimization algorithm. This algorithm repeatedly substitutes values for the interest rate at the current observation while keeping all other variables constant, and then forecasts the variance of the distribution using the mixture model. The process concludes when a value for $i^\star$ is identified that minimizes the objective.


In practice, central banks often avoid extreme changes in the interest rate and instead pursue interest rate smoothing. To emulate this behavior, the loss function is extended to include the volatility of the interest rate itself 

\begin{equation}
\underset{i_t}{\text{argmin}} \:\: w_i\cdot\sum_{h=1}^4 \delta^{h-1} BR_{t+h} + (1-w_i)\cdot (i_t-\bar{i}_t)^2,\text{ for } t=1,2,\dots.
\end{equation}

The weight parameter $w_i$ is calibrated to ensure that the volatility of the implied inflation risk shadow rate and the volatility of the FFR are the same over the sample, leading to $w_i=0.996$ for $\delta=0.95$. In particular, $w_i$ is calibrated using the following steps:

\begin{enumerate}
\item Set up a grid for $w_i\in[0,1]$.
\item Computed the implied policy rate path and its variance over the full sample for each grid point. 
\item Collect all variances for all paths under different grid points and subtract the variance of the FFR.
\item Interpolate the resulting curve. This results in a monotonically increasing curve with a single root.
\item Run a root finding procedure to find the $w_i$ that results in an implied interest rate that has the same variance as the FFR over the full sample. 
\end{enumerate}

\begin{figure}[H]
\centering
\caption{Implied Policy Rate Estimates} \label{fig:interest_rate}
\includegraphics[width=0.65\textwidth, trim={10cm 2cm 10cm 3cm}, clip]{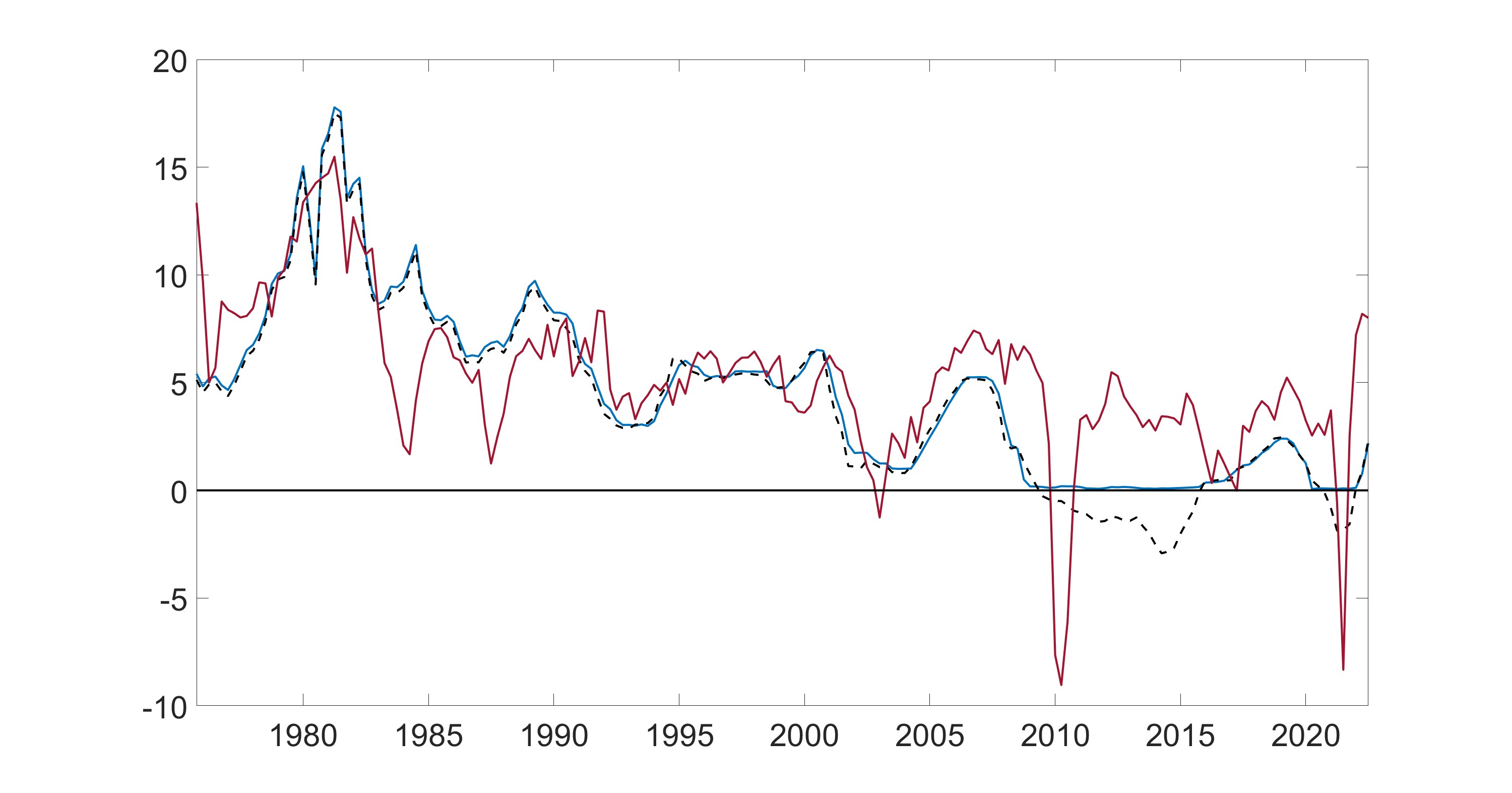}
\begin{flushleft}
\vspace{-.7cm}
\footnotesize \singlespacing \textit{Notes: The blue line indicates the FFR and the dashed black line indicates the WuXia shadow rate. The red line indicates the policy rate with interest rate smoothing that is implied by the risk measure obtained from the density regression model. The weight on smoothing is set such that the volatility of the predicted interest rate and the FFR are equal.}
\end{flushleft}
\end{figure}

The resulting smoothed interest rate (in red) is displayed in \autoref{fig:interest_rate}, alongside the FFR (in blue) and the shadow rate estimate in  \cite{wuMeasuringMacroeconomicImpact2016} (in dashed black). While the interest rates generally exhibit a high degree of correlation (0.71) and track similar dynamics, the smoothed inflation risk shadow rate suggests higher rates at the beginning of the sample, before the financial crisis, and during the subsequent low inflation period compared to the Federal Funds Rate (FFR). However, during the Volcker era and the great moderation, the rate aligns closely with the actual FFR. Notably, the smoothed rate implies deeper interest rate cuts during recessions, such as the financial crisis (-9.0\%) and pandemic (-8.4\%), and more aggressive rate adjustments during the post-pandemic high inflation period (approximately 8.0\%). 

Naturally, this exercise has several limitations. For example, it ignores other policy objectives, the indirect effect of interest rate changes through other economic variables, or the zero lower bound. Consequently, the interpretation should focus solely on the model and inflation risks, refraining from making judgements on optimal rate setting. Nonetheless, despite these constraints,  the risk measurement and modelling framework proposed in this paper implies interest rate movements that are broadly aligned with actual rate setting. In addition, the results suggest that the FED may indeed factor in inflation risk when determining its policy rates, offering avenues for future research.

\subsection{Additional Figures}\label{supp:add_figures}

\begin{figure}[H]
\centering
\caption{Annualized Quarterly Inflation Density, MCMC and VB estimates} \label{app:comparison_MCMC_VB}
\includegraphics[width=0.6\textwidth, trim={5.5cm 2cm 5cm 0cm}, clip]{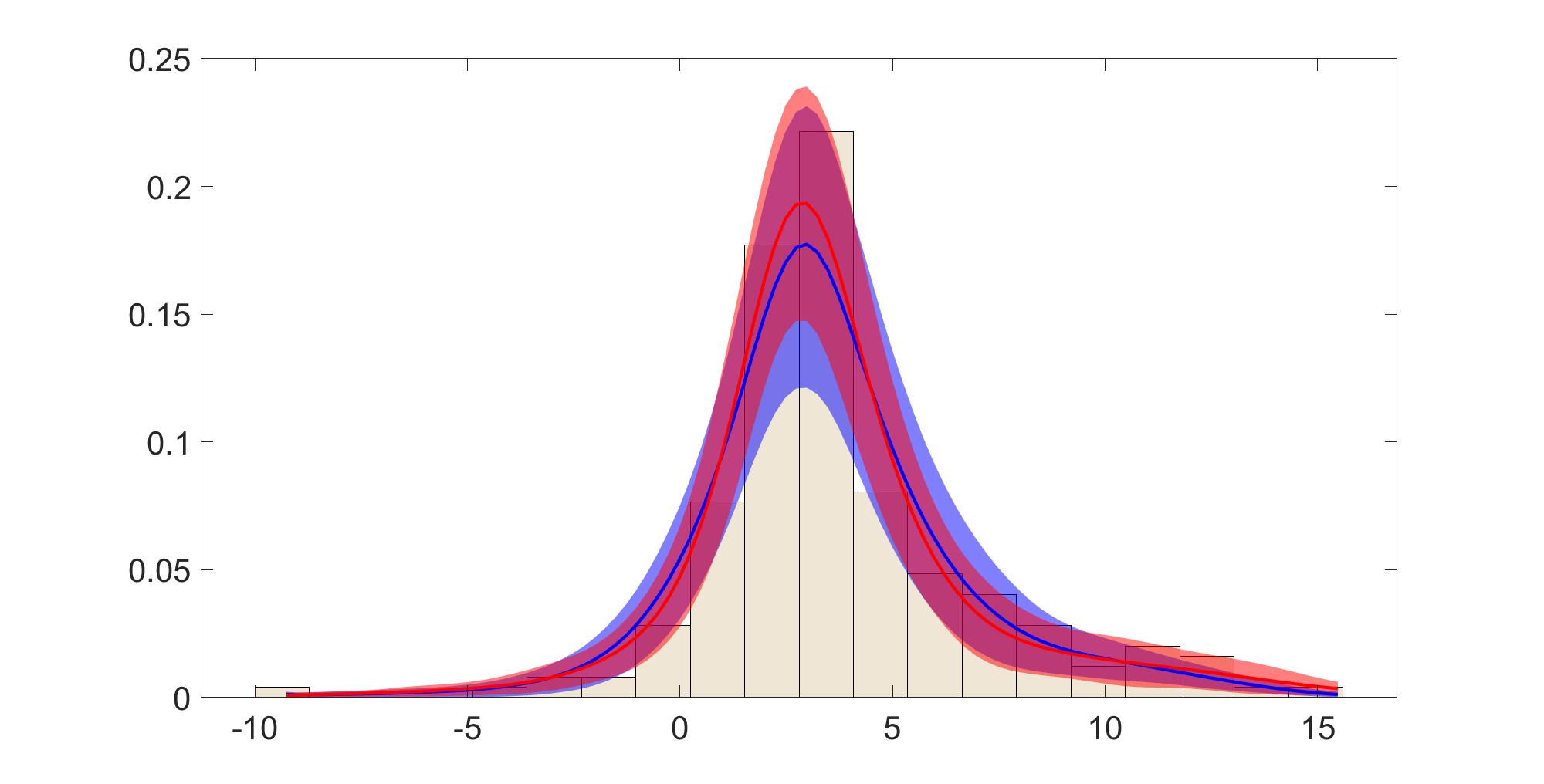}
\begin{flushleft}
\scriptsize \singlespacing \textit{Notes: The histograms display the observed values for annualized quarterly inflation from 1974Q1 to 2022Q3 (beige). The blue (red) colored line and shaded area show the mean density estimate and 68\% confidence bounds for MCMC (VB).}
\end{flushleft}
\end{figure}

 \begin{figure}[h]
\hspace{-4cm}\centering
\caption{Historical Downside and Upside Risk, $h=1$}\label{fig:risk_measures_app}
\subcaptionbox{DR for $\alpha=\beta=0$}
{\resizebox*{!}{0.25\textwidth}{\includegraphics[width=0.5\textwidth, trim={6cm 2cm 5cm 2cm},clip]{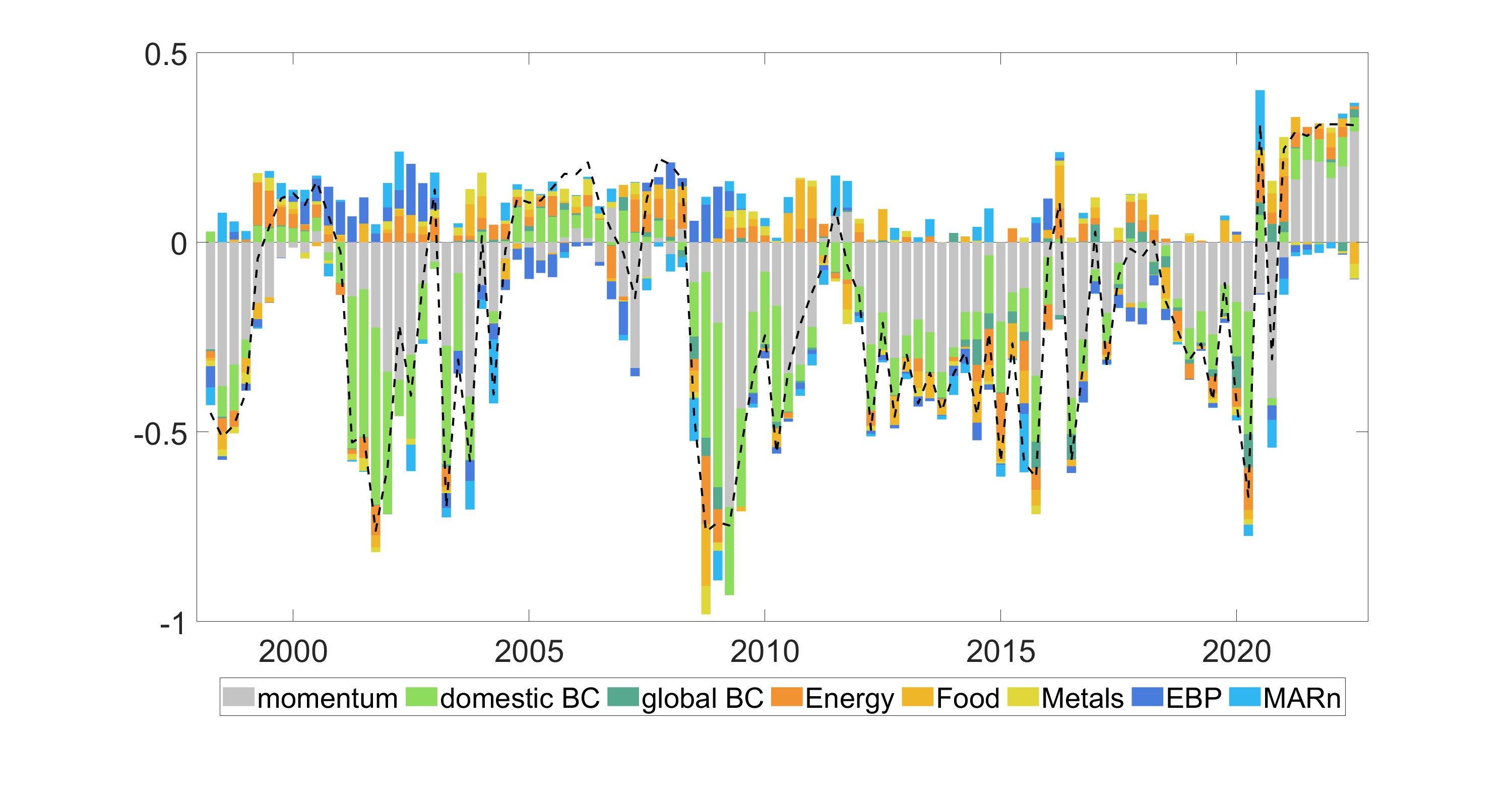}}}
\subcaptionbox{DR for $\alpha=\beta=2$}
{\resizebox*{!}{0.25\textwidth}{\includegraphics[width=0.5\textwidth, trim={6cm 2cm 5cm 2cm},clip]{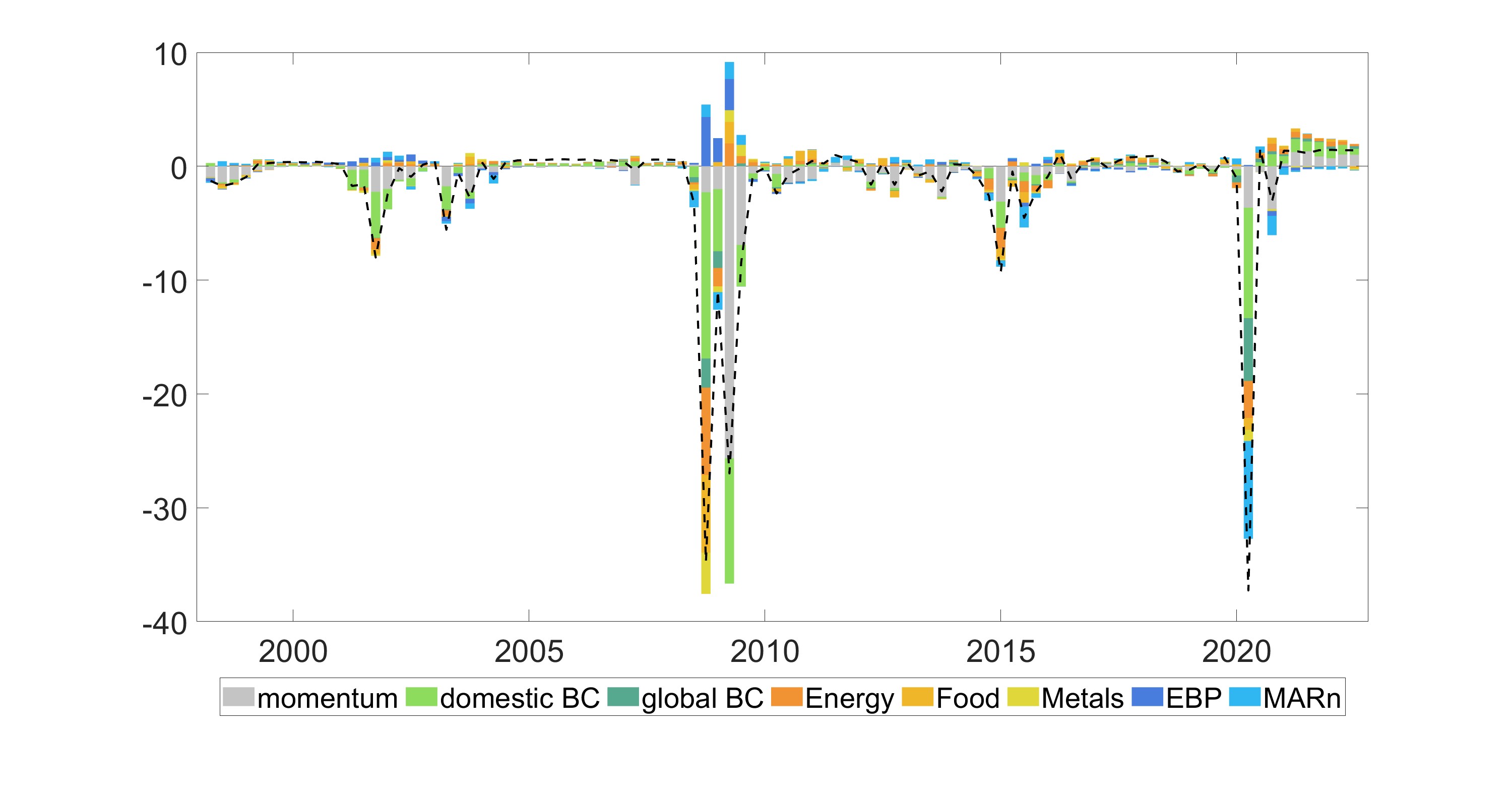}}}
\subcaptionbox{EIR for $\alpha=\beta=0$}
{\resizebox*{!}{0.25\textwidth}{\includegraphics[width=0.5\textwidth, trim={6cm 2cm 5cm 2cm},clip]{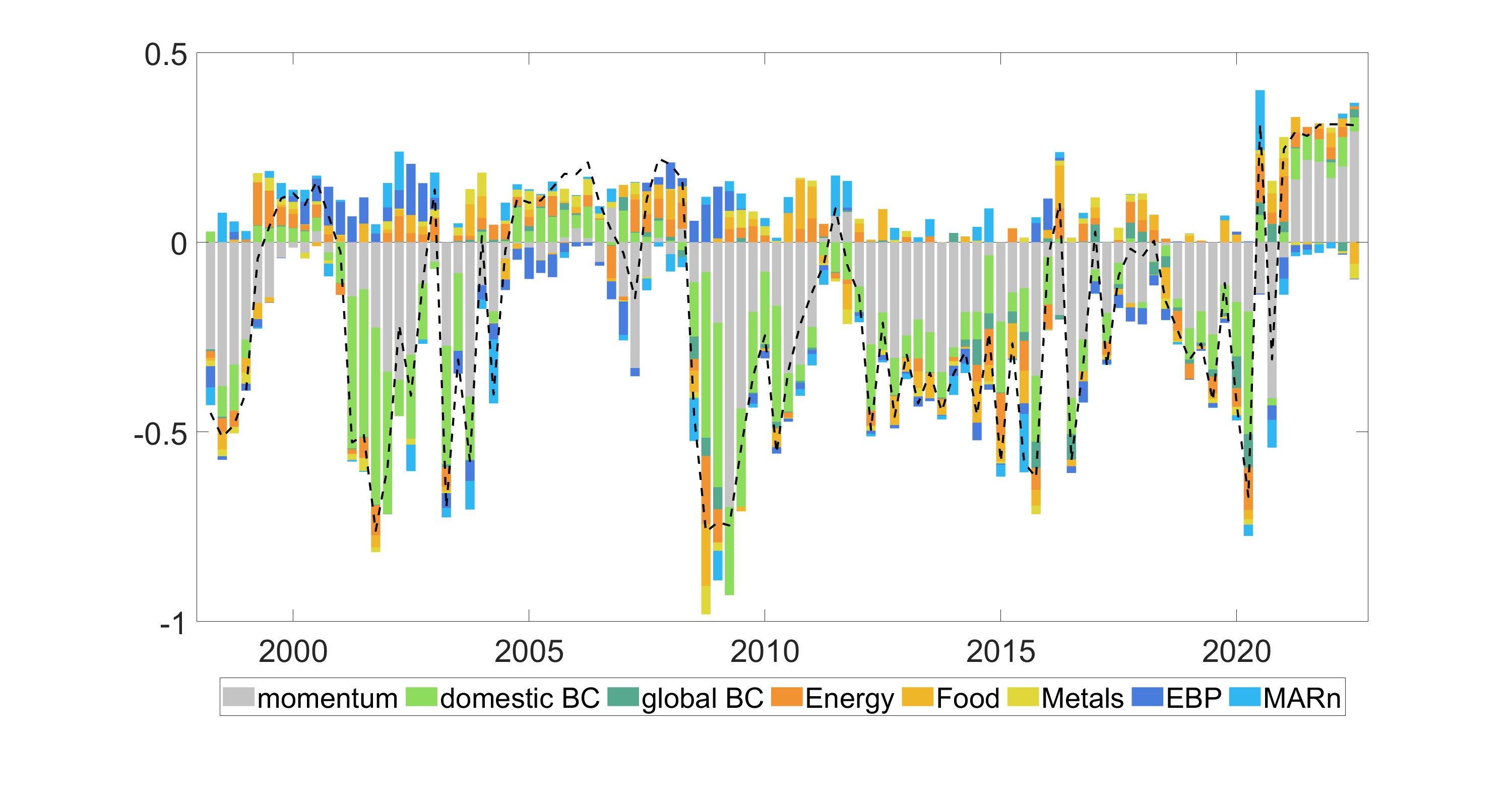}}}
\subcaptionbox{EIR for $\alpha=\beta=2$}
{\resizebox*{!}{0.25\textwidth}{\includegraphics[width=0.5\textwidth, trim={6cm 2cm 5cm 2cm},clip]{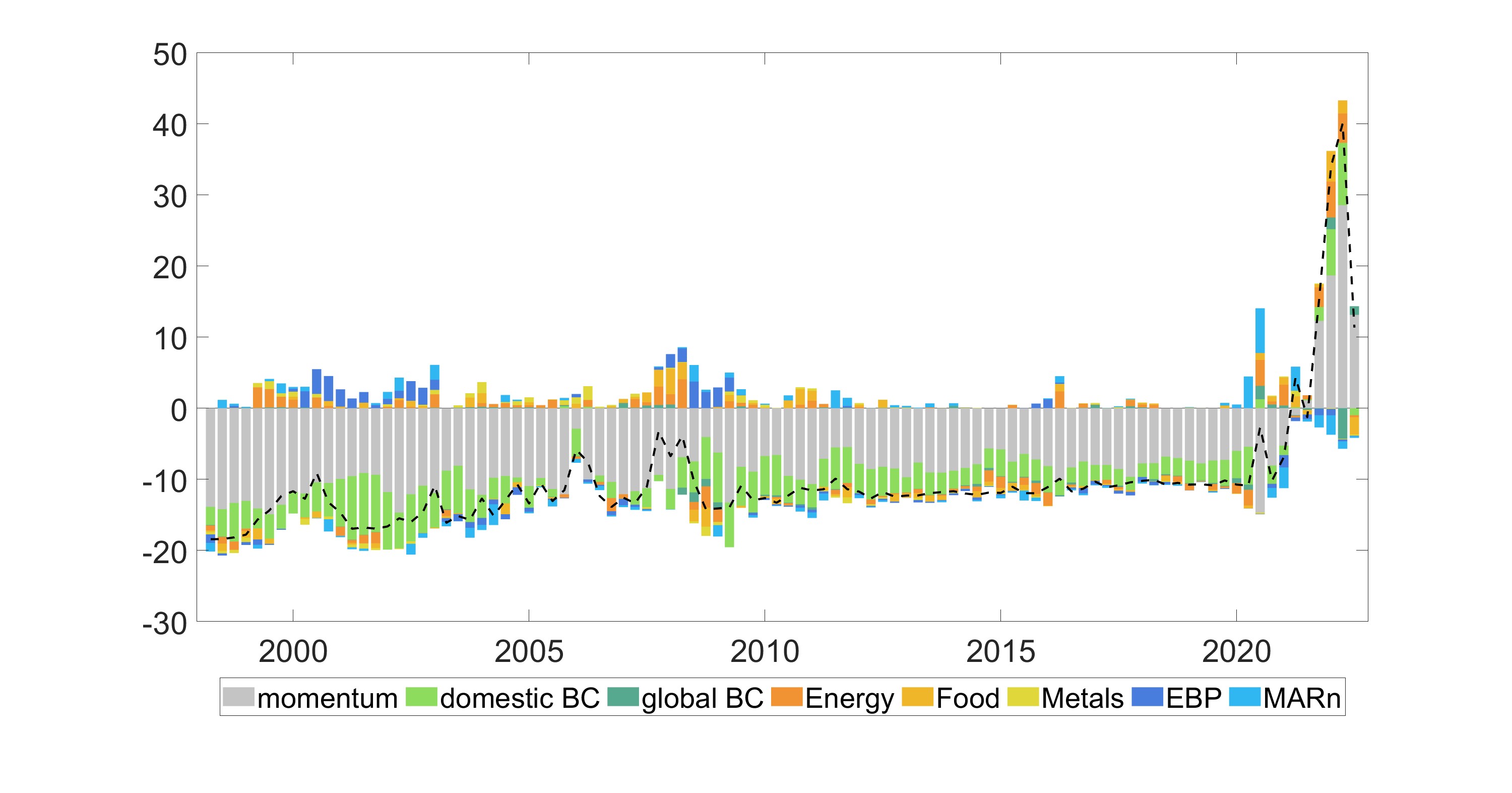}}}
\begin{flushleft}
\scriptsize \singlespacing \textit{Notes: The coloured bars indicate the contribution of the corresponding variable to the balance of risk for the respective parameter setting. The dashed line indicates the observed value for the balance of risk minus the sample average.}
\end{flushleft}
\end{figure} 
 
\begin{figure}[h]
\hspace{-4cm}\centering
\caption{Historical Downside and Upside Risk, $h=4$}\label{fig:risk_measures_app}
\subcaptionbox{DR for $\alpha=\beta=0$}
{\resizebox*{!}{0.25\textwidth}{\includegraphics[width=0.5\textwidth, trim={6cm 2cm 5cm 2cm},clip]{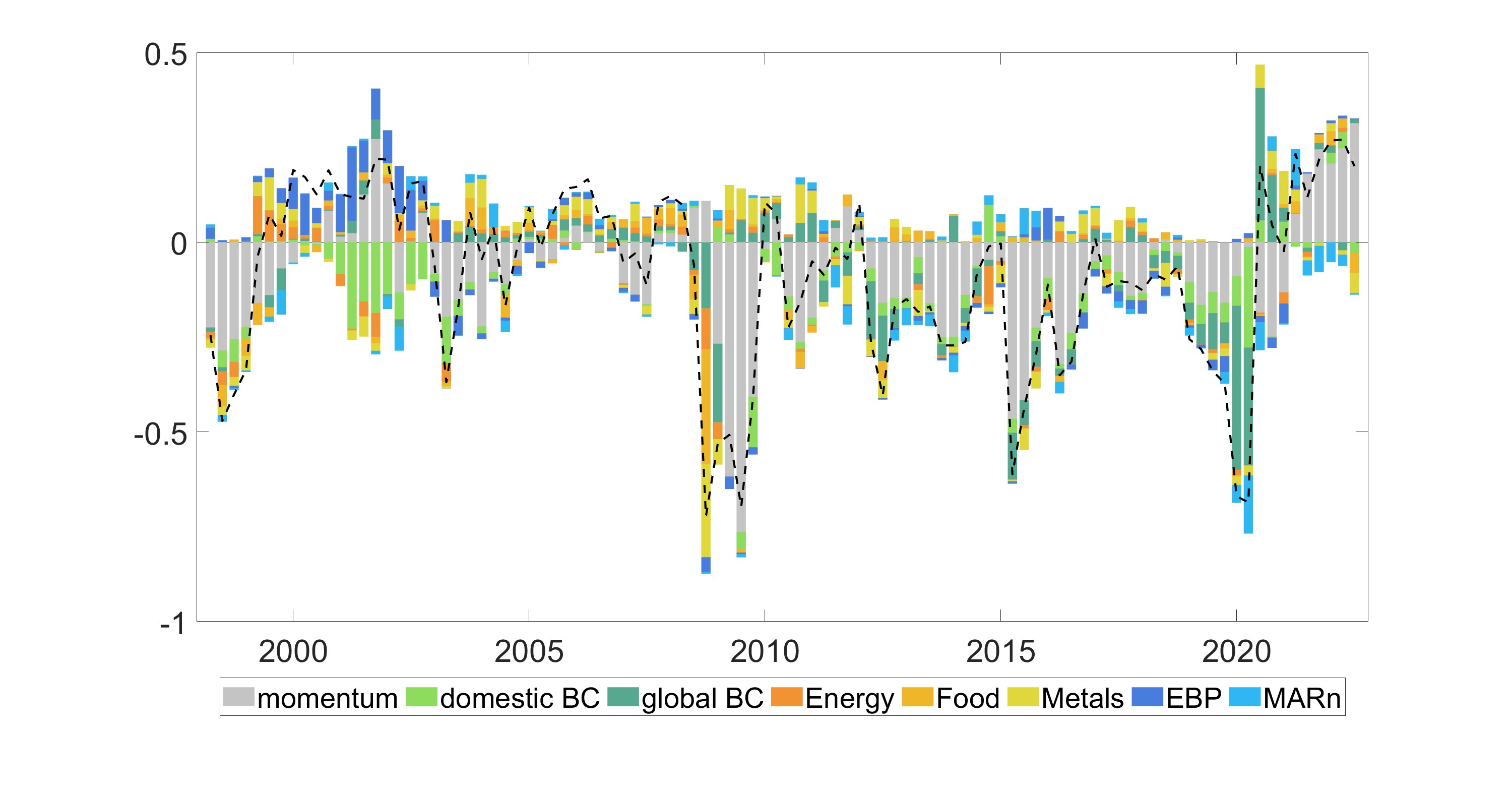}}}
\subcaptionbox{DR for $\alpha=\beta=2$}
{\resizebox*{!}{0.25\textwidth}{\includegraphics[width=0.5\textwidth, trim={6cm 2cm 5cm 2cm},clip]{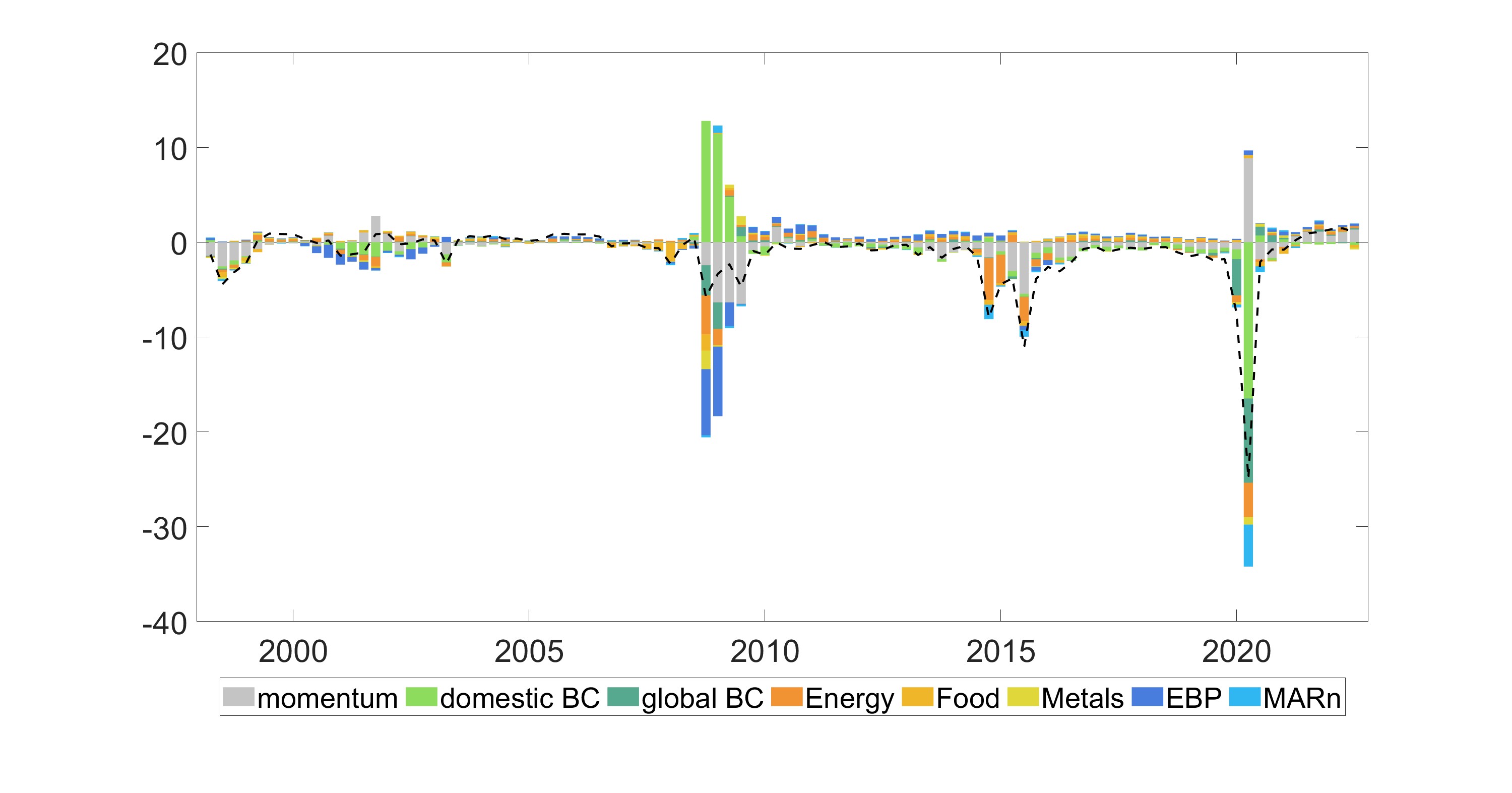}}}
\subcaptionbox{EIR for $\alpha=\beta=0$}
{\resizebox*{!}{0.25\textwidth}{\includegraphics[width=0.5\textwidth, trim={6cm 2cm 5cm 2cm},clip]{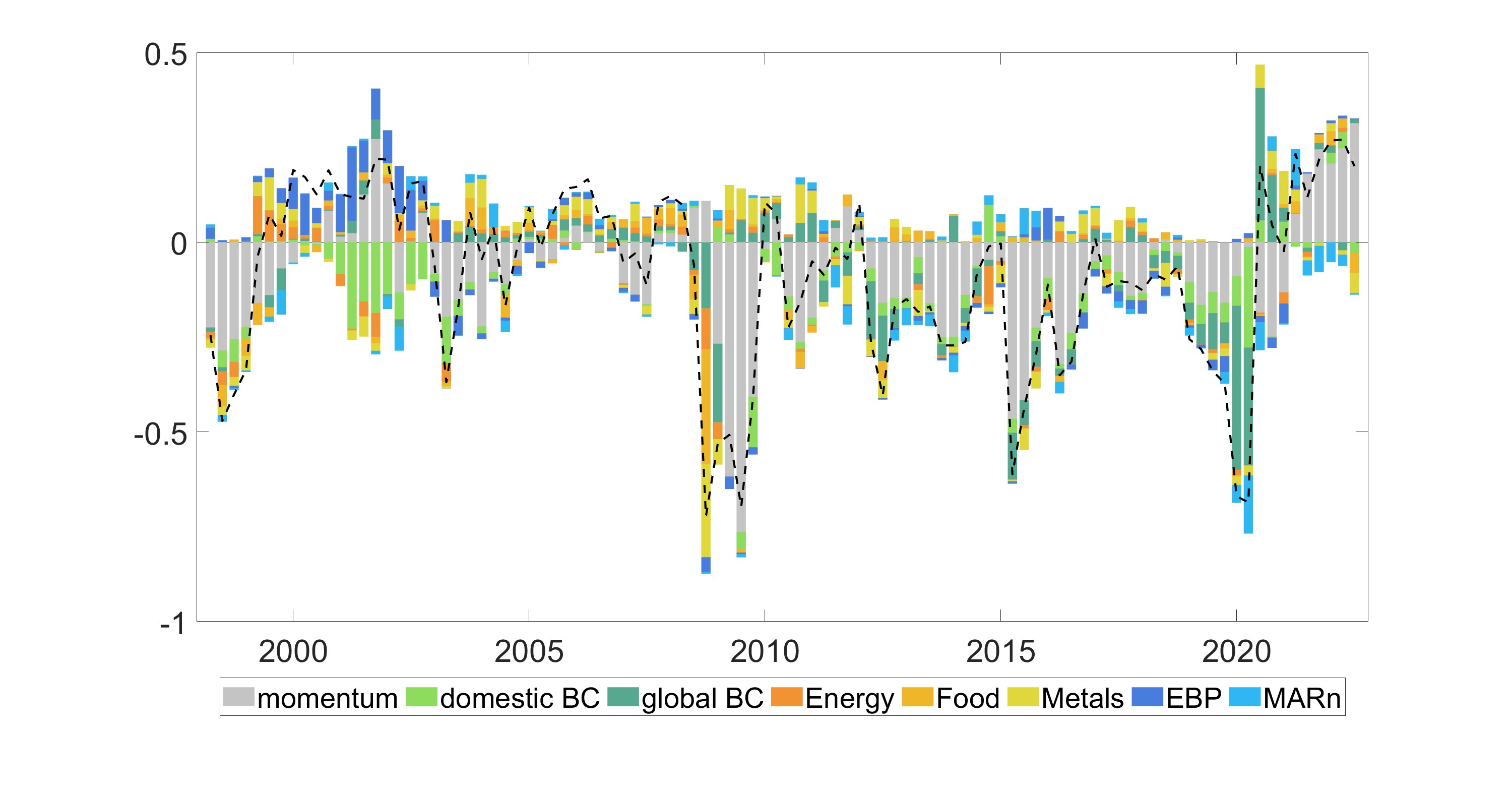}}}
\subcaptionbox{EIR for $\alpha=\beta=2$}
{\resizebox*{!}{0.25\textwidth}{\includegraphics[width=0.5\textwidth, trim={6cm 2cm 5cm 2cm},clip]{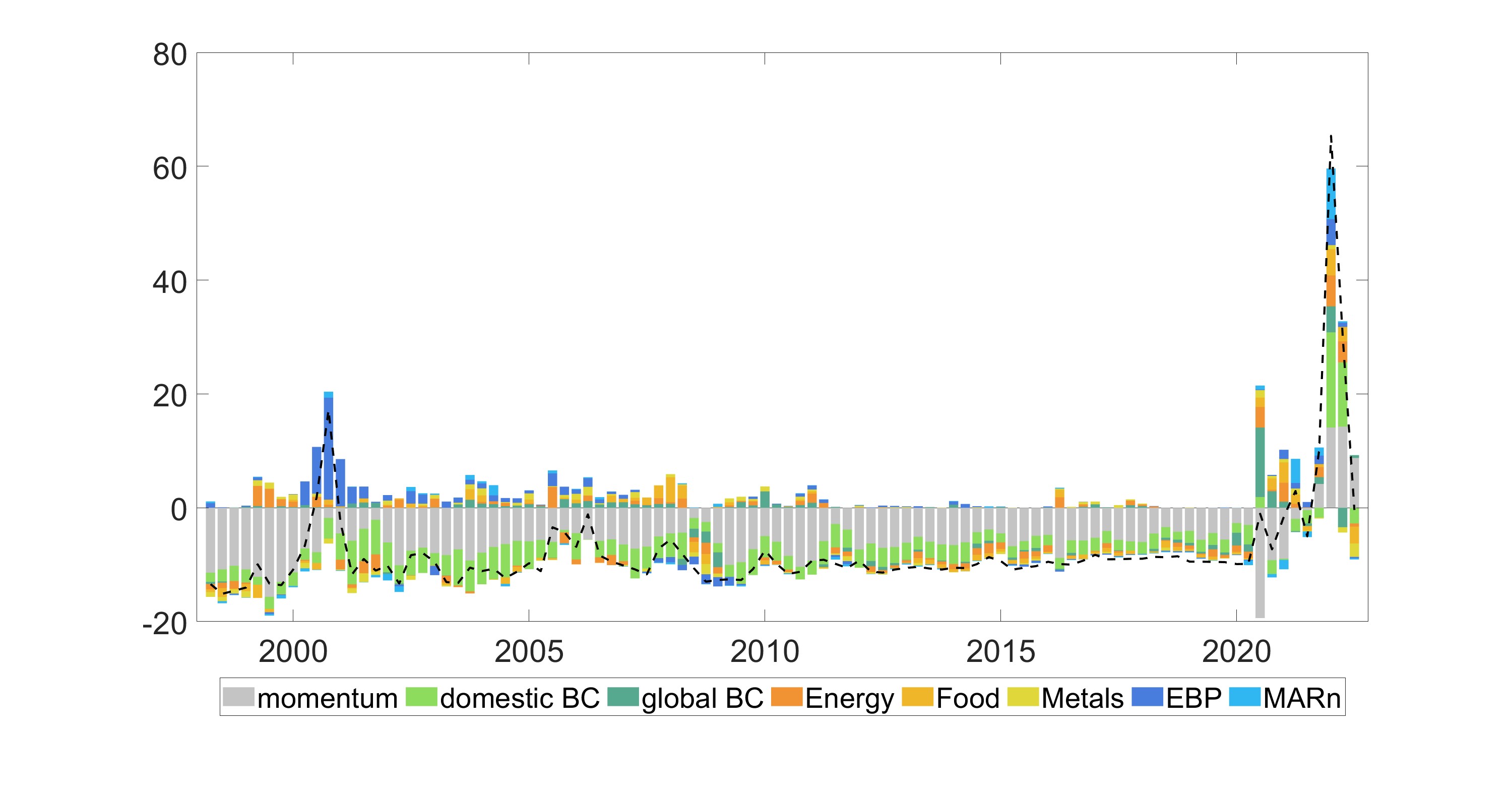}}}
\begin{flushleft}
\scriptsize \singlespacing \textit{Notes: The coloured bars indicate the contribution of the corresponding variable to the balance of risk for the respective parameter setting. The dashed line indicates the observed value for the balance of risk minus the sample average.}
\end{flushleft}
\end{figure}

%
%
%
%

  \clearpage
\subsection{Additional Tables}\label{supp:add_tables}
 \begin{table}[H]
\caption{Absolute RMSE }\label{tab:RMSE_abs}
\centering\resizebox{.7\textwidth}{!}{
\begin{threeparttable}
\begin{tabular}{lccccccc}\hline \hline
$h$ & DR        &     AR    & TVP-AR     &  SV-AR   &  TVPSV-AR &  T-AR  & QR\\\hline
1	&	2.037	&	2.312	&	2.856	&	2.297	&	3.388	&	2.291	&	2.302	\\
2	&	2.222	&	2.691	&	4.869	&	2.490	&	4.483	&	2.586	&	2.482	\\
3	&	2.397	&	2.809	&	2.826	&	2.446	&	5.562	&	2.669	&	2.548	\\
4	&	2.499	&	2.707	&	5.073	&	2.396	&	7.120	&	2.624	&	2.421	\\
5	&	2.483	&	2.560	&	3.477	&	2.392	&	3.989	&	2.440	&	2.360	\\
6	&	2.330	&	2.639	&	5.730	&	2.238	&	5.425	&	2.475	&	2.388	\\
7	&	2.411	&	2.850	&	4.679	&	2.319	&	7.089	&	2.575	&	2.465	\\
8	&	2.461	&	2.943	&	3.386	&	2.343	&	6.275	&	2.597	&	2.461	\\
9	&	2.458	&	2.960	&	3.892	&	2.444	&	5.201	&	2.652	&	2.511	\\
10	&	2.629	&	3.123	&	5.729	&	2.507	&	6.073	&	2.747	&	2.579	\\
11	&	2.797	&	3.212	&	6.919	&	2.718	&	9.953	&	2.766	&	2.652	\\
12	&	2.907	&	3.382	&	5.077	&	2.602	&	5.502	&	2.770	&	2.653	\\
\hline\hline
\end{tabular}
\footnotesize
\begin{tablenotes}
\item The table shows the absolute RMSE for the density regression and the benchmark models for the forecast horizons $h=\left\{ 1,\dots,12 \right\} $. RMSE for the QR model are based on forecasts for the median.
\end{tablenotes}
\end{threeparttable}
}
\end{table}


\begin{table}[H]
\caption{Absolute CRPS }\label{tab:CRPS_abs}
\centering\resizebox{.7\textwidth}{!}{
\begin{threeparttable}
\begin{tabular}{lcccccc}\hline \hline
$h$ & DR        &     AR    & TVP-AR     &  SV-AR   &  TVPSV-AR &  T-AR  \\\hline
1	&	7.590	&	7.692	&	7.913	&	7.587	&	8.646	&	8.377	\\
2	&	7.704	&	7.776	&	8.318	&	7.642	&	8.568	&	8.723	\\
3	&	8.658	&	8.423	&	8.840	&	8.560	&	10.310	&	9.487	\\
4	&	4.846	&	4.547	&	5.187	&	4.826	&	6.454	&	5.703	\\
5	&	1.619	&	1.338	&	2.213	&	1.540	&	3.047	&	2.056	\\
6	&	4.412	&	3.608	&	6.489	&	4.342	&	7.629	&	4.950	\\
7	&	3.682	&	2.806	&	5.152	&	3.678	&	6.915	&	4.097	\\
8	&	1.287	&	0.928	&	2.739	&	1.275	&	4.312	&	1.389	\\
9	&	0.712	&	0.782	&	1.420	&	0.751	&	2.748	&	0.762	\\
10	&	2.758	&	1.747	&	4.074	&	2.800	&	4.651	&	2.912	\\
11	&	2.111	&	1.371	&	3.565	&	2.436	&	5.046	&	2.318	\\
12	&	3.628	&	2.107	&	4.925	&	3.657	&	6.335	&	3.623	\\
\hline\hline
\end{tabular}
\footnotesize
\begin{tablenotes}
\item The table shows the absolute CRPS for the density regression and the AR, TVP-AR, SV-AR, TVPSV-AR, and T-AR for the forecast horizons $h=\left\{ 1,\dots,12 \right\} $. 
\end{tablenotes}
\end{threeparttable}
}
\end{table}


\begin{landscape} 
\begin{table}[H]
\caption{p-values for tests on PITs for all forecast horizons\\}\label{tab:dens}
\centering\resizebox{1.25\textwidth}{!}{
\begin{threeparttable}
\begin{tabular}{lccccccc|ccccccc|ccccccc}\hline \hline
& \multicolumn{3}{c}{Uniformity} & \multicolumn{2}{c}{Identical} & \multicolumn{2}{c}{Independent}& \multicolumn{3}{c}{Uniform} & \multicolumn{2}{c}{Identical} & \multicolumn{2}{c}{Independent} & \multicolumn{3}{c}{Uniform} & \multicolumn{2}{c}{Identical} & \multicolumn{2}{c}{Independent}\\
 & KS & AD & DH &  1$^{st}$ &  2$^{nd}$ &  1$^{st}$ &  2$^{nd}$& KS & AD & DH &  1$^{st}$ &  2$^{nd}$ &  1$^{st}$ &  2$^{nd}$& KS & AD & DH &  1$^{st}$ &  2$^{nd}$ &  1$^{st}$ &  2$^{nd}$\\
 & \multicolumn{7}{c|}{$H=1$}  & \multicolumn{7}{c}{$H=2$} & \multicolumn{7}{c}{$H=3$}\\
DR	&	0.751	&	0.703	&	0.208	&	1.000	&	0.628	&	0.978	&	0.752	&	0.432	&	0.366	&	0.154	&	0.239	&	0.119	&	0.175	&	0.951	&	0.280	&	0.363	&	0.484	&	0.136	&	0.103	&	0.035	&	0.860	\\
AR	&	0.556	&	0.375	&	0.086	&	0.562	&	0.114	&	0.730	&	0.010	&	0.246	&	0.397	&	0.001	&	0.847	&	0.422	&	0.005	&	0.000	&	0.070	&	0.009	&	0.000	&	0.835	&	0.350	&	0.001	&	0.005	\\
TVP-AR	&	0.046	&	0.001	&	0.683	&	0.155	&	0.000	&	0.467	&	0.000	&	0.021	&	0.000	&	0.874	&	0.818	&	0.383	&	0.178	&	0.001	&	0.001	&	0.000	&	0.570	&	0.441	&	0.895	&	0.014	&	0.015	\\
SV-AR	&	0.562	&	0.391	&	0.115	&	1.000	&	1.000	&	0.538	&	0.669	&	0.485	&	0.007	&	0.003	&	0.874	&	0.347	&	0.008	&	0.070	&	0.367	&	0.004	&	0.048	&	0.554	&	0.080	&	0.025	&	0.000	\\
TVPSV-AR	&	0.000	&	0.000	&	0.000	&	1.000	&	1.000	&	0.124	&	0.124	&	0.000	&	0.000	&	0.000	&	0.858	&	0.858	&	0.473	&	0.473	&	0.000	&	0.000	&	0.000	&	0.117	&	0.117	&	0.019	&	0.019	\\
T-AR	&	0.025	&	0.014	&	0.000	&	0.657	&	0.208	&	0.585	&	0.310	&	0.008	&	0.011	&	0.011	&	0.361	&	0.064	&	0.000	&	0.036	&	0.001	&	0.005	&	0.004	&	0.735	&	0.290	&	0.000	&	0.046	\\
QR	&	0.000	&	0.000	&	0.000	&	1.000	&	0.740	&	0.428	&	0.019	&	0.000	&	0.000	&	0.000	&	0.843	&	0.450	&	0.001	&	0.000	&	0.000	&	0.000	&	0.000	&	0.173	&	0.038	&	0.001	&	0.002	\\
[1ex]
 & \multicolumn{7}{c|}{$H=4$}  & \multicolumn{7}{c}{$H=5$} & \multicolumn{7}{c}{$H=6$} \\
DR	&	0.316	&	0.407	&	0.741	&	0.775	&	0.591	&	0.213	&	0.359	&	0.260	&	0.071	&	0.479	&	0.879	&	0.765	&	0.009	&	0.204	&	0.530	&	0.403	&	0.060	&	0.757	&	0.711	&	0.244	&	0.713	\\
AR	&	0.071	&	0.017	&	0.021	&	1.000	&	0.712	&	0.001	&	0.012	&	0.021	&	0.004	&	0.031	&	0.876	&	0.730	&	0.000	&	0.003	&	0.002	&	0.000	&	0.182	&	0.359	&	0.824	&	0.002	&	0.000	\\
TVP-AR	&	0.000	&	0.000	&	0.364	&	0.096	&	0.310	&	0.016	&	0.167	&	0.011	&	0.000	&	0.634	&	0.371	&	0.181	&	0.007	&	0.000	&	0.000	&	0.000	&	0.034	&	0.093	&	0.057	&	0.000	&	0.252	\\
SV-AR	&	0.196	&	0.000	&	0.269	&	0.864	&	0.349	&	0.010	&	0.064	&	0.106	&	0.002	&	0.138	&	0.253	&	0.069	&	0.003	&	0.047	&	0.110	&	0.002	&	0.167	&	0.449	&	0.127	&	0.049	&	0.043	\\
TVPSV-AR	&	0.000	&	0.000	&	0.000	&	0.104	&	0.104	&	0.007	&	0.007	&	0.000	&	0.000	&	0.000	&	0.347	&	0.347	&	0.016	&	0.016	&	0.000	&	0.000	&	0.000	&	1.000	&	1.000	&	0.183	&	0.183	\\
T-AR	&	0.000	&	0.002	&	0.000	&	0.550	&	0.137	&	0.053	&	0.235	&	0.000	&	0.000	&	0.000	&	0.822	&	0.399	&	0.160	&	0.197	&	0.000	&	0.000	&	0.000	&	0.853	&	0.597	&	0.227	&	0.030	\\
QR	&	0.000	&	0.000	&	0.000	&	0.429	&	0.133	&	0.004	&	0.009	&	0.000	&	0.000	&	0.000	&	0.350	&	0.104	&	0.001	&	0.036	&	0.000	&	0.000	&	0.000	&	0.481	&	0.372	&	0.007	&	0.028	\\
[1ex]
 & \multicolumn{7}{c|}{$H=7$}  & \multicolumn{7}{c}{$H=8$} & \multicolumn{7}{c}{$H=9$}\\
DR	&	0.356	&	0.067	&	0.667	&	0.816	&	0.853	&	0.003	&	0.777	&	0.196	&	0.058	&	0.901	&	0.685	&	0.811	&	0.043	&	0.601	&	0.203	&	0.395	&	0.476	&	0.177	&	0.682	&	0.010	&	0.207	\\
AR	&	0.000	&	0.000	&	0.112	&	0.109	&	0.518	&	0.001	&	0.002	&	0.000	&	0.000	&	0.067	&	0.083	&	0.344	&	0.002	&	0.000	&	0.000	&	0.000	&	0.010	&	0.045	&	0.197	&	0.000	&	0.000	\\
TVP-AR	&	0.025	&	0.000	&	0.404	&	0.170	&	0.502	&	0.031	&	0.052	&	0.000	&	0.000	&	0.047	&	0.019	&	0.018	&	0.051	&	0.780	&	0.011	&	0.000	&	0.337	&	0.000	&	0.000	&	0.005	&	0.031	\\
SV-AR	&	0.193	&	0.003	&	0.484	&	0.425	&	0.117	&	0.002	&	0.000	&	0.416	&	0.003	&	0.328	&	0.442	&	0.175	&	0.017	&	0.010	&	0.423	&	0.001	&	0.586	&	0.369	&	0.131	&	0.001	&	0.000	\\
TVPSV-AR	&	0.000	&	0.000	&	0.000	&	0.601	&	0.601	&	0.120	&	0.120	&	0.000	&	0.000	&	0.000	&	0.322	&	0.322	&	0.380	&	0.380	&	0.000	&	0.000	&	0.000	&	0.541	&	0.541	&	0.347	&	0.347	\\
T-AR	&	0.000	&	0.000	&	0.003	&	1.000	&	0.642	&	0.012	&	0.000	&	0.000	&	0.001	&	0.001	&	0.394	&	0.026	&	0.003	&	0.001	&	0.000	&	0.000	&	0.000	&	0.370	&	0.112	&	0.002	&	0.001	\\
QR	&	0.000	&	0.000	&	0.000	&	0.508	&	0.834	&	0.011	&	0.025	&	0.000	&	0.000	&	0.000	&	0.430	&	0.799	&	0.008	&	0.017	&	0.000	&	0.000	&	0.000	&	0.448	&	0.798	&	0.001	&	0.049	\\
[1ex]
 & \multicolumn{7}{c|}{$H=10$}  & \multicolumn{7}{c}{$H=11$} & \multicolumn{7}{c}{$H=12$}\\
DR	&	0.663	&	0.517	&	0.110	&	0.398	&	0.253	&	0.001	&	0.001	&	0.269	&	0.366	&	0.420	&	0.118	&	0.090	&	0.002	&	0.000	&	0.497	&	0.408	&	0.298	&	0.112	&	0.046	&	0.000	&	0.000	\\
AR	&	0.000	&	0.000	&	0.000	&	0.014	&	0.038	&	0.000	&	0.000	&	0.000	&	0.000	&	0.002	&	0.000	&	0.000	&	0.000	&	0.000	&	0.000	&	0.000	&	0.005	&	0.000	&	0.000	&	0.000	&	0.000	\\
TVP-AR	&	0.057	&	0.000	&	0.618	&	0.122	&	0.178	&	0.000	&	0.436	&	0.001	&	0.000	&	0.108	&	0.000	&	0.000	&	0.010	&	0.241	&	0.000	&	0.000	&	0.028	&	0.000	&	0.000	&	0.000	&	0.000	\\
SV-AR	&	0.362	&	0.000	&	0.344	&	0.491	&	0.389	&	0.002	&	0.000	&	0.241	&	0.001	&	0.457	&	0.477	&	0.251	&	0.002	&	0.007	&	0.597	&	0.001	&	0.234	&	0.190	&	0.104	&	0.015	&	0.001	\\
TVPSV-AR	&	0.000	&	0.000	&	0.000	&	0.000	&	0.000	&	0.000	&	0.000	&	0.000	&	0.000	&	0.000	&	0.144	&	0.144	&	0.757	&	0.757	&	0.000	&	0.000	&	0.000	&	0.000	&	0.000	&	0.293	&	0.293	\\
T-AR	&	0.000	&	0.000	&	0.000	&	0.161	&	0.054	&	0.000	&	0.000	&	0.000	&	0.000	&	0.000	&	0.044	&	0.021	&	0.000	&	0.000	&	0.000	&	0.000	&	0.001	&	0.014	&	0.016	&	0.000	&	0.000	\\
QR	&	0.001	&	0.000	&	0.000	&	0.665	&	0.511	&	0.000	&	0.004	&	0.000	&	0.000	&	0.000	&	0.343	&	0.205	&	0.000	&	0.000	&	0.000	&	0.000	&	0.000	&	0.162	&	0.058	&	0.000	&	0.000	\\
\hline\hline
\end{tabular}
\footnotesize
\begin{tablenotes}
\item The table contains the p-values for several test statistics: KS=Kolmogorov-Smirnov, AD=Anderson-Darling, DH=Doornik-Hansen, Andrews, Ljung-Box. For the Andrews and Ljung-Box test, 1$^{st}$ and 2$^{nd}$ denote the results for the first and second moment, respectively. $H$ indicates the forecast horizon. Bold values mark p-values greater than 5\%.
\end{tablenotes}
\end{threeparttable}
}
\end{table}  
\end{landscape}

\end{document}